\documentclass[12pt,openany,onecolumn]{book}
\usepackage{fancyheadings}
\usepackage{fullpage}
\usepackage{amsmath}
\usepackage{amsthm} 
\usepackage{algorithm}

\ifx\pdftexversion\undefined
  \usepackage[colorlinks,linkcolor=black,filecolor=black,
  citecolor=black,urlcolor=black,pdfstartview=FitH]{hyperref}
\else
  \usepackage[colorlinks,linkcolor=blue,filecolor=blue,
  citecolor=blue,urlcolor=blue,pdfstartview=FitH]{hyperref}
\fi

\usepackage{epsfig}
\usepackage{graphicx}
\usepackage{cite}
\usepackage{amsmath}
\usepackage{amssymb}
\usepackage{cancel}

\newcommand{\comment}[1]  {}
\def\BE{\begin{equation}}
\def\EE{\end{equation}}
\def\BEA{\begin{eqnarray}}
\def\EEA{\end{eqnarray}}

\newcommand{\pdpd}[2]{\frac{\partial ^{2} #1}{\partial #2 ^{2}}}

\DeclareMathOperator*{\argmax}{arg\,max}
\newtheorem{thm}{Theorem}
\newtheorem{lem}[thm]{Lemma}
\newtheorem{alg}{Algorithm}
\newtheorem{prop}[thm]{Proposition}
\newtheorem{defn}[thm]{Definition}
\newtheorem{assm}[thm]{Assumption}

\newtheorem{corol}[thm]{Corollary}
\newtheorem{remark}[thm]{Remark}

\newtheorem{lemma}{Lemma}
\newcommand{\BEAS}{\begin{eqnarray*}}
\newcommand{\EEAS}{\end{eqnarray*}}
\newcommand{\BEQ}{\begin{equation}}
\newcommand{\EEQ}{\end{equation}}
\newcommand{\reals}{{\mbox{\bf R}}}
\newcommand{\dom}{\mathop{\bf dom}}
\newcommand{\ones}{\mathbf 1}

\newcommand{\diag}{\mathop{\bf diag}}
\newcommand{\R}{\mathbb{R}}

\newcommand\ie{{\textsl{i.e.\,}}}
\newcommand\eg{{\textsl{e.g.\,}}}
\newcommand\etal{{\textsl{et al.\,}}}
\newcommand\G{{\mathcal{G}}}
\newcommand\vb{{\bf b}}
\newcommand\vc{{\bf c}}

\newcommand\ve{{\bf e}}

\newcommand\vg{{\bf g}}
\newcommand\vh{{\bf h}}

\newcommand\vk{{\bf k}}

\newcommand\vn{{\bf n}}

\newcommand\vr{{\bf r}}

\newcommand\vw{{\bf w}}
\newcommand\vx{{\bf x}}
\newcommand\vy{{\bf y}}
\newcommand\vz{{\bf z}}
\newcommand\mA{{\bf A}} 
\newcommand\mB{{\bf B}}
\newcommand\mC{{\bf C}}
\newcommand\mD{{\bf D}}

\newcommand\mF{{\bf F}}

\newcommand\mI{{\bf I}}
\newcommand\mJ{{\bf J}}
\newcommand\mK{{\bf K}}

\newcommand\mM{{\bf M}}

\newcommand\mP{{\bf P}}

\newcommand\mR{{\bf R}}
\newcommand\mS{{\bf S}}

\newcommand\mV{{\bf V}}

\newcommand\mX{{\bf X}}

\newcommand\mZ{{\bf Z}}

\def\ceil#1{\lceil #1 \rceil}

\newcommand{\ignore}[1]{}

\begin{document}

\title{\bf\huge  Gaussian Belief Propagation:\\ Theory and Application}

\author{Thesis for the degree of\\\\
DOCTOR~~of~~PHILOSOPHY\\\\
by\\\\
{\bf Danny Bickson}
\\\\
\\\\
{\sc submitted to the senate of} \\
{\sc \Large The Hebrew University of Jerusalem} }
\date{$1^{st}$ Version: October 2008. \\
$2^{nd}$ Revision: May 2009.}
\renewcommand{\thepage}{\roman{page}}
\maketitle \thispagestyle{empty}
%




\section*{This work was carried out under the supervision of\\
\textit{Prof. Danny Dolev and Prof. Dahlia Malkhi }}
\thispagestyle{fancy}
\clearpage

\centerline{\LARGE Acknowledgements }
\vspace{5mm}
I would first like to thank my advisors, Prof.\! Danny Dolev and Prof. \!Dahlia Malkhi.
Danny Dolev encouraged me to follow this interesting research direction, had infinite
time to meet and our brainstorming sessions where always valuable and enjoyable.
Dahlia Malkhi encouraged me to do a Ph.D., worked with me closely at the first part of my Ph.D.,
and was always energetic and inspiring. My time as in Intern in Microsoft Research, Silicon Valley, will never be forgotten, in the period  where my research skills where immensely improved.

I would like to thank Prof. Yair Weiss, for teaching highly interesting courses and introducing me
to the graphical models world. Also for continuous support in answering millions of questions.

Prof.\! Scott Kirkpatrick introduced me to the world of statistical physics, mainly using the Evergrow
project. It was a pleasure working with him, specifically watching his superb management skills
which could defeat every obstacle. In this project, I've also met Prof.\! Erik Aurell from SICS/KTH
and we had numerous interesting discussions about Gaussians, the bread-and-butter of statistical physics.

I am lucky that I had the opportunity to work with Dr.\! Ori Shental from USCD. Ori introduced me into the world
of information theory and together we had a fruitful joint research.

Further thanks to Prof.~\!Yuval Shavitt from Tel Aviv university, for serving in my Ph.D. committee, and
for fruitful discussions.

Support vector regression work was done when I was intern in IBM Research Haifa Lab. Thanks to Dr. \!Elad Yom-Tov and Dr. \!Oded Margalit for their encouragement and for our enjoyable joint work.
The linear programming and Kalman filter work was done
with the great encouragement of Dr. Gidon Gershinsky from IBM\ Haifa Reseach Lab. 

I thank Prof. Andrea Montanari for sharing his multiuser detection code.

I would like to thank Dr.\! Misha Chetkov and Dr.\! Jason K. Johnson for inviting me to visit Los Alamos National Lab, which resulted in the convergence fix results, reported in this work.

Further encouragement I  got from Prof.\! Stephen Boyd, Stanford University.

Finally I would like to thank my wife Ravit, for her continuous support.

\clearpage

\section*{Abstract}
\thispagestyle{fancy}
The canonical problem of solving a system of linear equations
arises in numerous contexts in information theory, communication
theory, and related fields. In this contribution, we develop a
solution based upon Gaussian belief propagation (GaBP) that does
not involve direct matrix inversion. The iterative nature of our
approach allows for a distributed message-passing implementation
of the solution algorithm. In the first part of this thesis, we
address the properties of the GaBP solver. We characterize the rate of convergence,
enhance its message-passing efficiency by introducing a broadcast version, discuss
its relation to classical solution methods including numerical
examples. We present a new method for forcing the GaBP algorithm to converge to the correct solution
for  arbitrary column dependent matrices.

In the second part we give five applications to
illustrate the applicability of the GaBP algorithm to very large
computer networks: Peer-to-Peer rating, linear detection, distributed computation of support vector regression, efficient computation of Kalman filter and  distributed linear programming. Using
extensive simulations on up to 1,024 CPUs in parallel using IBM
Bluegene supercomputer we demonstrate the attractiveness and
applicability of the GaBP algorithm, using real network
topologies with up to millions of nodes and hundreds of millions
of communication links. We further relate to several other
algorithms and explore their connection to the GaBP algorithm.

\thispagestyle{fancy}
\tableofcontents

\clearpage

\pagenumbering{arabic}
\setcounter{page}{1}

\chapter{Introduction}
Solving a linear system of equations \mbox{$\mA\vx=\vb$} is one of
the most fundamental problems in algebra, with countless
applications in the mathematical sciences and engineering.
In this thesis, we propose an efficient distributed iterative algorithm
for solving systems of linear equations. 

The problem of solving a linear system of equations is described as follows.
Given an observation vector \mbox{$\vb\in\mathbb{R}^{m}$} and the data
matrix \mbox{$\mA\in\mathbb{R}^{m\times n}$} ($m\ge
n\in\mathbb{Z}$), a unique solution,
\mbox{$\vx=\vx^{\ast}\in\mathbb{R}^{n}$}, exists if and only if
the data matrix $\mA$ has full column rank. Assuming a nonsingular matrix $\mA$, the system of equations
can be solved either directly or in an iterative manner. Direct
matrix inversion methods, such as Gaussian elimination (LU
factorization,~\cite{BibDB:BookMatrix}-Ch. 3) or band Cholesky
factorization (\cite{BibDB:BookMatrix}-Ch. 4), find the solution
with a finite number of operations, typically, for a dense
$n\times n$ matrix, of the order of $n^{3}$. The former is
particularly effective for systems with unstructured dense data
matrices, while the latter is typically used for structured dense
systems.

Iterative methods~\cite{BibDB:BookAxelsson} are inherently
simpler, requiring only additions and multiplications, and have
the further advantage that they can exploit the sparsity of the
matrix $\mA$ to reduce the computational complexity as well as the
algorithmic storage requirements~\cite{BibDB:BookSaad}. By
comparison, for large, sparse and amorphous data matrices, the
direct methods are impractical due to the need for excessive
matrix reordering operations. 

The main drawback of the iterative approaches is that, under
certain conditions, they converge only asymptotically to the exact
solution $\vx^{\ast}$~\cite{BibDB:BookAxelsson}. Thus, there is
the risk that they may converge slowly, or not at all. In
practice, however, it has been found that they often converge to
the exact solution or a good approximation after a relatively
small number of iterations.

A powerful and efficient iterative algorithm, belief propagation
(BP,~\cite{BibDB:BookPearl}), also known as the sum-product
algorithm, has been very successfully used to solve, either
exactly or approximately, inference problems in probabilistic
graphical models~\cite{BibDB:BookJordan}.

In this thesis, we reformulate the general problem of solving a
linear system of algebraic equations as a probabilistic inference
problem on a suitably-defined graph\footnote{Recently, we have found out
the work of Moallemi and Van Roy \cite{MinSum} which discusses the connection between
the Min-Sum message passing algorithm and solving quadratic programs.
Both works~\cite{Allerton,MinSum} were published in parallel, and the algorithms where derived independently, using different techniques. In Section \ref{MinSum} we discuss the connection between the two algorithms, and show
they are equivalent.}. Furthermore, for the first time, 
 a full step-by-step derivation of the GaBP algorithm from the 
belief propagation algorithm is provided.

As an important consequence, we demonstrate that Gaussian BP (GaBP) provides an
efficient, distributed approach to solving a linear system that
circumvents the potentially complex operation of direct matrix
inversion. Using the seminal work of Weiss and
Freeman~\cite{BibDB:Weiss01Correctness} and some recent related 
developments~\cite{BibDB:jmw_walksum_nips,BibDB:mjw_walksum_jmlr06,MinSum},
we address the convergence and exactness properties of the
proposed GaBP solver. 

This thesis is structured as follows Chapter \ref{chap2}  introduces the GaBP by providing a clean step-by-step derivation of the GaBP\ algorithm by substituting gaussian probability into Pearl's belief propagation update rules.

Starting from Chapter \ref{chap2a}, we present our novel contributions  in this domain. Chapter \ref{chap2a} presents our novel broadcast version of GaBP that reduces the number of unique messages on a dense graph from $O(n^2)$ to $O(n)$. This version allows for efficient implementation on communication networks that supports broadcast such as wireless and Ethernet.
(Example of  an efficient implementation of the broadcast version on top of 1,024 CPUs is reported in Chapter \ref{chap:svm}). We investigate the use of acceleration methods from linear algebra to be applied for GaBP.
We compare methodically GaBP\ algorithm to other linear iterative methods. Chapter \ref{chap2a} further provides theoretical analysis of GaBP convergence rate assuming diagonally dominant inverse covariance matrix.
This is the first time convergence rate of GaBP is characterized.

In Chapter \ref{chap3} we give numerical examples for illustrating the convergence
properties of the GaBP algorithm.

The GaBP algorithm, like the linear iterative methods, has sufficient conditions for convergence. When those sufficient conditions do not hold, the algorithm may diverge. To address this problem, Chapter \ref{chap4} presents our  novel construction for forcing convergence of the GaBP\ algorithm to the correct solution, for positive definite matrices, as well as for column dependent non-square matrices.
We believe that this result is one of the main novelties of this work, since it applies not only to the GaBP\ algorithm but to other linear iterative methods as well.

The second part of this work is mainly concentrated with applications for the GaBP\ algorithm. The first application we investigate (Chapter~\ref{PPNA}) is the
rating of nodes in a Peer-to-Peer network. We propose a unifying
family of quadratic cost functions to be used in Peer-to-Peer
ratings. We show that our approach is general, since it captures
many of the existing algorithms in the fields of visual layout,
collaborative filtering and Peer-to-Peer rating, among them Koren's
spectral layout algorithm, Katz's method, Spatial ranking,
Personalized PageRank and Information Centrality. Beside of the
theoretical interest in finding common basis of algorithms that
were not linked before, we allow a single efficient
implementation for computing those various rating methods, using
the GaBP solver. We provide simulation results on top of real life
topologies including the MSN Messenger social network.

In Chapter~\ref{linear_detection} we consider the problem of linear
detection using a decorrelator in a code-division multiple-access
(CDMA) system. Through the use of the iterative message-passing
formulation, we implement the decorrelator detector in a
distributed manner. This example allows us to quantitatively
compare the new GaBP solver with the classical iterative solution
methods that have been previously investigated in the context of a
linear implementation of CDMA
demodulation~\cite{grant99iterative,BibDB:TanRasmussen,BibDB:YenerEtAl}.
We show that the GaBP-based decorrelator yields faster convergence
than these conventional methods. We further extend the applicability of the GaBP\ solver to non-square column dependent matrices.

Third application from the field of machine learning is support
vector regression, described in Chapter~\ref{KRR}. We show how to
compute kernel ridge regression using our GaBP solver. For
demonstrating the applicability of our approach we used a cluster
of IBM BlueGene computers with up to 1,024 CPUs in parallel on
very large data sets consisting of millions of data points.
Up to date, this is the largest implementation of belief propagation ever performed. 

Fourth application is the efficient distributed calculation of Kalman filter, presented in  Chapter \ref{chap:Kalman}. We further provide some theoretical results that link the  Kalman filter algorithm to the Gaussian information bottleneck algorithm and the Affine-scaling interior point method.

Fifth application is the efficient distributed solution of linear programming problems using GaBP, presented in Chapter \ref{chap:LP}. As a case study, we discuss the network utility maximization problem and show that our construction has a better accuracy than previous methods, despite the fact it is distributed. 
We provide a large scale simulation  using networks of hundreds of thousands of nodes. 

Chapter \ref{chap:other_algos} identifies a family of previously proposed algorithms,  showing they are instances of GaBP.
This provides the ability to apply the vast knowledge about GaBP to those special cases, for example applying sufficient conditions for convergence as well as applying the convergence fix presented in Chapter \ref{chap:conv_fix}. 

\section{Material Covered in this Thesis}
This thesis is divided into two parts. The first part discusses
the theory of Gaussian belief propagation algorithm and covers the
following papers:~\cite{Allerton,ISIT1,IEEE,ITA08,ISIT09-1,ISIT09-3}.
The second part discusses several applications that were covered
in the following
papers:~\cite{ISIT2,p2p-rating,PPNA08,NIPS-workshop,ECCS08,Allerton08-1,Allerton08-2}.

Below we briefly outline some other related papers that did not fit
into the main theme of this thesis. We have looked at belief propagation at the
using discrete distribution as a basis for various distributed
algorithms: clustering~\cite{EWSN08}, data
placement~\cite{WDAS06,ECCS05} Peer-to-Peer streaming
media~\cite{p2p-streaming} and wireless
settings~\cite{BroadcastBP}. The other major topic we worked on is Peer-to-Peer networks,
especially content distribution networks (CDNs). The Julia content
distribution network is described on~\cite{Julia,WDAS04}. A
modified version of Julia using network coding~\cite{BitCode}.
Tulip is a Peer-to-Peer overlay that enables fast routing and
searching~\cite{Tulip}. The eMule protocol specification is found
on~\cite{eMule}. An implementation of a distributed testbed on
eight European clusters is found on~\cite{EverLab}.


\section{Preliminaries: Notations and Definitions}
We shall use the following linear-algebraic notations and
definitions. The operator $\{\cdot\}^{T}$ stands for a vector or
matrix transpose, the matrix $\mI_{n}$ is an $n\times n$ identity
matrix, while the symbols $\{\cdot\}_{i}$ and $\{\cdot\}_{ij}$
denote entries of a vector and matrix, respectively. Let
$\mM\in\mathbb{R}^{n\times n}$ be a real symmetric square matrix
and $\mA\in\mathbb{R}^{m\times n}$ be a real (possibly
rectangular) matrix. Let $\mathbf{1}$ denotes the all ones vector.

\begin{defn}[Pseudoinverse]
The Moore-Penrose pseudoinverse matrix of the matrix $\mA$,
denoted by $\mA^{\dag}$, is defined as \BE
\mA^{\dag}\triangleq{(\mA^{T}\mA)}^{-1}\mA^{T}. \EE
\end{defn}

\begin{defn}[Spectral radius]
The spectral radius of the matrix $\mM$, denoted by $\rho(\mM)$,
is defined to be the maximum of the absolute values of the
eigenvalues of $\mM$, \ie, \BE \rho(\mM)\triangleq\max_{1\leq
i\leq s}(|\lambda_{i}|)\,,\EE where $\lambda_{1},\ldots\lambda_{s}$
are the eigenvalues of the matrix $\mM$.
\end{defn}

\begin{defn}[Diagonal dominance]
The matrix $\mM$ is\\
\begin{enumerate}
  \item weakly diagonally dominant if \BE |M_{ii}|\ge\sum_{j\neq
  i}|M_{ij}|,\forall i\,,\EE
  \item strictly diagonally dominant if \BE |M_{ii}|>\sum_{j\neq
  i}|M_{ij}|,\forall i,\EE
  \item irreducibly diagonally dominant if $\mM$ is irreducible\footnote{
  A matrix is said to be reducible if there is a permutation
  matrix $\mP$ such that $\mP\mM\mP^{T}$ is block upper triangular. Otherwise, it is irreducible.},
  and \BE |M_{ii}|\ge\sum_{j\neq
  i}|M_{ij}|,\forall i\,,\EE with strict inequality for at least one
  $i$.
\end{enumerate}
\end{defn}

\begin{defn}[PSD]
The matrix $\mM$ is positive semi-definite (PSD) if and only if
for all non-zero real vectors $\vz\in\mathbb{R}^{n}$, \BE
    \vz^{T}\mM\vz\geq0.
\EE
\end{defn}

\begin{defn}[Residual]
For a real vector $\vx\in\mathbb{R}^{n}$, the residual,
$\vr=\vr(\vx)\in\mathbb{R}^{m}$, of a linear system is
$\vr=\mA\vx-\vb$.
\end{defn}
The standard norm of the residual, $||\vr||_{p}
(p=1,2,\ldots,\infty)$, is a good measure of the accuracy of a
vector $\vx$ as a solution to the linear system. In our
experimental study, the Frobenius norm (\ie, $p=2$) per equation
is used, $\tfrac{1}{m}||\vr||_{F}=\tfrac{1}{m}\sqrt{\sum_{i=1}^{m}r_{i}^{2}}$.

 \begin{defn}[Operator Norm] Given a matrix $\mM$ the operator norm $||M||_p$ is defined as  \[ ||M||_p \triangleq \sup_{\vx \ne 0} \frac{||\mM \vx||_p}{||\vx||_p}\,. \]
 \end{defn}

\begin{defn}
The condition number, $\kappa$, of the matrix $\mM$ is defined as
\BE \kappa_{p}\triangleq||\mM||_{p}||\mM^{-1}||_{p}.\EE For $\mM$
being a normal matrix (\ie, $\mM^{T}\mM=\mM\mM^{T}$), the
condition number is given by \BE
\kappa=\kappa_{2}=\Big|\frac{\lambda_{\textrm{max}}}{\lambda_{\textrm{min}}}\Big|,
\EE where $\lambda_{\textrm{max}}$ and $\lambda_{\textrm{min}}$
are the maximal and minimal eigenvalues of $\mM$, respectively.
\end{defn}

Even though a system is nonsingular it could be ill-conditioned.
Ill-conditioning means that a small perturbation in the data
matrix $\mA$, or the observation vector $\vb$, causes large
perturbations in the solution, $\vx^{\ast}$. This determines the
difficulty of solving the problem. The condition number is a good
measure of the ill-conditioning of the matrix. The better the
conditioning of a matrix the smaller the condition number.  The condition number of a non-invertible (singular) matrix
is set arbitrarily to infinity.

 \begin{defn}[Graph Laplacian\footnote{Note this definition is an extension of the Laplacian in the unweighed case, where all edges have weight 1.} \cite{koren-spectral}] \label{def:Laplacian} Given a matrix a weighted matrix $\mA$ describing a graph with $n$ nodes, the graph Laplacian $L$ is a symmetric matrix defined as follows:
\[ L_{i,j} =\begin{cases}
 i=j & deg(i)\\
 else &\ -w_{i,j} 
\end{cases} \]
where $deg(i) = \sum_{j \in N(i)} w_{ji}$ is the degree of node i.\end{defn}
It can be further shown \cite{koren-spectral} that given the Laplacian $L$, it holds that $\vx^TL\vx = \sum_{i<j}w_{ij}(x_i - x_j)^2.$ 
\section{Problem Formulation}
Let $\mA \in \mathbb{R}^{m\times n}$ ($m,n\in\mathbb{N}^{\ast}$)
be a full column rank, $m \times n$ real-valued matrix, with $m
\ge n$, and let $\vb \in \mathbb{R}^m$ be a real-valued vector.
Our objective is to efficiently find a solution $\vx^{\ast}$ to
the linear system of equations $\mA\vx=\vb$ given by\BE
\label{eq_solution} \vx^{\ast}=\mA^{\dag}\vb. \EE

\begin{assm}\label{assum_square}
The  matrix $\mA$ is square (\ie, $m=n$) and symmetric.
\end{assm}
For the case of square invertible matrices the pseudoinverse matrix is
nothing but the data matrix inverse, \ie,
\mbox{$\mA^{\dag}=\mA^{-1}$}. For any linear system of equations
with a unique solution, Assumption~\ref{assum_square} conceptually
entails no loss of generality, as can be seen by considering the
invertible system defined by the new symmetric (and PSD) matrix
${\mA^{T}}_{n\times m}\mA_{m\times n}\mapsto\mA_{n\times n}$ and
vector ${\mA^{T}}_{n\times m}\vb_{m\times 1}\mapsto\vb_{n\times
1}$. However, this transformation involves an excessive
computational complexity of $\mathcal{O}(n^{2}m)$ and
$\mathcal{O}(nm)$ operations, respectively. Furthermore, a sparse
data matrix may become dense due to the transformation, an
undesired property as far as complexity is concerned. Thus, we first
limit the discussion to the solution of the popular case of square
matrices. In Section~\ref{sec_new_const} the proposed GaBP solver is
extended to the more general case of linear systems with
rectangular $m\times n$ full rank matrices.

\chapter*{Part 1: Theory}
\addcontentsline{toc}{chapter}{Part 1: Theory}

\chapter[The GaBP Algorithm]{The GaBP-Based Solver Algorithm}
\label{sec_GaBP}
\label{chap2}
In this section, we show how to derive
the iterative, Gaussian BP-based algorithm that we propose for solving
the linear system \[ \mA_{n\times n}\vx_{n\times1}=\vb_{n\times1}. \]

\label{derivation}
\section[Linear Algebra to Inference]{From Linear Algebra to Probabilistic Inference}
We begin our derivation by defining an undirected graphical model
(\ie, a Markov random field), $\mathcal{G}$, corresponding to the
linear system of equations. Specifically, let
$\mathcal{G}=(\mathcal{X},\mathcal{E})$, where $\mathcal{X}$ is a
set of nodes that are in one-to-one correspondence with the linear
system's variables $\vx=\{x_{1},\ldots,x_{n}\}^{T}$, and where
$\mathcal{E}$ is a set of undirected edges determined by the
non-zero entries of the (symmetric) matrix $\mA$.


We will make use of the following
terminology and notation in the discussion of the GaBP algorithm.
Given the data matrix $\mA$ and the observation vector $\vb$, one
can write explicitly the Gaussian density function $p(\vx) \sim \exp(-\tfrac{1}{2}\vx^T\mA\vx+\vb^T\vx)$, and
its corresponding graph $\mathcal{G}$ consisting of edge
potentials ('compatibility functions') $\psi_{ij}$ and self
potentials (`evidence') $\phi_{i}$. These graph potentials are
 determined according to the following pairwise
factorization of the Gaussian function~(\ref{eq_G})\BE
p(\vx)\propto\prod_{i=1}^{n}\phi_{i}(x_{i})\prod_{\{i,j\}}\psi_{ij}(x_{i},x_{j})\,,\EE
resulting in \mbox{$\psi_{ij}(x_{i},x_{j})\triangleq
\exp(-\tfrac{1}{2}x_{i}A_{ij}x_{j})$} and
\mbox{$\phi_{i}(x_{i})\triangleq\exp\big(-\tfrac{1}{2}A_{ii}x_{i}^{2} +b_{i}x_{i}\big)$}.
The edges set $\{i,j\}$ includes all non-zero entries of $\mA$ for
which $i>j$. The set of graph nodes $\textrm{N}(i)$ denotes the
set of all the nodes neighboring the $i$th node (excluding node $i$). The set
$\textrm{N}(i)\backslash j$ excludes the node $j$ from
$\textrm{N}(i)$.

Using this graph, we can translate the problem of solving the
linear system from the algebraic domain to the domain of
probabilistic inference, as stated in the following theorem.

\begin{prop}\label{prop_3}
The computation of the solution vector $\vx^{\ast}=\mA^{-1}\vb$ is identical
to the inference of the vector of marginal means
\mbox{$\mathbf{\mu}\triangleq\{\mu_{1},\ldots,\mu_{n}\}$} over the
graph $\mathcal{G}$ with the associated joint Gaussian probability
density function  $p(\vx)\sim\mathcal{N}(\mu,\mA^{-1})$.
\end{prop}

\begin{figure*}[t] \center \fbox{\begin{minipage}{4.6in}
\footnotesize
\setlength{\baselineskip}{3.5mm} \noindent 
\centering{
\[ \mA \vx = \vb \]
\[ \Updownarrow \]
\[ \mA \vx - \vb = 0 \]
\[ \Updownarrow \]
\[ \min_{\vx}( \tfrac{1}{2} \vx^T\mA \vx - \vb^T\vx) \]
\[ \Updownarrow \]
\[ \max_\vx (- \tfrac{1}{2} \vx^T\mA \vx + \vb^T\vx) \]
\[ \Updownarrow \]
\[ \max_\vx  \exp( -\tfrac{1}{2} \vx^T\mA \vx + \vb^T\vx) \]
}
\normalsize
\end{minipage} }
\caption{Schematic outline of the of the proof to Proposition \ref{prop_3}.} \label{alg:ssByz-Q}
\end{figure*}

\begin{proof}[{\bf Proof}]
Another way of solving the set of linear equations
$\mA\vx-\vb=\mathbf{0}$ is to represent it by using a quadratic
form \mbox{$q(\vx)\triangleq\vx^{T}\mA\vx/2-\vb^{T}\vx$}. As the
matrix $\mA$ is symmetric, the derivative of the quadratic form
w.r.t. the vector $\vx$ is given by the vector $\partial
q/\partial\vx=\mA\vx-\vb$. Thus equating $\partial q/\partial
\vx=\mathbf{0}$ gives the global minimum $\vx^{\ast}$ of this
convex function, which is nothing but the desired solution to
$\mA\vx=\vb$.

Next, one can define the following joint Gaussian probability
density function \BE\label{eq_G}
p(\vx)\triangleq\mathcal{Z}^{-1}\exp{\big(-q(\vx)\big)}=\mathcal{Z}^{-1}\exp{(-\vx^{T}\mA\vx/2+\vb^{T}\vx)},\EE
where $\mathcal{Z}$ is a distribution normalization factor.
Denoting the vector $\mathbf{\mu}\triangleq\mA^{-1}\vb$, the
Gaussian density function can be rewritten as \BEA\label{eq_G2}
p(\vx)&=&\mathcal{Z}^{-1}\exp{(\mathbf{\mu}^{T}\mA\mathbf{\mu}/2)}\nonumber\\&\times&\exp{(-\vx^{T}\mA\vx/2+\mathbf{\mu}^{T}\mA\vx-\mathbf{\mu}^{T}\mA\mathbf{\mu}/2)}
\nonumber\\&=&\mathcal{\zeta}^{-1}\exp{\big(-(\vx-\mathbf{\mu})^{T}\mA(\vx-\mathbf{\mu})/2\big)}\nonumber\\&=&\mathcal{N}(\mathbf{\mu},\mA^{-1}),\EEA
where the new normalization factor
$\mathcal{\zeta}\triangleq\mathcal{Z}\exp{(-\mathbf{\mu}^{T}\mA\mathbf{\mu}/2)}$.
It follows that the target solution $\vx^{\ast}=\mA^{-1}\vb$ is
equal to $\mathbf{\mu}\triangleq\mA^{-1}\vb$, the mean vector of
the distribution $p(\vx)$, as defined above~(\ref{eq_G}).

Hence, in order to solve the system of linear equations we need to
infer the marginal densities, which must also be Gaussian,
\mbox{$p(x_{i})\sim\mathcal{N}(\mu_{i}=\{\mA^{-1}\vb\}_{i},P_{i}^{-1}=\{\mA^{-1}\}_{ii})$},
where $\mu_{i}$ and $P_{i}$ are the marginal mean and inverse
variance (sometimes called the precision), respectively.
\end{proof}

According to Proposition~\ref{prop_3}, solving a deterministic
vector-matrix linear equation translates to solving an inference
problem in the corresponding graph. The move to the probabilistic
domain calls for the utilization of BP as an efficient inference
engine.

\begin{remark}
Defining a jointly Gaussian probability density function,
immediately yields an implicit assumption on the positive
semi-definiteness of the precision matrix $\mA$, in addition to
the symmetry assumption. However, we would like to stress out that
this assumption emerges only for exposition purposes, so we can
use the notion of `Gaussian probability', but the derivation of
the GaBP solver itself does not use this assumption. See the
numerical example of the exact GaBP-based solution of a system
with a symmetric, but not positive semi-definite, data matrix
$\mA$ in Section~\ref{sec_nonPSD}.
\end{remark}
\section{Belief Propagation}
Belief propagation (BP) is equivalent to applying Pearl's local
message-passing algorithm~\cite{BibDB:BookPearl}, originally
derived for exact inference in trees, to a general graph even if
it contains cycles (loops). BP has been found to have outstanding
empirical success in many applications, \eg, in decoding Turbo
codes and low-density parity-check (LDPC) codes. The excellent
performance of BP in these applications may be attributed to the
sparsity of the graphs, which ensures that cycles in the graph are
long, and inference may be performed as if it were a tree.

The BP algorithm functions by passing real-valued messages across
edges in the graph and consists of two computational rules, namely
the `sum-product rule' and the `product rule'. In contrast to
typical applications of BP in coding theory~\cite{BibDB:AjiMcEliece}, our graphical
representation resembles to a pairwise Markov random
field\cite{BibDB:BookJordan} with a single type of propagating
messages, rather than a factor graph~\cite{BibDB:FactorGraph} with
two different types of messages, originated from either the
variable node or the factor node. Furthermore, in most graphical
model representations used in the information theory literature
the graph nodes are assigned with discrete values, while in this
contribution we deal with nodes corresponding to continuous
variables. Thus, for a graph $\mathcal{G}$ composed of potentials
$\psi_{ij}$ and $\phi_{i}$ as previously defined, the conventional
sum-product rule becomes an integral-product
rule~\cite{BibDB:Weiss01Correctness} and the message
$m_{ij}(x_j)$, sent from node $i$ to node $j$ over their shared
edge on the graph, is given by \BE\label{eq_contBP}
    m_{ij}(x_j)\propto\int_{x_i} \psi_{ij}(x_i,x_j) \phi_{i}(x_i)
\prod_{k \in \textrm{N}(i)\setminus j} m_{ki}(x_i) dx_{i}. \EE The
marginals are computed (as usual) according to the product rule
\BE\label{eq_productrule}
p(x_{i})=\alpha
\phi_{i}(x_{i})\prod_{k\in\textrm{N}(i)}m_{ki}(x_{i}), \EE where
the scalar $\alpha$  is a normalization constant. Note that the
propagating messages (and the graph potentials) do not have to
describe valid (\ie, normalized) density probability functions, as
long as the inferred marginals do.

\section[GaBP Algorithm]{The Gaussian BP Algorithm}\label{sec:GaBP}
Gaussian BP is a special case of continuous BP, where the
underlying distribution is Gaussian. The GaBP algorithm was originally introduced by Weiss \etal \cite{BibDB:Weiss01Correctness}. Weiss work do not detail the derivation of the GaBP\ algorithm. We believe that this derivation is important for the complete understanding of the GaBP algorithm. To this end, we derive the Gaussian
BP update rules by substituting Gaussian distributions into the
continuous BP update equations.

Given the data matrix $\mA$ and the observation vector $\vb$, one
can write explicitly the Gaussian density function,
$p(\vx) \sim \exp(-\tfrac{1}{2}\vx^T\mA\vx + \vb^T\vx)$, and its corresponding graph $\mathcal{G}$.
Using the graph definition and a certain (arbitrary) pairwise
factorization of the Gaussian function~(\ref{eq_G2}), the edge
potentials ('compatibility functions') and self potentials
(`evidence') $\phi_{i}$ are determined to be
\BEA\psi_{ij}(x_{i},x_{j})&\triangleq&
\exp(-\tfrac{1}{2}x_{i}A_{ij}x_{j}),\\
\phi_{i}(x_{i})&\triangleq&\exp\big(-\tfrac{1}{2}A_{ii}x_{i}^{2} + b_{i}x_{i}\big),\EEA
respectively. Note that by completing the square, one can observe
that \BE\label{eq_PhiGauss}
\phi_{i}(x_{i})\propto\mathcal{N}(\mu_{ii}=b_{i}/A_{ii},P_{ii}^{-1}=A_{ii}^{-1}).\EE
The graph topology is specified by the structure of the matrix
$\mA$, \ie the edges set $\{i,j\}$ includes all non-zero entries
of $\mA$ for which $i>j$.

Before describing the inference algorithm performed over the
graphical model, we make the elementary but very useful
observation that the product of Gaussian densities over a common
variable is, up to a constant factor, also a Gaussian density.

\begin{lem}\label{Lemma_Gaussian}
Let $f_{1} (x)$ and $f_{2} (x)$ be the probability density
functions of a Gaussian random variable with two possible
densities $\mathcal{N}(\mu_{1},P_{1}^{-1})$ and
$\mathcal{N}(\mu_{2},P_{2}^{-1})$, respectively. Then their
product, $f(x)=f_1(x) f_2(x)$ is, up to a constant factor, the
probability density function of a Gaussian random variable with
distribution $\mathcal{N}(\mu,P^{-1})$, where \BEA
\mu&=&P^{-1}(P_{1}\mu_{1}+P_{2}\mu_{2})\label{eq_productmean},\\
P^{-1}&=&(P_{1}+P_{2})^{-1}\label{eq_productprec}. \EEA
\end{lem}
\begin{proof}[\bf Proof]
Taking the product of the two Gaussian probability density
functions \BE
    f_{1} (x)f_{2} (x)=\frac{\sqrt{P_{1}P_{2}}}{2\pi}\exp{\Big(-\big(P_{1}(x-\mu_{1})^{2}+P_{2}(x-\mu_{2})^{2}\big)/2\Big)}
\EE and completing the square, one gets \BE
    f_{1} (x)f_{2} (x)=\frac{C\sqrt{P}}{2\pi}\exp{\big(-P(x-\mu)^{2}/2\big)},
\EE with \BEA P&\triangleq&P_{1}+P_{2},\\
                \mu&\triangleq&P^{-1}(\mu_{1}P_{1}+\mu_{2}P_{2})
\EEA and the scalar constant determined by\BE
    C\triangleq\sqrt{\frac{P}{P_{1}P_{2}}}\exp{\Big(\tfrac{1}{2}\big(P_{1}\mu_{1}^{2}(P^{-1}P_{1}-1)+P_{2}\mu_{2}^{2}(P^{-1}P_{2}-1)+2P^{-1}P_{1}P_{2}\mu_{1}\mu_{2}\big)\big)}.
\EE Hence, the product of the two Gaussian densities is
$C\cdot\mathcal{N}(\mu,P^{-1})$.
\end{proof}
\begin{figure}[h!]
\begin{center}
    \includegraphics[width=0.4\textwidth]{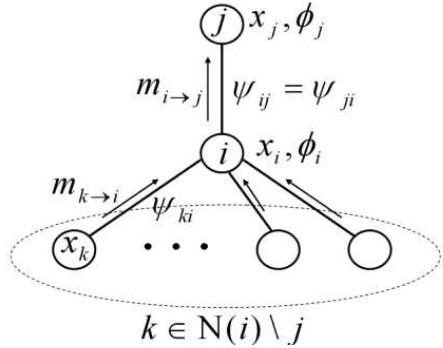}
\caption{Belief propagation message flow}
\label{fig_flow}
\end{center}

\end{figure}

Fig.~\ref{fig_flow} plots a portion of a certain graph, describing
the neighborhood of node $i$. Each node (empty circle) is
associated with a variable and self potential $\phi$, which is a
function of this variable, while edges go with the pairwise
(symmetric) potentials $\Psi$. Messages are propagating along the
edges on both directions (only the messages relevant for the
computation of $m_{ij}$ are drawn in Fig.~\ref{fig_flow}). Looking
at the right hand side of the integral-product
rule~(\ref{eq_contBP}), node $i$ needs to first calculate the
product of all incoming messages, except for the message coming
from node $j$. Recall that since $p(\vx)$ is jointly Gaussian, the
factorized self potentials
$\phi_{i}(x_i)\propto\mathcal{N}(\mu_{ii},P_{ii}^{-1})$
(\ref{eq_PhiGauss}) and similarly all messages
$m_{ki}(x_i)\propto\mathcal{N}(\mu_{ki},P_{ki}^{-1})$ are of
Gaussian form as well.

As the terms in the product of the incoming messages and the self
potential in the integral-product rule~(\ref{eq_contBP}) are all a
function of the same variable, $x_{i}$ (associated with the node
$i$), then, according to the multivariate extension of
Lemma~\ref{Lemma_Gaussian}, \BE\label{eq_product} \mathcal{N}(\mu_{i\backslash j},P_{i\backslash j}^{-1}) \propto \phi_{i}(x_i)
\prod_{k \in \textrm{N}(i) \backslash j} m_{ki}(x_i)\,. \EE
Applying the multivariate version of the product precision
expression in~(\ref{eq_productprec}), the update rule for the
inverse variance is given by (over-braces denote the origin of
each of the terms) \BE\label{eq_prec} P_{i\backslash j} =
\overbrace{P_{ii}}^{\phi_{i}(x_i)} + \sum_{k\in \textrm{N}(i)
\backslash j} \overbrace{P_{ki}}^{m_{ki}(x_i)}, \EE where
$P_{ii}\triangleq A_{ii}$ is the inverse variance a-priori
associated with node $i$, via the precision of $\phi_{i}(x_{i})$,
and $P_{ki}$ are the inverse variances of the messages
$m_{ki}(x_i)$. Similarly using~(\ref{eq_productmean}) for the
multivariate case, we can calculate the mean \BE\label{eq_mean}
 \mu_{i\backslash j} = P_{i\backslash j}^{-1}\Big(\overbrace{P_{ii}\mu_{ii}}^{\phi_{i}(x_i)} +
\sum_{{k} \in \textrm{N}(i) \backslash j}
\overbrace{P_{ki}\mu_{ki}}^{m_{ki}(x_i)}\Big), \EE where
$\mu_{ii}\triangleq b_{i}/A_{ii}$ is the mean of the self
potential and $\mu_{ki}$ are the means of the incoming messages.

Next, we calculate the remaining terms of the message
$m_{ij}(x_j)$, including the integration over~$x_{i}$.
\BEA\label{eq_contBP2}
    m_{ij}(x_j)&\propto&\int_{x_{i}} \psi_{ij}(x_i,x_j) \phi_{i}(x_i)
\prod_{k \in \textrm{N}(i)\setminus j} m_{ki}(x_i)
dx_{i}\\&\propto&\int_{x_{i}}\overbrace{\exp{(-x_{i}A_{ij}x_{j})}}^{\psi_{ij}(x_{i},x_{j})}\overbrace{\exp{(-P_{i\backslash
j}(x_{i}^{2}/2-\mu_{i\backslash j}x_{i}))}}^{\phi_{i}(x_i)
\prod_{k \in \textrm{N}(i) \backslash j}
m_{ki}(x_i)}dx_{i}\\&=&\int_{x_{i}}\exp{((-P_{i\backslash
j}x_{i}^{2}/2)+(P_{i\backslash j}\mu_{i\backslash
j}-A_{ij}x_{j})x_{i})}dx_{i}\\\label{eq_exponent}&\propto&\exp{((P_{i\backslash
j}\mu_{i\backslash j}-A_{ij}x_{j})^{2}/(2P_{i\backslash
j}))}\\&\propto&\mathcal{N}(\mu_{ij}=-P_{ij}^{-1}A_{ij}\mu_{i\backslash
j},P_{ij}^{-1} = -A_{ij}^{-2}P_{i \backslash
j}^{-1})\label{eq_IntegrationResult},\EEA where the
exponent~(\ref{eq_exponent}) is obtained by using the Gaussian
integral~(\ref{eq_GaussianIntegral}):

\BE\label{eq_GaussianIntegral}
\int_{-\infty}^{\infty}\exp{(-ax^{2}+bx)}dx=\sqrt{\pi/a}\exp{(b^{2}/4a)},
\EE we find that the messages $m_{ij}(x_j)$ are proportional to
normal distribution with precision and mean
\BE\label{eq_prec_message} P_{ij} = -A_{ij}^2P_{i \backslash
j}^{-1}\,, \EE \BE \mu_{ij}\label{eq_mean_message} =
-P_{ij}^{-1}A_{ij}\mu_{i\backslash j}\,. \EE These two scalars
represent the messages propagated in the Gaussian BP-based
algorithm.

Finally, computing the product rule~(\ref{eq_productrule}) is
similar to the calculation of the previous
product~(\ref{eq_product}) and the resulting mean~(\ref{eq_mean})
and precision~(\ref{eq_prec}), but including all incoming
messages. The marginals are inferred by normalizing the result of
this product. Thus, the marginals are found to be Gaussian
probability density functions $\mathcal{N}(\mu_{i},P_{i}^{-1})$
with precision and mean \BEA P_{i}
= \overbrace{P_{ii}}^{\phi_{i}(x_i)} + \sum_{k\in \textrm{N}(i)} \overbrace{P_{ki}}^{m_{ki}(x_i)},\\
\mu_{i} = P_{i\backslash
j}^{-1}\Big(\overbrace{P_{ii}\mu_{ii}}^{\phi_{i}(x_i)} + \sum_{{k}
\in \textrm{N}(i)}
\overbrace{P_{ki}\mu_{ki}}^{m_{ki}(x_i)}\Big)\label{eq_marginal_mean},
\EEA respectively. The derivation of the GaBP-based solver
algorithm is concluded by simply substituting the explicit derived expressions
of $P_{i\backslash j}$~(\ref{eq_prec}) into
$P_{ij}$~(\ref{eq_prec_message}), $\mu_{i\backslash
j}$~(\ref{eq_mean}) and $P_{ij}$~(\ref{eq_prec_message}) into
$\mu_{ij}$~(\ref{eq_mean_message}) and $P_{i\backslash
j}$~(\ref{eq_prec}) into $\mu_{i}$~(\ref{eq_marginal_mean}).

The message passing in the GaBP solver can be performed subject to
any scheduling. We refer to two conventional messages updating
rules: parallel (flooding or synchronous) and serial (sequential,
asynchronous) scheduling. In the parallel scheme, messages are
stored in two data structures: messages from the previous
iteration round, and messages from the current round. Thus,
incoming messages do not affect the result of the computation in
the current round, since it is done on messages that were received
in the previous iteration round. Unlike this, in the serial
scheme, there is only one data structure, so incoming messages in
this round, change the result of the computation. In a sense it is
exactly like the difference between the Jacobi and Gauss-Seidel
algorithms, to be discussed in the following. Some in-depth
discussions about parallel vs. serial scheduling in the BP context
(the discrete case)\ can be found in the work of Elidan \etal~\cite{BibDB:ElidanEtAl}.

\begin{table}[h!]
\begin{alg}\label{alg_GaBP}\end{alg}\centerline{
\begin{tabular}{|lll|}
  \hline&&\\
  \texttt{1.} & \emph{\texttt{Initialize:}} & $\checkmark$\quad\texttt{Set the neighborhood} $\textrm{N}(i)$ \texttt{to include}\\&&\quad\quad$\forall k\neq i \exists A_{ki}\neq0$.\\&& $\checkmark$\quad\texttt{Set the scalar fixes}\\&&$\quad\quad P_{ii}=A_{ii}$ \texttt{and} $\mu_{ii}=b_{i}/A_{ii}$, $\forall i$.\\
  && $\checkmark$\quad\texttt{Set the initial $\textrm{N}(i)\ni k\rightarrow i$ scalar messages}\\&&\quad\quad $P_{ki}=0$ \texttt{and} $\mu_{ki}=0$.\\&& $\checkmark$\quad \texttt{Set a convergence threshold} $\epsilon$.\\
  {\texttt{2.}} & {\emph{\texttt{Iterate:}} }&  $\checkmark$\quad\texttt{Propagate the $\textrm{N}(i)\ni k\rightarrow i$ messages}\\&&$\quad\quad P_{ki}$ \texttt{and} $\mu_{ki}$, $\forall i$ \texttt{(under certain scheduling)}.\\&&
  $\checkmark$\quad\texttt{Compute the} $\textrm{N}(j)\ni i\rightarrow j$ \texttt{scalar messages} \\&& $\quad\quad P_{ij} = -A_{ij}^{2}/\big(P_{ii}+\sum_{{k}\in\textrm{N}(i) \backslash j}
P_{ki}\big)$,\\&& $\quad\quad\mu_{ij} =
\big(P_{ii}\mu_{ii}+\sum_{k \in \textrm{N}(i) \backslash j}
P_{ki}\mu_{ki}\big)/A_{ij}$.\\
  {\texttt{3.}} & {\emph{\texttt{Check:}}} & $\checkmark$\quad\texttt{If the messages} $P_{ij}$ \texttt{and} $\mu_{ij}$ \texttt{did not}\\&&\quad\quad\texttt{converge (w.r.t. $\epsilon$),} \texttt{return to
    Step 2.}\\&&$\checkmark$\quad\texttt{Else, continue to Step 4.}\\
  {\texttt{4.}} & {\emph{\texttt{Infer:}}} & $\checkmark$\quad\texttt{Compute the marginal means}\\&&{\quad\quad$\mu_{i}=\big(P_{ii}\mu_{ii}+\sum_{k \in
\textrm{N}(i)}P_{ki}\mu_{ki}\big)/\big(P_{ii}+\sum_{{k}\in\textrm{N}(i)}
P_{ki}\big)$, $\forall i$.}\\&&$(\checkmark$\quad\texttt{Optionally compute the marginal precisions}\\&&\quad\quad$P_{i}=P_{ii}+\sum_{k\in\textrm{N}(i)}P_{ki}\quad)$\\
  {\texttt{5.}} & {\emph{\texttt{Solve:}}} & $\checkmark$\quad\texttt{Find the solution}\\&& {$\quad\quad x_{i}^{\ast}=\mu_{i}$, $\forall i$.}\\&&\\\hline
\end{tabular}
}\newpage
\end{table}

\section{Max-Product Rule}
\label{MaxProductRule}
A well-known alternative version to the sum-product BP is the
max-product (a.k.a. min-sum) algorithm~\cite{Max-product}. In this variant of
BP, maximization operation is performed rather than
marginalization, \ie, variables are eliminated by taking maxima
instead of sums. For Trellis trees (\eg, graphically representing
convolutional codes or ISI channels), the conventional sum-product
BP algorithm boils down to performing the BCJR
algorithm\cite{BibDB:BCJR}, resulting in the most probable symbol,
while its max-product counterpart is equivalent to the Viterbi
algorithm~\cite{Viterbi}, thus inferring the most probable sequence of
symbols~\cite{BibDB:FactorGraph}.

In order to derive the max-product version of the proposed GaBP
solver, the integral (sum)-product rule~(\ref{eq_contBP}) is
replaced by a new rule\BE\label{eq_contBP2}
    m_{ij}(x_j)\propto\argmax_{x_i} \psi_{ij}(x_i,x_j) \phi_{i}(x_i)
\prod_{k \in \textrm{N}(i)\setminus j} m_{ki}(x_i) dx_{i}. \EE
Computing $m_{ij}(x_j)$ according to this max-product rule, one
gets
\BEA
    m_{ij}(x_j)&\propto&\argmax_{x_{i}} \psi_{ij}(x_i,x_j) \phi_{i}(x_i)
\prod_{k \in \textrm{N}(i)\setminus j}
m_{ki}(x_i)\\&\propto&\argmax_{x_{i}}\overbrace{\exp{(-x_{i}A_{ij}x_{j})}}^{\psi_{ij}(x_{i},x_{j})}\overbrace{\exp{(-P_{i\backslash
j}(x_{i}^{2}/2-\mu_{i\backslash j}x_{i}))}}^{\phi_{i}(x_i)
\prod_{k \in \textrm{N}(i) \backslash j}
m_{ki}(x_i)}\\&=&\argmax_{x_{i}}\exp{((-P_{i\backslash
j}x_{i}^{2}/2)+(P_{i\backslash j}\mu_{i\backslash
j}-A_{ij}x_{j})x_{i})}. \EEA
Hence, $x_{i}^{\textrm{max}}$, the value of $x_{i}$ maximizing the product $\psi_{ij}(x_i,x_j)
\phi_{i}(x_i) \prod_{k \in \textrm{N}(i)\setminus j} m_{ki}(x_i)$
is given by equating its derivative w.r.t. $x_{i}$ to zero,
yielding \BE x_{i}^{\textrm{max}}=\frac{P_{i\backslash
j}\mu_{i\backslash j}-A_{ij}x_{j}}{P_{i\backslash j}}. \EE
Substituting $x_{i}^{\textrm{max}}$ back into the product, we get
\BEA m_{ij}(x_j)&\propto&\exp{((P_{i\backslash j}\mu_{i\backslash
j}-A_{ij}x_{j})^{2}/(2P_{i\backslash
j}))}\\&\propto&\mathcal{N}(\mu_{ij}=-P_{ij}^{-1}A_{ij}\mu_{i\backslash
j},P_{ij}^{-1} = -A_{ij}^{-2}P_{i \backslash j}),\EEA which is
identical to the result obtained when eliminating $x_{i}$ via
integration~(\ref{eq_IntegrationResult}).
\BE
m_{ij}(x_j)\propto\mathcal{N}(\mu_{ij}=-P_{ij}^{-1}A_{ij}\mu_{i\backslash
j},P_{ij}^{-1} = -A_{ij}^{-2}P_{i \backslash j}), \EE which is
identical to the messages derived for the sum-product
case~(\ref{eq_prec_message})-(\ref{eq_mean_message}). Thus
interestingly, as opposed to ordinary (discrete) BP, the following
property of the GaBP solver emerges.
\begin{corol}
The max-product \eqref{eq_contBP2} and sum-product \eqref{eq_contBP}
versions of the GaBP solver are identical.
\end{corol}

\section[Properties]{Convergence and Exactness}
\label{sec:converge-exact} In ordinary BP, convergence does not
entail exactness of the inferred probabilities, unless the graph
has no cycles.   Luckily, this is not the case for the GaBP
solver. Its underlying Gaussian nature yields a direct connection
between convergence and exact inference. Moreover, in contrast to
BP the convergence of GaBP is not limited for tree or sparse
graphs and can occur even for dense (fully-connected) graphs,
adhering to certain rules discussed in the following.

We can use results from the literature on probabilistic inference
in graphical
models~\cite{BibDB:Weiss01Correctness,BibDB:jmw_walksum_nips,BibDB:mjw_walksum_jmlr06}
to determine the convergence and exactness properties of the
GaBP-based solver. The following two theorems establish sufficient
conditions under which GaBP is guaranteed to converge to the exact
marginal means.

\begin{thm}{\cite[Claim 4]{BibDB:Weiss01Correctness}}
\label{dd_theorem_weiss}
If the matrix $\mA$ is strictly diagonally dominant, then GaBP
converges and the marginal means converge to the true means.
\end{thm}

This sufficient condition was recently relaxed to include a wider
group of matrices.

\begin{thm}{\cite[Proposition 2]{BibDB:jmw_walksum_nips}}
\label{spectral_radius_thm}
If the spectral radius of the matrix $\mA$ satisfies \BE
\rho(|\mI_{n}-\mA|)<1, \EE then GaBP converges and the marginal
means converge to the true means. (The assumption here is that the matrix $\mA$ is first
normalized by multiplying with $\mD^{-1/2}\mA\mD^{-1/2}$, where $\mD = \diag(\mA)$.)
\end{thm}

A third and weaker sufficient convergence condition (relative to Theorem \ref{spectral_radius_thm})
which characterizes the convergence of the variances is given in
\cite[Theorem 2]{MinSum}: For each row in the matrix $\mA$, if $A_{ii}^2 > \Sigma_{j \ne i} A_{ij}^2$ then
the variances converge. Regarding the means, additional condition related to Theorem \ref{spectral_radius_thm} is given. \\

There are many examples of linear systems that violate these
conditions, for which GaBP converges to the exact means. In
particular, if the graph corresponding to the system is acyclic
(\ie, a tree), GaBP yields the exact marginal means (and
variances~\cite{BibDB:Weiss01Correctness}), regardless of the
value of the spectral radius of the matrix~\cite{BibDB:Weiss01Correctness}. 
In contrast to conventional iterative methods derived
from linear algebra, understanding the conditions for exact
convergence remain intriguing open problems.

\chapter{GaBP Algorithm Properties}
\label{chap2a}
Starting this chapter, we present our novel contributions  in this GaBP domain. We provide a theoretical analysis of GaBP convergence rate assuming diagonally dominant inverse covariance matrix.
This is the first time convergence rate of GaBP is characterized.
Chapter \ref{chap2a} further presents our novel broadcast version of GaBP, which reduces the number of unique messages on a dense graph from $O(n^2)$ to $O(n)$. This version allows for efficient implementation on communication networks that supports broadcast such as wireless and Ethernet.
(Example of  an efficient implementation of the broadcast version on top of 1,024 CPUs is reported in Chapter \ref{chap:svm}). We investigate the use of acceleration methods from linear algebra to be applied for GaBP
and compare methodically GaBP\ algorithm to other linear iterative methods. 
\section[Upper Bound on Rate]{Upper Bound on Convergence Rate} \label{convergence} \label{conv-speed}
\label{chap:conv_rate}
In this section we give an upper bound on  convergence rate of
the GaBP algorithm. As far as we know this is the first
theoretical result bounding the convergence speed of the GaBP
algorithm.

Our upper bound is based on the work of Weiss \etal~\cite[Claim
4]{Weiss}, which proves the correctness of the mean computation.
Weiss uses the pairwise potentials form\footnote{Weiss assumes
variables with zero means. The mean value does not affect convergence speed.}, where
\BEA p(\vx) &\propto& \Pi_{i,j} \psi_{ij}(x_i,x_j) \Pi_i \psi_i(x_i)\,, \nonumber \\
 \psi_{i,j}(x_i, x_j) &\triangleq& \exp(-\tfrac{1}{2} [x_i\  x_j]^T \mV_{ij} [x_i\ x_j])\,. \nonumber \\
 \psi_{i,i}(x_i) &\triangleq& \exp(-\tfrac{1}{2} x_i^T \mV_{ii} x_i)\,. \nonumber
\EEA
We further assume scalar variables. Denote the entries of the inverse pairwise covariance matrix $\mV_{ij}$ and the inverse covariance matrix $\mV_{ii}$ as:
\BE
 \mV_{ij}  \equiv  \left(%
\begin{array}{cc}
  \tilde{a}_{ij} & \tilde{b}_{ij} \\
  \tilde{b}_{ji} & \tilde{c}_{ij} \\
\end{array}%
\right)\,, \ \ \ \ \ \ \ \ \mV_{ii} = (\tilde{a_{ii}})\,. \nonumber \EE Assuming the optimal solution is $\vx^{*}$, for a
desired accuracy $\epsilon||\vb||_{\infty}$ where
$||\vb||_{\infty} \equiv \max_i |\vb_i|$, and $\vb$ is the shift
vector, \cite[Claim
4]{Weiss} proves that the GaBP algorithm converges to an accuracy of
$|x^* - x_t| < \epsilon||\vb||_{\infty}$ after at most $t = \ceil{{
\log(\epsilon)}/{ \log(\beta) }}$ rounds, where $\beta =
\max_{ij}|\tilde{b}_{ij}/\tilde{c}_{ij}|$.

The problem with applying Weiss' result directly to our model is
that we are working with different parameterizations. We use the
{\em information form} $ p(\vx) \propto \exp(-\tfrac{1}{2}\vx^T\mA\vx +
\vb^T\vx). $ The decomposition of the matrix $\mA$ into pairwise
potentials is not unique. In order to use Weiss' result, we
propose such a decomposition. Any decomposition from the information
form to the pairwise potentials form should be subject to the
following constraints \cite{Weiss} \ \BE \overbrace{\tilde{b}_{ij} }^{\mbox{Pairwise form}}= \overbrace{a_{ij}}^{\mbox{Information form}}
, \ \ \ \ \ \nonumber \EE 
which means that the inverse covariance in the pairwise model should be equivalent to inverse covariance in the information form. 
\BE \overbrace{ \tilde{a_{ii}}+\sum_{j \in N(i)} \tilde{c}_{ij}}^{\mbox{Pairwise form}} = \overbrace{a_{ii}}^{\mbox{Information form}}\,. \nonumber \EE
The second constraints says that the sum of node $i$'s inverse variance (in both the self potentials and edge potentials) should be identical to the inverse variance in the information form.

We propose to
initialize the pairwise potentials as following. Assuming the
matrix $\mA$ is diagonally dominant, we define $\varepsilon_i$ to
be the non negative gap \BE \varepsilon_i \triangleq |a_{ii}| -
\sum_{j\in N(i)}|a_{ij}| > 0\,, \label{varepsilon} \nonumber \EE and the
following decomposition \BE \tilde{b}_{ij} = a_{ij} , \ \ \ \
\tilde{a}_{ij} = c_{ij} + \varepsilon_i / |N(i)|\,, \nonumber \EE where
$|N(i)|$ is the number of graph neighbors of node $i$. Following
Weiss, we define $\gamma$ to be \BE \gamma = \max_{i,j}
\frac{|\tilde{b}_{ij}|}{|\tilde{c}_{ij}|} = \max_{i,j}\frac{|a_{ij}|}{|a_{ij}| +
\varepsilon_i /|N(i)|} = \nonumber \max_{i,j} \frac{1}{1 + (
\varepsilon_i) /(|a_{ij}||N(i)|)} < 1\,. \label{gamma} \EE In total, we get
that for a desired accuracy of $\epsilon||\vb||_{\infty}$ we need
to iterate  for $t = \ceil{ {\log(\epsilon)}/{ \log(\gamma)}}$
rounds. Note that this is an upper bound and in practice we indeed
have observed a much faster convergence rate.

The computation of the parameter $\gamma$ can be easily done in a
distributed manner: Each node locally computes $\varepsilon_i$,
and $\gamma_i = \max_{j} {1}/{( 1 + |a_{ij}|\varepsilon_i/N(i)
)}$. Finally, one maximum operation is performed globally, $\gamma
= \max_{i} \gamma_i$.

\section[Convergence Acceleration]{Convergence Acceleration}
\label{sec:accel}
Further speed-up of GaBP can
be achieved by adapting known acceleration techniques from linear
algebra, such Aitken's method and Steffensen's
iterations~\cite{BibDB:BookHenrici}. Consider a sequence
$\{\vx_{n}\}$ where $n$ is the iteration number, and $\vx$ is the marginal probability computed by GaBP. Further assume that $\{\vx_n\}$ linearly
converge to the limit $\hat{\vx}$, and $\vx_n \ne \hat{\vx}$ for $n
\ge 0$. According to Aitken's method, if there exists a real
number $a$ such that $|a|<1 $ and \mbox{$\lim_{n \rightarrow
\infty}(\vx_n-\hat{\vx})/(\vx_{n-1} - \hat{\vx}) = a$}, then the sequence
$\{ \vy_n\}$ defined by
\[ \vy_n = \vx_n - \frac{(\vx_{n+1} -\vx_n)^2}{\vx_{n+2} - 2\vx_{n+1} + \vx_n} \]
converges to $\hat{\vx}$ faster than $\{ \vx_n \}$ in the sense that
\mbox{$\lim_{n \rightarrow \infty} |(\hat{\vx} - \vy_n)/(\hat{\vx} -
\vx_n)| = 0$}. Aitken's method can be viewed as a generalization of
over-relaxation, since one uses values from three, rather than
two, consecutive iteration rounds. This method can be easily
implemented in GaBP as every node computes values based only on
its own history.

Steffensen's iterations incorporate Aitken's method. Starting with
$\vx_{n}$, two iterations are run to get $\vx_{n+1}$ and $\vx_{n+2}$.
Next, Aitken's method is used to compute $\vy_{n}$, this value
replaces the original $\vx_{n}$, and GaBP is executed again to get a
new value of $\vx_{n+1}$. This process is repeated iteratively until
convergence. \comment{Table~\ref{tab_2} demonstrates the speed-up
of GaBP obtained by using these acceleration methods, in
comparison with that achieved by the similarly modified Jacobi
algorithm.\footnote{Application of Aitken and Steffensen's methods
for speeding-up the convergence of standard (non-BP) iterative
solution algorithms in the context of MUD was introduced by Leibig
\etal~\cite{LDF05}.}} We remark that, although the convergence
rate is improved with these enhanced algorithms, the region of
convergence of the accelerated GaBP solver remains unchanged.
Chapter \ref{chap3} gives numerical examples to illustrate the proposed acceleration method performance.
\comment{
\subsection{Approximation for Asymmetric Data Matrices}\label{sec_asymmetric}

\footnote{For the case of  the approximate solution of asymmetric
linear systems is discussed in Section~\ref{sec_asymmetric}}.

if we want to keep sparsity. }

\section[GaBP Broadcast]{GaBP Broadcast Variant}
\comment{For a dense data matrix $\mA$, The GaBP solver
algorithm~\ref{alg_GaBP} can be easily implemented in a distributed
fashion. Each node $i$ on the corresponding graph receives as an
input the $i$'th row (or column) of the symmetric data matrix
$\mA$ and the scalar observation $b_{i}$. In each iteration, for
every non-zero entry $A_{ij}$, a message containing two real
numbers, $\mu_{ij}$ and $P_{ij}$, is sent from node $i$ to node
$j$ along their shared edge.

For a dense matrix $\mA$ each node out of the $n$ nodes sends a
unique message to every other node on the fully-connected graph.
This recipe results in a total of $\mathcal{o}(n^2)$ messages per
iteration round.

The algorithm can be easily implemented in a distributed fashion.
Each node $i$ on the corresponding graph receives as an input the
$i$'th row (or column) of the data matrix $\mA$ and the scalar
observation $b_{i}$. In each iteration, for every non-zero entry
$A_{ij}$, a message containing two real numbers, $\mu_{ij}$ and
$P_{ij}$, is sent from node $i$ to node $j$ along their shared
edge.}

For a dense matrix $\mA$ each
node out of the $n$ nodes sends a unique message to every other
node on the fully-connected graph. This recipe results in a total
of $n^{2}$ messages per iteration round.

The computational complexity of the GaBP solver as described in
Algorithm~\ref{alg_GaBP} for a dense linear system, in terms of
operations (multiplications and additions) per iteration round. is
shown in Table~\ref{tab_complexity}. In this case, the total
number of required operations per iteration is
$\mathcal{O}(n^{3})$. This number is obtained by evaluating the
number of operations required to generate a message multiplied by
the number of messages. Based on the summation expressions for the
propagating messages $P_{ij}$ and $\mu_{ij}$, it is easily seen
that it takes $\mathcal{O}(n)$ operations to compute such a
message. In the dense case, the graph is fully-connected resulting
in $\mathcal{O}(n^{2})$ propagating messages.

In order to estimate the total number of operations required for
the GaBP algorithm to solve the linear system, we have to evaluate
the number of iterations required for convergence. It is
known~\cite{BibDB:BookBertsekasTsitsiklis} that the number of iterations required for an
iterative solution method is $\mathcal{O}(f(\kappa))$, where
$f(\kappa)$ is a function of the condition number of the data
matrix $\mA$. Hence the total complexity of the GaBP solver can be
expressed by $\mathcal{O}(n^3)\times\mathcal{O}(f(\kappa))$. The
analytical evaluation of the convergence rate function $f(\kappa)$
is a challenging open problem. However, it can be upper bounded by
$f(\kappa)<\kappa$. Furthermore, based on our experimental study,
described in Section~\ref{sec_results}, we can conclude that
$f(\kappa)\leq\sqrt{\kappa}$. This is because typically the GaBP algorithm converges faster than the SOR algorithm. An upper bound on the number of iterations of the SOR algorithm is $\sqrt{\kappa}.$ Thus, the total complexity of the
GaBP solver in this case is
$\mathcal{O}(n^{3})\times\mathcal{O}(\sqrt{\kappa})$. For
well-conditioned (as opposed to ill-conditioned) data matrices the
condition number is $\mathcal{O}(1)$. Thus, for well-conditioned
linear systems the total complexity is $\mathcal{O}(n^3)$, \ie,
the complexity is cubic, the same order of magnitude as for direct solution
methods, like Gaussian elimination.

At first sight, this result may be considered disappointing, with
no complexity gain w.r.t. direct matrix inversion. Luckily, the
GaBP implementation as described in Algorithm~\ref{alg_GaBP} is a
naive one, thus termed naive GaBP. In this implementation we did
not take into account the correlation between the different
messages transmitted from a certain node $i$. These messages,
computed by summation, are distinct from one another in only two
summation terms.

\begin{table}
\centerline{
\begin{tabular}{|l|c|c|c|}
  \hline
  \textbf{Algorithm}
  & \textbf{Operations per msg} & \textbf{msgs} & \textbf{Total operations per iteration}\\\hline\hline & & &\\
  Naive GaBP (Algorithm~\ref{alg_GaBP}) & $\mathcal{O}(n)$ & $\mathcal{O}(n^{2})$ & $\mathcal{O}(n^{3})$ \\\hline & & & \\
  Broadcast GaBP (Algorithm~\ref{alg_GaBP_Broadcast}) & $\mathcal{O}(n)$ & $\mathcal{O}(n)$ & $\mathcal{O}(n^{2})$\\
  \hline
\end{tabular}
}\vspace{0.5cm} \caption{Computational complexity of the GaBP
solver for dense $n\times n$ matrix $\mA$.}\label{tab_complexity}
\end{table}

\comment{The number of messages passed on the graph can be reduced
significantly down to $n$ messages per round by using a similar
construction to Bickson \etal~\cite{BroadcastBP}. This
modification is described in the following corollary:
\begin{corol}[GaBP solver with reduced message passing]\label{corol_5}
Instead of sending a message composed of the pair of $\mu_{ij}$
and $P_{ij}$, a node broadcasts aggregated sums, and consequently
each node can retrieve locally the $P_{i\backslash
j}$~(\ref{eq_prec}) and $\mu_{i\backslash j}$~(\ref{eq_mean}) from
the sums by means of a subtraction:
\end{corol}}

\begin{table}[h!]
\begin{alg}\label{alg_GaBP_Broadcast}\end{alg}\centerline{
\begin{tabular}{|lll|}
  \hline&&\\
  \texttt{1.} & \emph{\texttt{Initialize:}} & $\checkmark$\quad\texttt{Set the neighborhood} $\textrm{N}(i)$ \texttt{to include}\\&&\quad\quad$\forall k\neq i \exists A_{ki}\neq0$.\\&& $\checkmark$\quad\texttt{Set the scalar fixes}\\&&$\quad\quad P_{ii}=A_{ii}$ \texttt{and} $\mu_{ii}=b_{i}/A_{ii}$, $\forall i$.\\
  && $\checkmark$\quad\texttt{Set the initial $i\rightarrow\textrm{N}(i)$ broadcast messages}\\&&\quad\quad $\tilde{P_{i}}=0$ \texttt{and} $\tilde{\mu}_{i}=0$.\\&&$\checkmark$\quad\texttt{Set the initial $\textrm{N}(i)\ni k\rightarrow i$ internal scalars}\\&&\quad\quad $P_{ki}=0$ \texttt{and} $\mu_{ki}=0$.\\&& $\checkmark$\quad \texttt{Set a convergence threshold} $\epsilon$.\\
  {\texttt{2.}} & {\emph{\texttt{Iterate:}} }&  $\checkmark$\quad\texttt{Broadcast the aggregated sum messages}\\&&$\quad\quad \tilde{P}_{i}=P_{ii}+\sum_{{k}\in\textrm{N}(i)}
P_{ki}$,\\&&
$\quad\quad\tilde{\mu}_{i}=\tilde{P_{i}}^{-1}(P_{ii}\mu_{ii}+\sum_{k
\in \textrm{N}(i)} P_{ki}\mu_{ki})$, $\forall
i$\\&&\quad\quad\texttt{(under certain scheduling)}.\\&&
  $\checkmark$\quad\texttt{Compute the} $\textrm{N}(j)\ni i\rightarrow j$ \texttt{internal scalars} \\&& $\quad\quad P_{ij} = -A_{ij}^{2}/(\tilde{P}_{i}-P_{ji})$,\\
  &&$\quad\quad\mu_{ij}=(\tilde{P_{i}}\tilde{\mu_{i}}-P_{ji}\mu_{ji})/A_{ij}$.\\
  {\texttt{3.}} & {\emph{\texttt{Check:}}} & $\checkmark$\quad\texttt{If the internal scalars} $P_{ij}$ \texttt{and} $\mu_{ij}$ \texttt{did not}\\&&\quad\quad\texttt{converge (w.r.t. $\epsilon$),} \texttt{return to
    Step 2.}\\&&$\checkmark$\quad\texttt{Else, continue to Step 4.}\\
  {\texttt{4.}} & {\emph{\texttt{Infer:}}} & $\checkmark$\quad\texttt{Compute the marginal means}\\&&{\quad\quad$\mu_{i}=\big(P_{ii}\mu_{ii}+\sum_{k \in
\textrm{N}(i)}P_{ki}\mu_{ki}\big)/\big(P_{ii}+\sum_{{k}\in\textrm{N}(i)}
P_{ki}\big)=\tilde{\mu}_{i}$, $\forall i$.}\\&&$(\checkmark$\quad\texttt{Optionally compute the marginal precisions}\\&&\quad\quad$P_{i}=P_{ii}+\sum_{k\in\textrm{N}(i)}P_{ki}=\tilde{P}_{i}\quad)$\\
  {\texttt{5.}} & {\emph{\texttt{Solve:}}} & $\checkmark$\quad\texttt{Find the solution}\\&& {$\quad\quad x_{i}^{\ast}=\mu_{i}$, $\forall i$.}\\&&\\\hline
\end{tabular}}
\end{table}

Instead of sending a message composed of the pair of $\mu_{ij}$
and $P_{ij}$, a node can broadcast the aggregated sums \BEA
\tilde{P}_{i}&=&P_{ii}+\sum_{{k}\in\textrm{N}(i)}
P_{ki},\\\tilde{\mu}_{i}&=&\tilde{P}_{i}^{-1}(P_{ii}\mu_{ii}+\sum_{k
\in \textrm{N}(i)} P_{ki}\mu_{ki}). \EEA Consequently, each node
can retrieve locally the $P_{i\backslash j}$~(\ref{eq_prec}) and
$\mu_{i\backslash j}$~(\ref{eq_mean}) from the sums by means of a
subtraction \BEA P_{i\backslash
j}&=&\tilde{P}_{i}-P_{ji}\,,\\\mu_{i\backslash
j}&=&\tilde{\mu}_{i}-P_{i \backslash j}^{-1}P_{ji}\mu_{ji}\,.\EEA
The rest of the algorithm remains the same. On dense graphs, the broadcast version sends
$O(n)$ messages per round, instead of $O(n^2)$ messages in the GaBP algorithm.
This construction is typically useful when implementing the GaBP\ algorithm in communication networks that support broadcast (for example Ethernet and wireless networks),\ where the messages are
sent in broadcast anyway. See for example \cite{NBP,Shiff}. Chapter \ref{chap:svm} brings an example of large scale implementation of our broadcast variant using 1,024 CPUs. 

\section{The GaBP-Based Solver and Classical Solution\\ Methods}
\subsection{Gaussian Elimination}

\begin{prop}\label{prop_GaBPGE}
The GaBP-based solver (Algorithm~\ref{alg_GaBP}) for a system of
linear equations represented by a tree graph is identical to the
renowned Gaussian elimination algorithm (a.k.a. LU
factorization,~\cite{BibDB:BookBertsekasTsitsiklis}).
\end{prop}

\begin{proof}[{\bf Proof}]
Consider a set of $n$ linear equations with $n$ unknown variables,
a unique solution and a tree graph representation. We aim at
computing the unknown variable associated with the root node.
Without loss of generality as the tree can be drawn with any of
the other nodes being its root. Let us enumerate the nodes in an
ascending order from the root to the leaves (see, e.g.,
Fig. \ref{fig_tree}).

\begin{center}
\begin{figure}[h!]
\begin{center}
  \includegraphics[width=100pt]{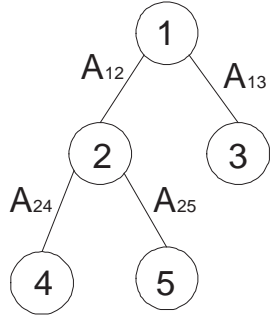}\\
  \caption{Example topology of a tree with 5 nodes}\label{fig_tree}
\end{center}
\end{figure}
\end{center}

As in a tree, each child node (\ie, all nodes but the root) has
only one parent node and based on the top-down ordering, it can be
easily observed that the tree graph's corresponding data matrix
$\mA$ must have one and only one non-zero entry in the upper
triangular portion of its columns. Moreover, for a leaf node this
upper triangular entry is the only non-zero off-diagonal entry in
the whole column. See, for example, the data matrix associated
with the tree graph depicted in Fig~\ref{fig_tree} \BE \left(
  \begin{array}{ccccc}
    A_{11} & \mathbf{A_{12}} & \underline{\mathbf{A_{13}}} & 0 & 0 \\
    A_{12} & A_{22} & 0 & \underline{\mathbf{A_{24}}} & \underline{\mathbf{A_{25}}} \\
    A_{13} & 0 & A_{33} & 0 & 0 \\
    0 & A_{24} & 0 & A_{44} & 0 \\
    0 & A_{25} & 0 & 0 & A_{55} \\
  \end{array}
\right), \EE where the non-zero upper triangular entries are in
bold and among these the entries corresponding to leaves are
underlined.

Now, according to GE we would like to lower triangulate the matrix
$\mA$. This is done by eliminating these entries from the leaves
to the root. Let $l$ be a leaf node, $i$ be its parent and $j$ be
its parent ($l$'th node grandparent). Then, the $l$'th row is
multiplied by $-A_{li}/A_{ll}$ and added to the $i$'th row. in
this way the $A_{li}$ entry is being eliminated. However, this
elimination, transforms the $i$'th diagonal entry to be
$A_{ii}\rightarrow A_{ii}-A_{li}^{2}/A_{ll}$, or for multiple
leaves connected to the same parent $A_{ii}\rightarrow
A_{ii}-\sum_{l\in\textrm{N}(i)\j}A_{li}^{2}/A_{ll}$. In our
example, \BE \left(
  \begin{array}{ccccc}
    A_{11} & A_{12} & 0 & 0 & 0 \\
    A_{12} & A_{22}-A_{13}^2/A_{33}-A_{24}^2/A_{44}-A_{25}^2/A_{55} & 0 & 0 & 0 \\
    A_{13} & 0 & A_{33} & 0 & 0 \\
    0 & A_{24} & 0 & A_{44} & 0 \\
    0 & A_{25} & 0 & 0 & A_{55} \\
  \end{array}
\right). \EE

Thus, in a similar manner, eliminating the parent $i$ yields the
multiplication of the $j$'th diagonal term by
$-A_{ij}^{2}/(A_{ii}-\sum_{l\in\textrm{N}(i)\j}A_{li}^{2}/A_{ll})$.
Recalling that $P_{ii}=A_{ii}$, we see that the last expression is
identical to the update rule of $P_{ij}$ in GaBP. Again, in our
example \BE \left(
  \begin{array}{ccccc}
    B & 0 & 0 & 0 & 0 \\
    0 & C & 0 & 0 & 0 \\
    A_{13} & 0 & A_{33} & 0 & 0 \\
    0 & A_{24} & 0 & A_{44} & 0 \\
    0 & A_{25} & 0 & 0 & A_{55} \\
  \end{array}
\right), \EE where $B =
A_{11}-A_{12}^{2}/(A_{22}-A_{13}^2/A_{33}-A_{24}^2/A_{44}-A_{25}^2/A_{55})$,
$C=A_{22}-A_{13}^2/A_{33}-A_{24}^2/A_{44}-A_{25}^2/A_{55}$. Now
the matrix is fully lower triangulated. To put differently in
terms of GaBP, the $P_{ij}$ messages are subtracted from the
diagonal $P_{ii}$ terms to triangulate the data matrix of the
tree. Performing the same row operations on the right hand side
column vector $\vb$, it can be easily seen that we equivalently
get that the outcome of the row operations is identical to the
GaBP solver's $\mu_{ij}$ update rule. These updadtes/row
operations can be repeated, in the general case, until the matrix
is lower triangulated.

Now, in order to compute the value of the unknown variable
associated with the root node, all we have to do is divide the
first diagonal term by the transformed value of $b_{1}$, which is
identical to the infer stage in the GaBP solver (note that by
definition all the nodes connected to the root are its children,
as it does not have parent node). In the example \BE
x_{1}^{\ast}=\frac{A_{11}-A_{12}^{2}/(A_{22}-A_{13}^2/A_{33}-A_{24}^2/A_{44}-A_{25}^2/A_{55})}{b_{11}-A_{12}/(b_{22}-A_{13}/A_{33}-A_{24}/A_{44}-A_{25}/A_{55})}\,.
\EE
\noindent Note that the rows corresponding to leaves remain unchanged.

To conclude, in the tree graph case, the `iterative' stage (stage
2 on algorithm~\ref{alg_GaBP}) of the GaBP solver actually
performs lower triangulation of the matrix, while the `infer'
stage (stage 4) reducers to what is known as forward substitution.
Evidently, using an opposite ordering, one can get the
complementary upper triangulation and back substitution,
respectively.
\end{proof}

It is important to note, that based on this proposition, the GaBP
solver can be viewed as GE run over an unwrapped version (\ie, a
computation tree as defined in \cite{BibDB:Weiss01Correctness}) of a general loopy graph.

\subsection{Iterative Methods}
Iterative methods that can be expressed in the simple form \BE\label{eq_iter}
\vx^{(t)}=\mB\vx^{(t-1)}+\vc, \EE where neither the iteration matrix $\mB$
nor the vector $\vc$ depend upon the iteration number $t$, are
called stationary iterative methods. In the following, we discuss
three main stationary iterative methods: the Jacobi method, the
Gauss-Seidel (GS) method and the successive overrelaxation (SOR)
method. The GaBP-based solver, in the general case, can not be
written in this form, thus can not be categorized as a stationary
iterative method.

\begin{prop}
\cite{BibDB:BookBertsekasTsitsiklis}
Assuming $\mI-\mB$ is invertible, then the iteration~\ref{eq_iter}
converges (for any initial guess, $\vx^{(0)}$).
\end{prop}

\subsection{Jacobi Method}
The Jacobi method (Gauss, 1823, and Jacobi
1845,\cite{BibDB:BookAxelsson}), a.k.a. the simultaneous iteration
method, is the oldest iterative method for solving a square linear system of
equations $\boldsymbol{A}\vx = \vb$. The method assumes that $\forall_i\mbox{  } A_{ii} \ne 0$.  It's complexity is $\mathcal{O}(n^{2})$ per iteration.
There are two know sufficient convergence conditions for the Jacobi method.
The first condition holds when the matrix $\mA$ is strictly or irreducibly diagonally dominant.
The second condition holds when $\rho(\boldsymbol{D}^{-1}(\boldsymbol{L}+\boldsymbol{U})) < 1.$
Where $\boldsymbol{D} = \diag(\boldsymbol{A})$, $\boldsymbol{L},\boldsymbol{U}$ are upper and lower triangular matrices of $\boldsymbol{A}$.

\begin{prop}\label{prop_GaBPJ}
The GaBP-based solver (Algorithm~\ref{alg_GaBP})
\begin{enumerate}
  \item with inverse variance messages  set to zero, \ie, $P_{ij}=0, i\in\textrm{N}(j),\forall{j}$;
  \item incorporating the message received from node $j$ when computing the message to be sent from node $i$ to node $j$, \ie replacing $k\in\textrm{N}(i)\backslash j$ with $k\in\textrm{N}(i)$,
\end{enumerate}
is identical to the Jacobi iterative method.
\end{prop}
\begin{proof}[{\bf Proof}]
Setting the precisions to zero, we get in
correspondence to the above derivation, \BEA
P_{i\backslash j}&=&P_{ii}=A_{ii},\\
P_{ij}\mu_{ij}&=&-A_{ij}\mu_{i\backslash j},\\
\label{eq_marginal_J}\mu_{i}&=&A_{ii}^{-1}(b_{i}-\sum_{k\in\textrm{N}(i)}A_{ki}\mu_{k\backslash
i}). \EEA Note that the inverse relation between $P_{ij}$ and
$P_{i\backslash j}$~(\ref{eq_prec_message}) is no longer valid in
this case.

Now, we rewrite the mean $\mu_{i\backslash j}$~(\ref{eq_mean})
without excluding the information from node $j$, \BE
\mu_{i\backslash
j}=A_{ii}^{-1}(b_{i}-\sum_{k\in\textrm{N}(i)}A_{ki}\mu_{k\backslash
i}). \EE Note that $\mu_{i\backslash j}=\mu_{i}$, hence
the inferred marginal mean $\mu_{i}$~(\ref{eq_marginal_J}) can be
rewritten as \BE \mu_{i}=A_{ii}^{-1}(b_{i}-\sum_{k\neq
i}A_{ki}\mu_{k}), \EE where the expression for all neighbors of
node $i$ is replaced by the redundant, yet identical, expression
$k\neq i$. This fixed-point iteration is identical to the renowned
Jacobi method, concluding the proof.
\end{proof}

The fact that Jacobi iterations can be obtained
as a special case of the GaBP solver further indicates the
richness of the proposed algorithm. 
%
%

\chapter{Numerical Examples}\label{sec_results}\label{chap3}
In this chapter we report experimental study of three numerical examples: toy linear system, 2D Poisson equation and symmetric non-PSD matrix. In all examples, but the Poisson's
equation~\ref{eq_poisson}, $\vb$ is assumed to be an $m$-length all-ones
observation vector. For fairness in comparison, the initial guess
in all experiments, for the various solution methods under
investigation, is taken to be the same and is arbitrarily set to
be equal to the value of the vector $\vb$. The stopping criterion
in all experiments determines that for all propagating messages
(in the context the GaBP solver) or all $n$ tentative solutions
(in the context of the compared iterative methods) the absolute
value of the difference should be less than $\epsilon\leq10^{-6}$.
As for terminology, in the following performing GaBP with parallel
(flooding or synchronous) message scheduling is termed `parallel
GaBP', while GaBP with serial (sequential or asynchronous) message
scheduling is termed `serial GaBP'.

\section[Toy Linear System]{Numerical Example: Toy Linear System: $3\times3$
Equations}
Consider the following $3\times3$ linear system \BE
\underbrace{\left(
                                                 \begin{array}{lll}
                                                   A_{xx}=1 & A_{xy}=-2 & A_{xz}=3 \\
                                                   A_{yx}=-2 & A_{yy}=1 & A_{yz}=0 \\
                                                   A_{zx}=3 & A_{zy}=0 & A_{zz}=1 \\
                                                 \end{array}
                                               \right)}_{\mA}\underbrace{\left(
                                                        \begin{array}{r}
                                                          x \\
                                                          y \\
                                                          z \\
                                                        \end{array}
                                                      \right)}_{\vx}=\underbrace{\left(
                                                                \begin{array}{r}
                                                                  -6 \\
                                                                  0 \\
                                                                  2 \\
                                                                \end{array}
                                                              \right)}_{{\vb}}. \EE
We would like to find the solution to this system,
$\vx^{\ast}=\{x^{\ast},y^{\ast},z^{\ast}\}^{T}$. Inverting the
data matrix $\mA$, we directly solve  \BE \underbrace{\left(
                                                                     \begin{array}{r}
                                                                       x^{\ast} \\
                                                                       y^{\ast} \\
                                                                       z^{\ast} \\
                                                                     \end{array}
                                                                   \right)}_{\vx^{\ast}}=
\underbrace{\left(
                                                 \begin{array}{rrr}
                                                   -1/12 & -1/6 & 1/4 \\
                                                   -1/6 & 2/3 & 1/2 \\
                                                   1/4 & 1/2 & 1/4 \\
                                                 \end{array}
                                               \right)}_{\mA^{-1}}\underbrace{\left(
                                                        \begin{array}{r}
                                                          -6 \\
                                                          0 \\
                                                          2 \\
                                                        \end{array}
                                                      \right)}_{\vb}=\left(
                                                                       \begin{array}{r}
                                                                         1 \\
                                                                         2 \\
                                                                        -1 \\
                                                                       \end{array}
                                                                     \right)
. \EE

\begin{figure}[h!]\label{fig_1}
\begin{center}
    \includegraphics[width=0.4\textwidth]{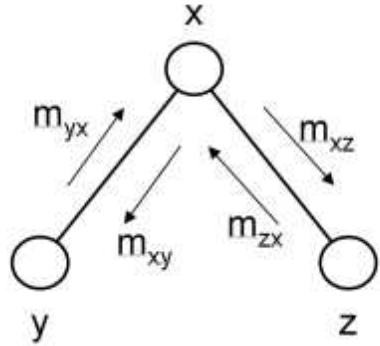}
  \caption{A tree topology with three nodes}
  \label{fig_tree_topo}
\end{center}
\end{figure}

Alternatively, we can now run the GaBP solver.
Fig.~\ref{fig_tree_topo} displays the graph, corresponding to the
data matrix $\mA$, and the message-passing flow. As
$A_{yz}=A_{zy}=0$, this graph is a cycle-free tree, thus GaBP is
guaranteed to converge in a finite number of rounds. As
demonstrated in the following, in this example GaBP converges only
in two rounds, which equals the tree's diameter. Each propagating
message, $m_{ij}$, is described by two scalars $\mu_{ij}$ and
$P_{ij}$, standing for the mean and precision of this
distribution. The evolution of the propagating means and
precisions, until convergence, is described in
Table~\ref{tab_evolve}, where the notation $t=0,1,2,3$ denotes the
iteration rounds. Converged values are written in bold.

\begin{table}[h!]
\begin{center}
$\begin{array}{|l|l|r|r|r|r|}
    \hline \textrm{Message}&\textrm{Computation}&\textrm{t=0}&\textrm{t=1}&\textrm{t=2}&\textrm{t=3}\\\hline\hline
  P_{xy} & -A_{xy}^{2}/(P_{xx}+P_{zx}) & 0 & -4& 1/2 & \mathbf{1/2} \\\hline
  P_{yx} & -A_{yx}^{2}/(P_{yy}) & 0 & -4 & \mathbf{-4} & \mathbf{-4} \\\hline
  P_{xz} & -A_{xz}^{2}/(P_{zz}) & 0 & -9 & 3 & \mathbf{3} \\\hline
  P_{zx} & -A_{zx}^{2}/(P_{xx}+P_{yx}) & 0 & -9 & \mathbf{-9} & \mathbf{-9} \\\hline
  \mu_{xy} & (P_{xx}\mu_{xx}+P_{zx}\mu_{zx})/A_{xy} & 0 & 3 & 6 & \mathbf{6} \\\hline
  \mu_{yx} & P_{yy}\mu_{yy}/A_{yx} & 0 & \mathbf{0} & \mathbf{0} & \mathbf{0} \\\hline
  \mu_{xz} & (P_{xx}\mu_{xx}+P_{yx}\mu_{yx})/A_{xz} & 0 & -2 & \mathbf{-2} & \mathbf{-2} \\\hline
  \mu_{zx} & P_{zz}\mu_{zz}/A_{zx} & 0 & 2/3 & \mathbf{2/3} & \mathbf{2/3}\\\hline
\end{array}$
\label{tab_evolve}
\end{center}
\caption{Evolution of means and precisions on a tree with three
nodes}
\end{table}
Next, following the GaBP solver algorithm, we infer the marginal
means. For exposition purposes we also present in
Table~\ref{tentative_means} the tentative solutions at each
iteration round.

\begin{table}[h!]
\begin{center}
$
\begin{array}{|l|l|r|r|r|r|}
    \hline \textrm{Solution}&\textrm{Computation}&\textrm{t=0}&\textrm{t=1}&\textrm{t=2}&\textrm{t=3}\\\hline\hline
  \mu_{x} & \big(P_{xx}\mu_{xx}+P_{zx}\mu_{zx}+P_{yx}\mu_{yx}\big)/\big(P_{xx}+P_{zx}+P_{yx}\big) & -6 & 1 & \mathbf{1} & \mathbf{1} \\\hline
  \mu_{y} & \big(P_{yy}\mu_{yy}+P_{xy}\mu_{xy}\big)/\big(P_{yy}+P_{xy}\big) & 0 & 4 & 2 & \mathbf{2}\\\hline
  \mu_{z} & \big(P_{zz}\mu_{zz}+P_{xz}\mu_{xz}\big)/\big(P_{zz}+P_{xz}\big) & 2 & -5/2& -1 & \mathbf{-1}\\\hline
\end{array}.
$ \comment{ \BEA
\mu_{x}&=&\big(P_{xx}\mu_{xx}+P_{zx}\mu_{zx}+P_{yx}\mu_{yx}\big)/\big(P_{xx}+P_{zx}+P_{yx}\big)=1,\\
\mu_{y}&=&\big(P_{yy}\mu_{yy}+P_{xy}\mu_{xy}\big)/\big(P_{yy}+P_{xy}\big)=2,\\
\mu_{z}&=&\big(P_{zz}\mu_{zz}+P_{xz}\mu_{xz}\big)/\big(P_{zz}+P_{xz}\big)=-1,
\EEA}
\end{center}
\caption{Tentative means computed on each iteration until
convergence} \label{tentative_means}
\end{table}

Thus, as expected, the GaBP solution
$\vx^{\ast}=\{x^{\ast}=1,y^{\ast}=2,z^{\ast}=-1\}^{T}$ is
identical to what is found using the direct approach. Note that
as the linear system is described by a tree graph, then for this
particular case, the inferred precision is also exact \BEA
P_{x}&=&P_{xx}+P_{yx}+P_{zx}=-12,\\
P_{y}&=&P_{yy}+P_{xy}=3/2,\\
P_{z}&=&P_{zz}+P_{xz}=4.\\
\EEA and gives
$\{P_{x}^{-1}=\{\mA^{-1}\}_{xx}=-1/12,P_{y}^{-1}=\{\mA^{-1}\}_{yy}=2/3,P_{z}^{-1}=\{\mA^{-1}\}_{zz}=1/4\}^{T}$,
\ie the true diagonal values of the data matrix's inverse,
$\mA^{-1}$.

\section[Non PSD Example]{Numerical Example: Symmetric Non-PSD Data
Matrix}\label{sec_nonPSD}

Consider the case of a linear system with a symmetric, but non-PSD
data matrix \BE
\left(%
\begin{array}{ccc}
  1 & 2 & 3 \\
  2 & 2 & 1 \\
  3 & 1 & 1 \\
\end{array}%
\right). \EE

Table~\ref{tab_nonPSD} displays the number of iterations required
for convergence for the iterative methods under consideration. The
classical methods diverge, even when aided with acceleration
techniques. This behavior (at least without the acceleration) is
not surprising in light of Theorem~\ref{spectral_radius_thm}. Again we observe that
serial scheduling of the GaBP solver is superior parallel
scheduling and that applying Steffensen iterations reduces the
number of iterations in $45\%$ in both cases. Note that SOR cannot
be defined when the matrix is not PSD. By definition CG works only
for symmetric PSD matrices. Because the solution is a saddle point
and not a minimum or maximum.

\begin{table}[h!]
\centerline{
\begin{tabular}{|c|r|}
  \hline
  \textbf{Algorithm}
  & \textbf{Iterations} $t$\\
  \hline\hline & \\
  Jacobi,GS,SR,Jacobi+Aitkens,Jacobi+Steffensen & $-$\\\hline & \\
  \textbf{Parallel GaBP} & \textbf{38} \\\hline & \\
  \textbf{Serial GaBP} & \textbf{25} \\\hline & \\
  \textbf{Parallel GaBP+Steffensen} & \textbf{21} \\\hline & \\
  \textbf{Serial GaBP+Steffensen} & \textbf{14} \\
  \hline
\end{tabular}
}\vspace{0.5cm}\caption{Symmetric non-PSD $3\times3$ data matrix.
Total number of iterations required for convergence (threshold
$\epsilon=10^{-6}$) for GaBP-based solvers vs. standard
methods.}\label{tab_nonPSD}
\end{table}

\begin{figure}[h!]\label{fig_S}
\begin{center}
    \includegraphics[width=0.5\textwidth]{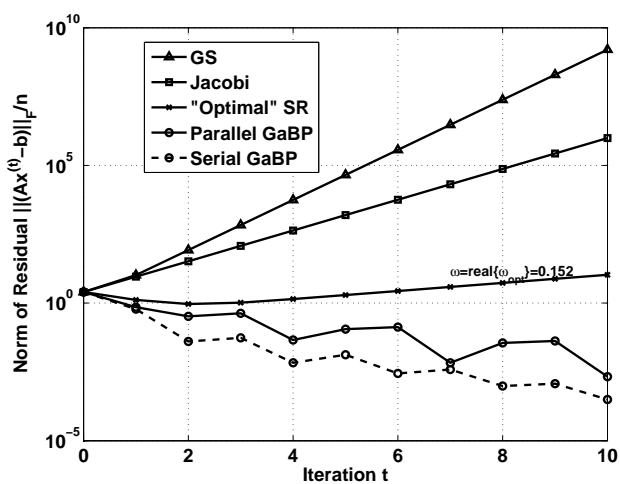}
  \caption{Convergence rate for a $3\times3$ symmetric non-PSD data matrix.
  The Frobenius norm of the residual per equation, $||\mA\vx^{t}-b||_{F}/n$, as a function of the iteration $t$
  for GS (triangles and solid line), Jacobi (squares and solid line), SR (stars and solid line), parallel GaBP (circles and solid line)
  and serial GaBP (circles and dashed line) solvers.}
\end{center}
\end{figure}

\begin{figure}[t!]
\begin{minipage}[b]{0.5\linewidth}
\centering
     \includegraphics[width=\textwidth]{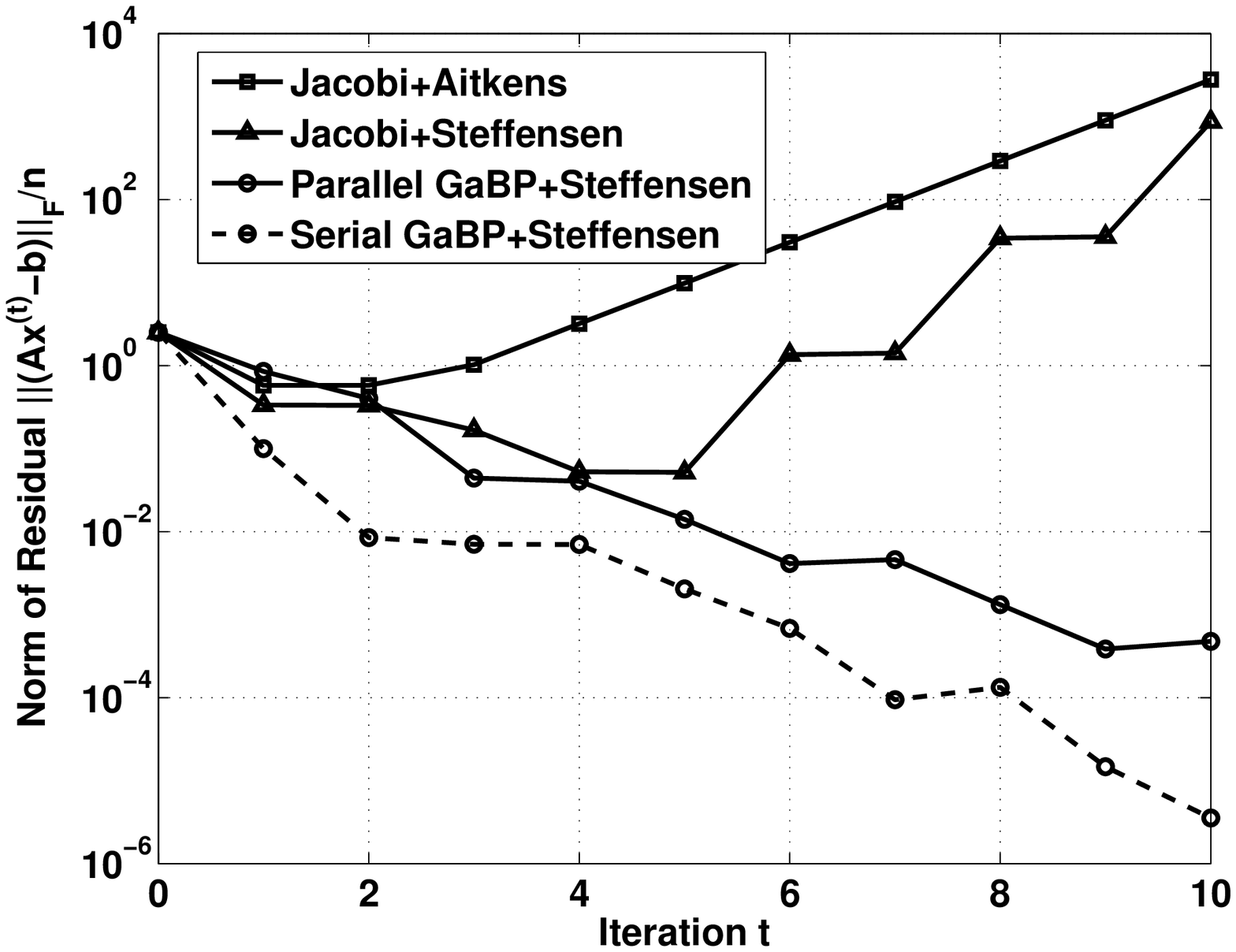}
 \label{fig_S_accel}
\end{minipage}
\begin{minipage}[b]{0.5\linewidth}
\centering
   \includegraphics[width=\textwidth]{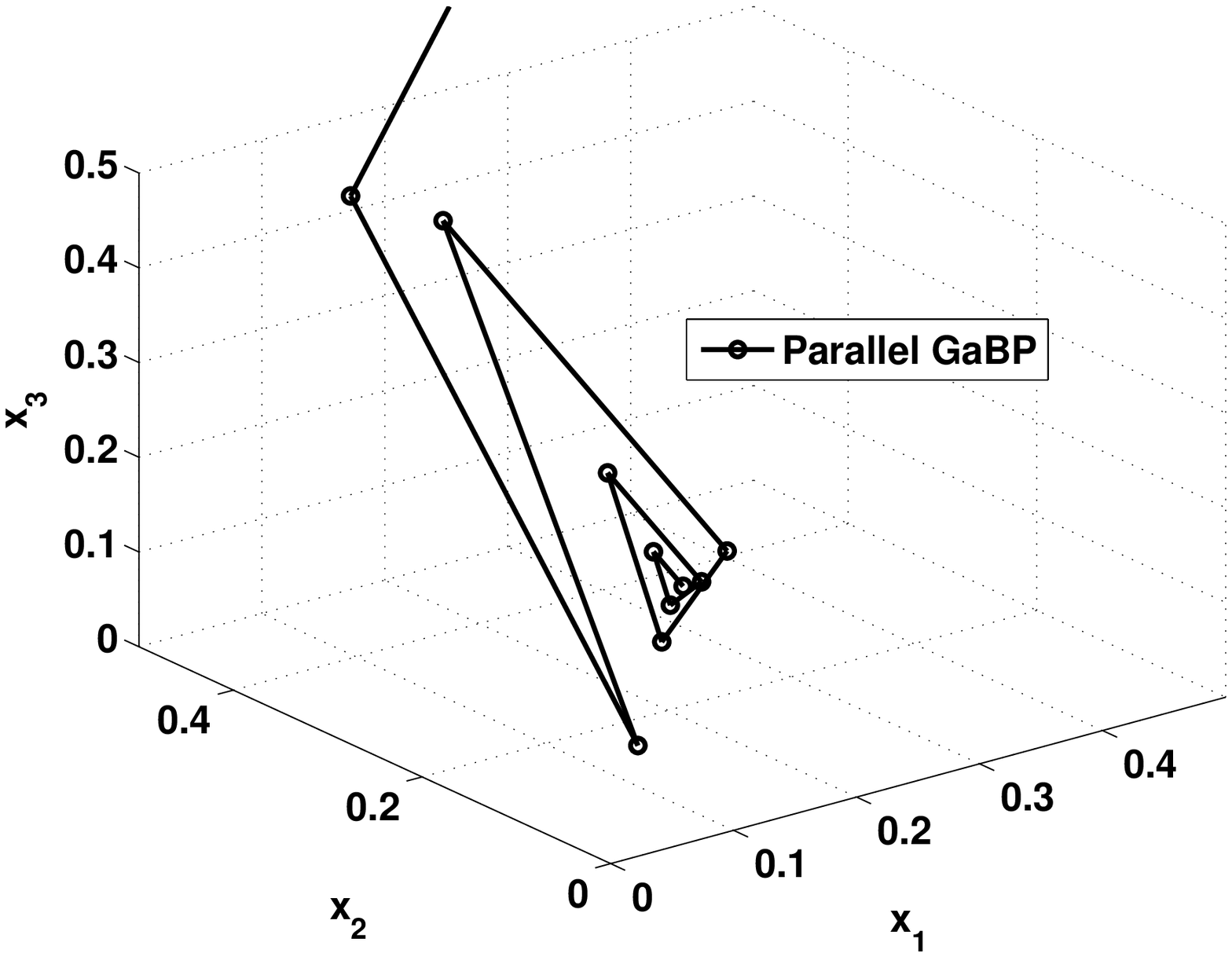}
 \label{fig_S_spiral}
\end{minipage}
\caption{The left graph depicts accelerated convergence rate for a $3\times3$ symmetric non-PSD data matrix.
  The Frobenius norm of the residual per equation, $||\mA\vx^{t}-b||_{F}/n$, as a function of the iteration $t$
  for Aitkens (squares and solid line) and Steffensen-accelerated (triangles and solid line) Jacobi method, parallel GaBP (circles and solid line)   and serial GaBP (circles and dashed line) solvers accelerated by Steffensen iterations.
  The right graph shows a visualization of parallel GaBP on the same problem, drawn in $\mathbb{R}^{3}$.}
\end{figure}

\section[2D Poisson's]{Application Example: 2-D Poisson's Equation} One of the
most common partial differential equations (PDEs) encountered in
various areas of exact sciences and engineering (\eg, heat flow,
electrostatics, gravity, fluid flow, quantum mechanics,
elasticity) is Poisson's equation. In two dimensions, the equation
is \BE \Delta u(x,y)=f(x,y), \label{eq_poisson} \EE for $\{x,y\}\in\Omega$, where \BE
\Delta{(\cdot)}=\pdpd{(\cdot)}{x}+\pdpd{(\cdot)}{y}. \EE is the
Laplacian operator and $\Omega$ is a bounded domain in
$\mathbb{R}^{2}$. The solution is well defined only under boundary
conditions, \ie, the value of $u(x,y)$ on the boundary of $\Omega$
is specified. We consider the simple (Dirichlet) case of
$u(x,y)=0$ for \{x,y\} on the boundary of $\Omega$. This equation
describes, for instance, the steady-state temperature of a uniform
square plate with the boundaries held at temperature $u=0$, and
$f(x,y)$ equaling the external heat supplied at point $\{x,y\}$.

\begin{figure}[h!]\label{fig_P_accel}
\begin{center}
    \includegraphics[width=0.5\textwidth]{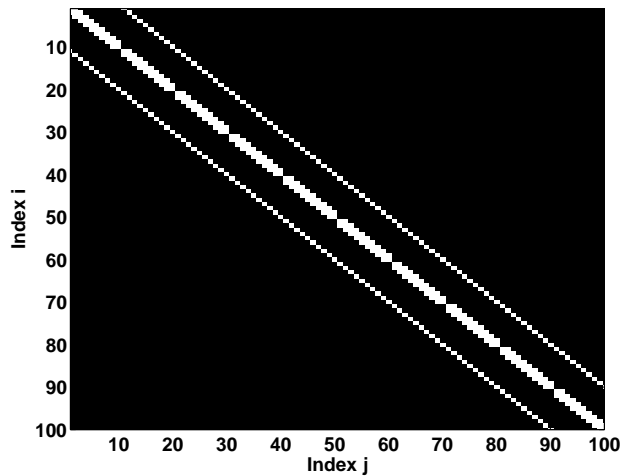}
  \caption{Image of the corresponding sparse data matrix for the 2-D discrete Poisson's PDE with $p=10$. Empty (full) squares denote non-zero (zero) entries.}
  \label{2D_Poisson}
\end{center}
\end{figure}

The poisson's PDE can be discretized by using finite differences.
An $p+1\times p+1$ square grid on $\Omega$ with size (arbitrarily)
set to unity is used, where $h\triangleq1/(p+1)$ is the grid
spacing. We let $U(i,j)$, $\{i,j=0,\ldots,p+1\}$, be the
approximate solution to the PDE at $x=ih$ and $y=jh$.
Approximating the Laplacian by \BEA \Delta
U(x,y)&=&\pdpd{(U(x,y))}{x}+\pdpd{(U(x,y))}{y}\nonumber\\&\approx&\frac{U(i+1,j)-2U(i,j)+U(i-1,j)}{h^{2}}+\frac{U(i,j+1)-2U(i,j)+U(i,j-1)}{h^{2}}\nonumber,
\EEA one gets the system of $n=p^{2}$ linear equations with $n$
unknowns \BE 4U(i,j)-U(i-1,j)-U(i+1,j)-U(i,j-1)-U(i,j+1)=b(i,j)
\forall i,j=1,\ldots,p, \EE where
$b(i,j)\triangleq-f(ih,jh)h^{2}$, the scaled value of the function
$f(x,y)$ at the corresponding grid point $\{i,j\}$. Evidently, the
accuracy of this approximation to the PDE increases with $n$.

Choosing a certain ordering of the unknowns $U(i,j)$, the linear
system can be written in a matrix-vector form. For example, the
natural row ordering (\ie, enumerating the grid points
left$\rightarrow$right, bottom$\rightarrow$up) leads to a linear
system with $p^{2}\times p^{2}$ sparse data matrix $\mA$. For
example, a Poisson PDE with $p=3$ generates the following
$9\times9$ linear system
\BE\underbrace{\left(%
\begin{array}{ccc|ccc|ccc}
  4 & -1 &  & -1 &  &  &  &\\
  -1 & 4 & -1 &  & -1 &  &  &\\
   & -1 & 4  &  &  & -1 &  &\\\hline
  -1 &  &  & 4 & -1 &  & -1 &\\
   & -1 &  & -1 & 4 & -1 &  & -1\\
   &  & -1 &  & -1 & 4 &  & &-1\\\hline
   &  &  & -1 &  &  & 4 &-1  &\\
   &  &  &  & -1 &  & -1 & 4 &-1\\
   &  &  &  &  & -1 &  & -1 & 4\\
\end{array}%
\right)}_{\mA}\underbrace{\left(%
\begin{array}{c}
  U(1,1) \\ U(2,1) \\ U(3,1) \\ U(1,2) \\ U(2,2) \\ U(3,2)  \\ U(1,3) \\ U(2,3) \\ U(3,3)\\
\end{array}%
\right)}_{\vx}=\underbrace{\left(%
\begin{array}{c}
b(1,1) \\ b(2,1) \\ b(3,1) \\ b(1,2) \\ b(2,2) \\ b(3,2)
\\ b(1,3) \\ b(2,3) \\ b(3,3)\\\end{array}%
\right)}_{\vb},\EE where blank data matrix $\mA$ entries denote
zeros.

\begin{table}[h!]
\centerline{
\begin{tabular}{|c|r|}
  \hline
  \textbf{Algorithm}
  & \textbf{Iterations} $t$\\
  \hline\hline & \\
  Jacobi & $354$\\\hline & \\
  GS & $136$\\\hline & \\
  Optimal SOR & $37$\\\hline & \\
  \textbf{Parallel GaBP} & \textbf{134} \\\hline & \\
  \textbf{Serial GaBP} & \textbf{73} \\\hline & \\
  \textbf{Parallel GaBP+Aitkens} & \textbf{25} \\\hline & \\
  \textbf{Parallel GaBP+Steffensen} & \textbf{56} \\\hline & \\
  \textbf{Serial GaBP+Steffensen} & \textbf{32} \\\hline
\end{tabular}
}\vspace{0.5cm}\caption{2-D discrete Poisson's PDE with $p=3$ and
$f(x,y)=-1$. Total number of iterations required for convergence
(threshold $\epsilon=10^{-6}$) for GaBP-based solvers vs. standard
methods.}\label{tab_2D_Poisson}
\end{table}

\begin{figure}[h!]\label{fig_P_accel}
\begin{center}
    \includegraphics[width=0.5\textwidth]{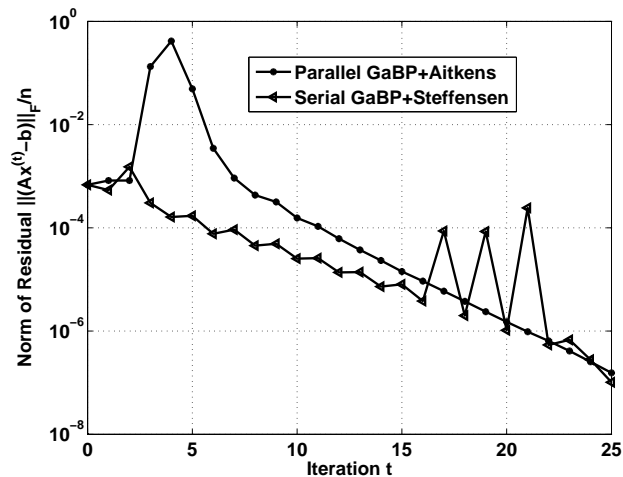}
  \caption{Accelerated convergence rate for the 2-D discrete Poisson's PDE with $p=10$ and $f(x,y)=-1$. The Frobenius norm of the residual. per equation, $||\mA\vx^{t}-b||_{F}/n$, as a function of the iteration $t$ for parallel GaBP solver accelrated by Aitkens method ($\times$-marks and solid line) and serial GaBP solver accelerated by Steffensen iterations (left triangles and dashed line).}
  \end{center}
\end{figure}
Hence, now we can solve the discretized 2-D Poisson's PDE by
utilizing the GaBP algorithm. Note that, in contrast to the other
examples, in this case the GaBP solver is applied for solving a
sparse, rather than dense, system of linear equations.

In order to evaluate the performance of the GaBP solver, we choose
to solve the 2-D Poisson's equation with discretization of $p=10$.
The structure of the corresponding $100\times100$ sparse data
matrix is illustrated in Fig.~\ref{2D_Poisson}.

\begin{table}
\centerline{
\begin{tabular}{|c|r|}
  \hline
  \textbf{Algorithm}
  & \textbf{Iterations} $t$\\
  \hline\hline & \\
  Jacobi,GS,SR,Jacobi+Aitkens,Jacobi+Steffensen & $-$\\\hline & \\
  \textbf{Parallel GaBP} & \textbf{84} \\\hline & \\
  \textbf{Serial GaBP} & \textbf{30} \\\hline & \\
  \textbf{Parallel GaBP+Steffensen} & \textbf{43} \\\hline & \\
  \textbf{Serial GaBP+Steffensen} & \textbf{17} \\
  \hline
\end{tabular}
}\vspace{0.5cm}\caption{Asymmetric $3\times3$ data matrix. total
number of iterations required for convergence (threshold
$\epsilon=10^{-6}$) for GaBP-based solvers vs. standard
methods.}\label{tab_Asym}
\end{table}

\begin{figure}[t!]
\begin{minipage}[b]{0.5\linewidth}
\centering
    \includegraphics[width=\textwidth]{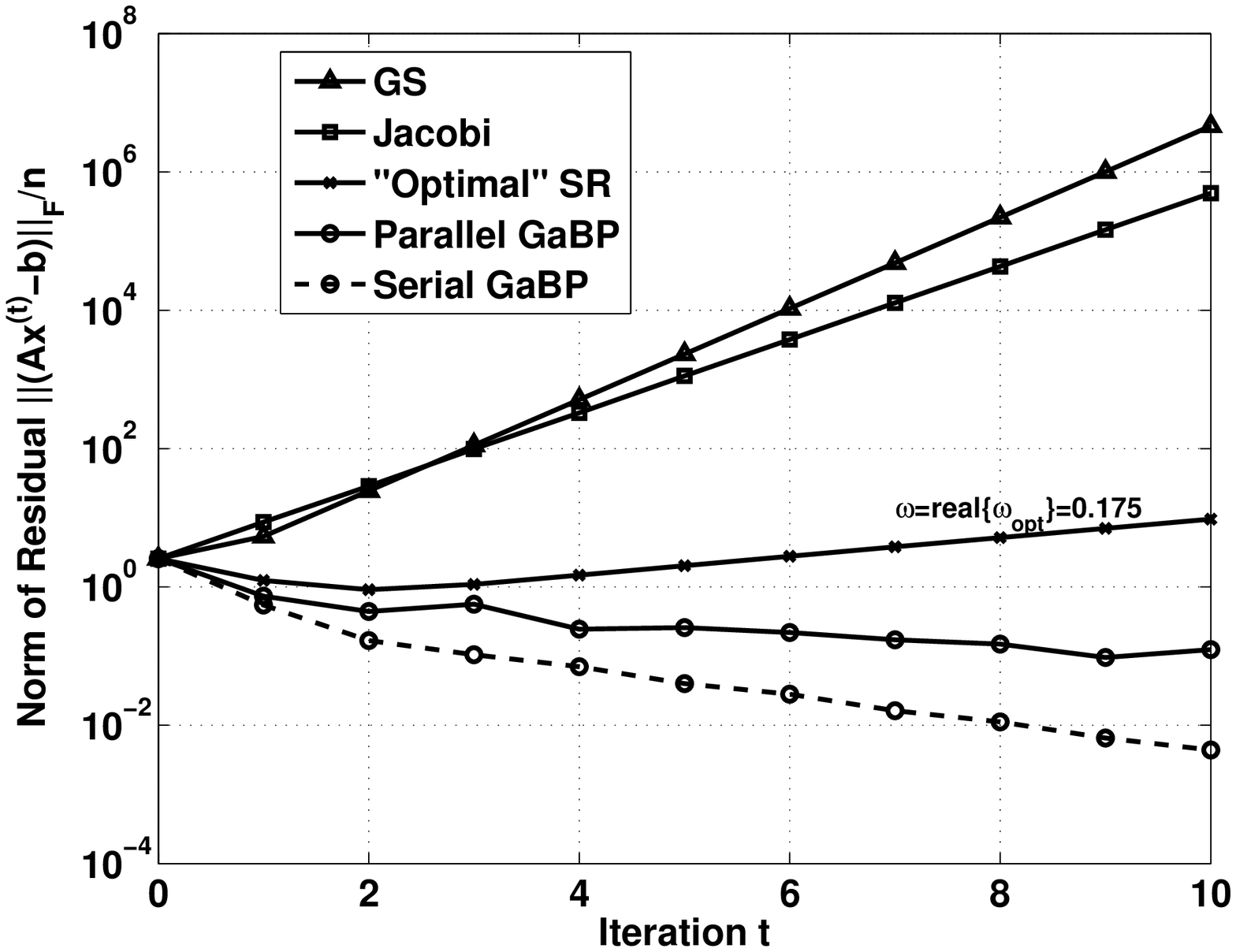}
 \label{fig_Asym}
\end{minipage}
\begin{minipage}[b]{0.5\linewidth}
\centering
   \includegraphics[width=\textwidth]{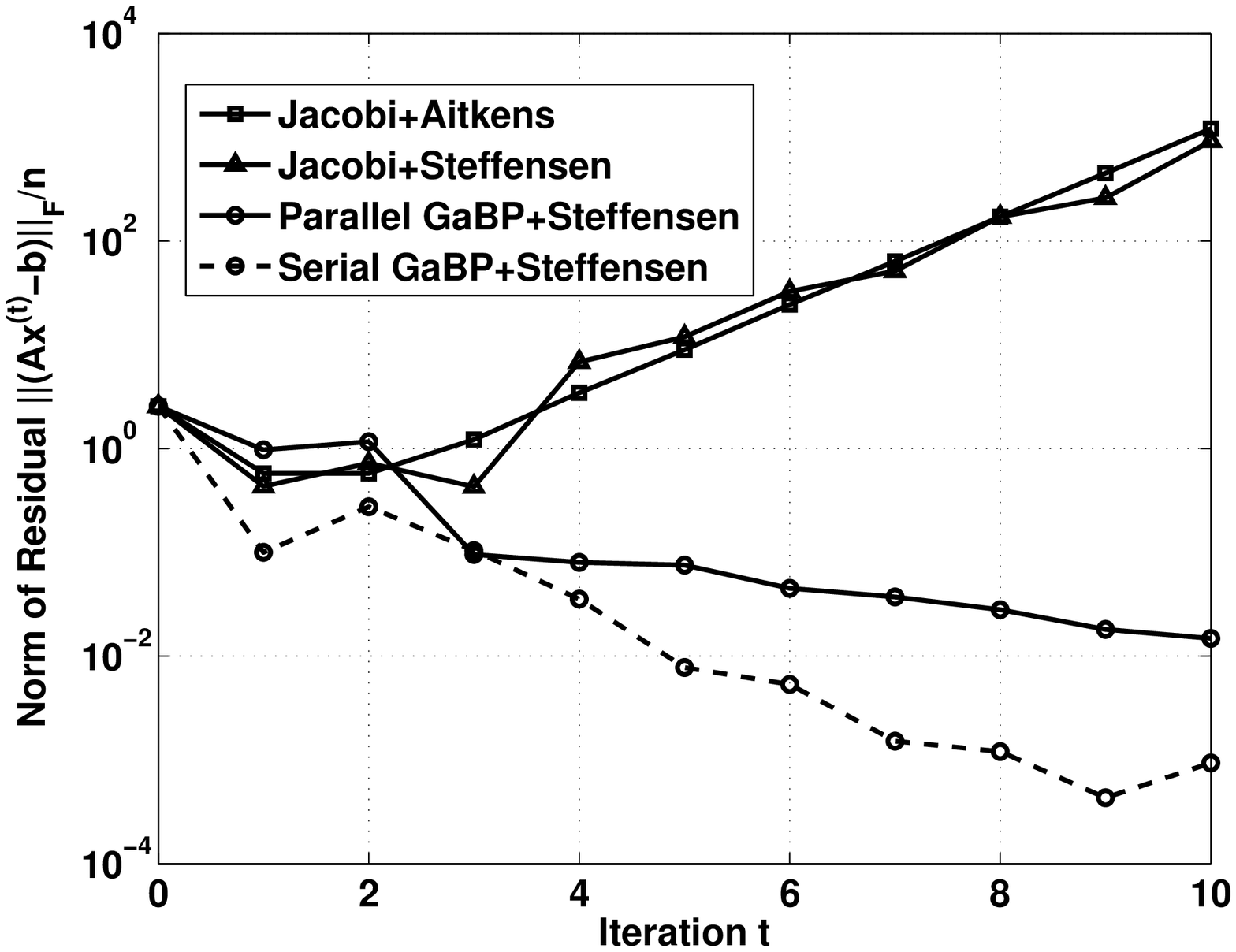}
 \label{fig_Asym_accel}
\end{minipage}
\caption{Convergence of an asymmetric $3 \times 3$ matrix. }
\end{figure}

\begin{figure}[h!]\label{fig_Asym_spiral}
\begin{center}
    \includegraphics[width=0.5\textwidth]{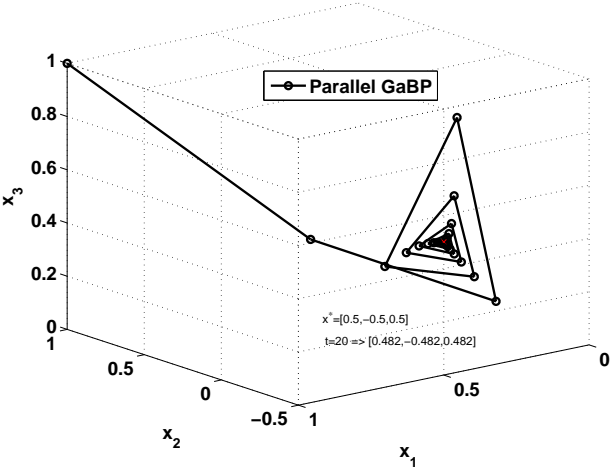}
  \caption{Convergence of a $3 \times 3$ asymmetric matrix, using 3D plot.}
\end{center}
\end{figure}

\chapter[Convergence Fix]{Fixing convergence of GaBP}\label{chap4}


\label{chap:conv_fix}
In this chapter, we present a novel construction that fixes the convergence of the
GaBP algorithm, for any Gaussian model with positive-definite information matrix
(inverse covariance matrix), even when the currently known sufficient
convergence conditions do not hold. We prove that our construction converges to
the correct solution. Furthermore, we consider how this method may be used to
solve for the least-squares solution of general linear systems. We defer experimental results evaluating the efficiency of the convergence fix to Chapter \ref{sec:fix-applied} in the context of linear detection. By using our convergence fix construction we are able to
show convergence in practical CDMA settings, where the original GaBP algorithm did
not converge, supporting a significantly higher number of users on each cell.


\section{Problem Setting}
\label{sec:problem_settings}

We wish to compute the \emph{maximum a posteriori} (MAP) estimate
of a random vector $x$ with Gaussian distribution (after conditioning
on measurements):
\begin{equation}
p(x) \propto \exp\{ -\tfrac{1}{2} x^T J x + h^T x \}\,, \label{eq:sys_prob}
\end{equation}
where $J \succ 0$ is a symmetric positive definite matrix (the information
matrix) and $h$ is the potential vector.  This problem is equivalent to solving
$J x = h$ for $x$ given $(h,J)$ or to solve the convex quadratic
optimization problem:
\begin{equation} \label{eq:objective_func}
\mbox{minimize} \;\; f(x) \triangleq \tfrac{1}{2} x^T J x - h^T x.
\end{equation}
We may assume without loss of generality (by rescaling variables) that
$J$ is normalized to have unit-diagonal, that is, $J \triangleq I-R$ with $R$
having zeros along its diagonal.  The off-diagonal entries of $R$ then
correspond to \emph{partial correlation coefficients} \cite{Lauritzen}. Thus,
the fill pattern of $R$ (and $J$) reflects the Markov structure
of the Gaussian distribution.  That is, $p(x)$ is Markov with respect to
the graph with edges $\G = \{(i,j) | r_{i,j} \neq 0 \}\,.$

If the model $J = I-R$ is \emph{walk-summable}
\cite{BibDB:jmw_walksum_nips,BibDB:mjw_walksum_jmlr06},
such that the spectral radius of $|R|=(|r_{ij}|)$ is
less than one ($\rho(|R|)<1$), then the method of GaBP may be used to solve this problem.  We
note that the walk-summable condition implies $I-R$ is positive definite.  An
equivalent characterization of the walk-summable condition is that $I-|R|$ is
positive definite.

This chapter presents  a method to solve non-walksummable models,
where $J = I-R$ is positive definite but $\rho(|R|) \ge 1$, using
GaBP.  There are two key ideas: (1) using diagonal loading to create a
perturbed model $J'=J+\Gamma$ which is walk-summable (such that the GaBP
may be used to solve $J'x=h$ for any $h$) and (2) using this
perturbed model $J'$ and convergent GaBP algorithm as a
\emph{preconditioner} in a simple iterative method to solve the
original non-walksummable model.

\section{Diagonal Loading}

We may always obtain a walk-summable model by \emph{diagonal
loading}.  This is useful as we can then solve a related system
of equations efficiently using Gaussian belief propagation.
For example, given a non-walk-summable model $J = I-R$ we
obtain a related walk-summable model $J_\gamma = J + \gamma I$ that is
walk-summable for large enough values of $\gamma$:

\begin{lemma} Let $J=I-R$ and $J' \triangleq
J+\gamma I = (1+\gamma)I-R$.  Let $\gamma > \gamma^*$
where
\begin{equation}
\gamma^* = \rho(|R|)-1\,.
\end{equation}
Then, $J'$ is walk-summable and GaBP based on $J'$ converges.
\end{lemma}

\begin{proof}
We normalize $J' = (1+\gamma) I - R$ to obtain
$J'_{\mathrm{norm}} = I - R'$ with $R' = (1+\gamma)^{-1} R$, which is
walk-summable if and only if $\rho(|R'|) < 1$.  Using $\rho(|R'|) =
(1+\gamma)^{-1} \rho(|R|)$ we obtain the condition $(1+\gamma)^{-1}
\rho(|R|) < 1$, which is equivalent to $\gamma >
\rho(|R|)-1$.
\end{proof}

It is also possible to achieve the same effect by adding a general
diagonal matrix $\Gamma$ to obtain a walk-summable model.  For
example, for all $\Gamma > \Gamma^*$, where $\gamma^*_{ii} =
J_{ii}-\sum_{j\neq i} |J_{ij}|$, it holds that $J+\Gamma$ is
diagonally-dominant and hence walk-summable (see
\cite{BibDB:mjw_walksum_jmlr06}).
More generally, we could allow $\Gamma$ to be any symmetric positive-definite
matrix satisfying the condition $I+\Gamma \succ |R|$.  However, only
the case of diagonal matrices is explored in this chapter.

\section{Iterative Correction Method}

Now we may use the diagonally-loaded model $J' = J+\Gamma$ to solve
$Jx = h$ for any value of $\Gamma \ge 0$.
The basic idea here is to use the diagonally-loaded matrix $J' =
J+\Gamma$ as a \emph{preconditioner} for solving the $Jx = h$ using
the iterative method:
\begin{equation} \label{eq:iterative_method}
\hat{x}^{(t+1)} = (J+\Gamma)^{-1} (h+\Gamma \hat{x}^{(t)})\,.
\end{equation}
Note that the effect of adding positive $\Gamma$ is to reduce the
size of the scaling factor $(J+\Gamma)^{-1}$ but we compensate for
this damping effect by adding a feedback term $\Gamma \hat{x}$ to
the input $h$. Each step of this iterative method may also be
interpreted as solving the following convex quadratic
optimization problem based on the objective $f(x)$ from (\ref{eq:objective_func}):
\begin{equation}
\hat{x}^{(t+1)} = \arg\min_x \left\{ f(x) + \tfrac{1}{2} (x-x^{(t)})^T \Gamma
(x-x^{(t)}) \right\}\,.
\end{equation}
This is basically a regularized version of Newton's method to minimize
$f(x)$, where we regularize the step-size
at each iteration. Typically, this regularization is used to
ensure positive-definiteness of the Hessian matrix when Newton's
method is used to optimize a non-convex function.  In contrast, we use it to ensure that
$J+\Gamma$ is walk-summable, so that the update step can be
computed via Gaussian belief propagation.  Intuitively, this will
always move us closer to the correct solution, but slowly if $\Gamma$ is
large. It is simple to demonstrate the following:

\begin{lemma} Let $J \succ 0$ and $\Gamma \succeq 0$. Then,
$\hat{x}^{(t)}$ defined by (\ref{eq:iterative_method}) converges
to $x^* = J^{-1} h$ for all initializations $\hat{x}^{(0)}$.
\end{lemma}

\emph{Comment.} The proof is given for a general (non-diagonal)
$\Gamma \succeq 0$.  For diagonal matrices, this
is equivalent to requiring $\Gamma_{ii} \ge 0$ for $i=1,\dots,n$.

\begin{proof} First, we note that there is only one possible
fixed-point of the algorithm and this is $x^* = J^{-1} h$.
Suppose $\bar{x}$ is a fixed point: $\bar{x} = (J+\Gamma)^{-1}
(h+\Gamma \bar{x})$.  Hence, $(J+\Gamma) \bar{x} = h+\Gamma \bar{x}$
and $J \bar{x} = h$. For non-singular $J$, we must then have $\bar{x}
= J^{-1} h$. Next, we show that the method converges.  Let $e^{(t)} =
\hat{x}^{(t)}-x^*$ denote the error of the $k$-th estimate.  The
error dynamics are then $e^{(t+1)} = (J+\Gamma)^{-1} \Gamma e^{(t)}$.
Thus, $e^{(t)} = ((J+\Gamma)^{-1}\Gamma)^k e^{(0)}$ and the error
converges to zero if and only if $\rho((J+\Gamma)^{-1} \Gamma) < 1$, or
equivalently $\rho(H)<1$, where $H = (J+\Gamma)^{-1/2} \Gamma
(J+\Gamma)^{-1/2} \succeq 0$ is a symmetric positive semi-definite
matrix.  Thus, the eigenvalues of $H$ are non-negative and we must
show that they are less than one.  It is simple to check that if
$\lambda$ is an eigenvalue of $H$ then $\frac{\lambda}{1-\lambda}$ is
an eigenvalue of $\Gamma^{1/2} J^{-1} \Gamma^{1/2} \succeq 0$.  This
is seen as follows: $Hx = \lambda x$, $(J+\Gamma)^{-1} \Gamma y =
\lambda y$ ($y = (J+\Gamma)^{-1/2} x$), $\Gamma y = \lambda
(J+\Gamma)y$, $(1-\lambda) \Gamma y = \lambda J y$, $J^{-1} \Gamma y =
\frac{\lambda}{1-\lambda} y$ and $\Gamma^{1/2} J^{-1} \Gamma^{1/2} z =
\frac{\lambda}{1-\lambda} z$ ($z = \Gamma^{1/2} y$) [note that $\lambda \neq
1$, otherwise $Jy=0$ contradicting $J \succ 0$].  Therefore
$\frac{\lambda}{1-\lambda} \ge 0$ and $0 \le \lambda < 1$.  Then
$\rho(H) < 1$, $e^{(t)} \rightarrow 0$ and $\hat{x}^{(t)} \rightarrow
  x^*$ completing the proof.
\end{proof}

Now, provided we also require that $J' = J+\Gamma$ is walk-summable, we may
compute $x^{(t+1)} = (J+\Gamma)^{-1} h^{(t+1)}$, where $h^{(t+1)} = h +
\Gamma \hat{x}^{(t)}$, by performing Gaussian belief propagation to
solve $J' x^{(t+1)} = h^{(t+1)}$.  Thus, we obtain a double-loop
method to solve $Jx = h$.  The inner-loop performs GaBP and the outer-loop
computes the next $h^{(t)}$. The overall procedure converges provided
the number of iterations of GaBP in the inner-loop is made large enough
to ensure a good solution to $J' x^{(t+1)} = h^{(t+1)}$.
Alternatively, we may compress this double-loop procedure
into a single-loop procedure by preforming just one
iteration of GaBP  message-passing per iteration of the outer loop.
Then it may become necessary to use the following damped update of $h^{(t)}$
with step size parameter $s \in (0,1)$:
\begin{eqnarray}
h^{(t+1)} &=& (1-s) h^{(t)} + s (h + \Gamma \hat{x}^{(t)}) \nonumber\\
&=& h + \Gamma ((1-s) \hat{x}^{(t-1)} + s \hat{x}^{(t)})\,,
\end{eqnarray}
This single-loop method converges for sufficiently small values
of $s$.  In practice, we have found good convergence with $s=\frac{1}{2}$.
This single-loop method can be more efficient than the double-loop
method.

\section{Extension to General Linear  Systems}

\label{sec_new_const} In this section, we efficiently extend the
applicability of the proposed double-loop construction for a general linear system of
equations (possibly over-constrained.)
Given a full column rank matrix $\tilde{J} \in \R^{n \times k}$, $n \ge k$,
and a shift vector $\tilde{h}$, we are interested in solving the
least squares problem $\min_x ||\tilde{J}x - \tilde{h}||^2_2$.  The naive
approach for using GaBP would be to take the information matrix
$\bar{J} \triangleq (\tilde{J}^T\tilde{J})$, and the shift vector
$\bar{h} \triangleq \tilde{J}^T\tilde{h}$.  Note that $\bar{J}$ is
positive definite and we can use GaBP to solve it.  The
MAP solution is \BE x =
\bar{J}^{-1}\bar{h} =(\tilde{J}^T\tilde{J})^{-1}\tilde{J}^Th\,,\label{pi} \EE which is the
pseudo-inverse solution.

Note, that the above construction has two drawbacks: first, we need
to explicitly compute $\bar{J}$ and $\bar{h}$, and second, $\bar{J}$
may not be sparse in case the original matrix $\tilde{J}$ is
sparse. To overcome this problem, following \cite{ISIT2}, we construct
a new symmetric data matrix $\bar{\bar{\mJ}}$ based on the arbitrary
rectangular matrix
$\tilde{\mJ}\in\mathbb{R}^{n\times k}$ \BE
\label{newJ} \bar{\bar{\mJ}}\triangleq\left(
  \begin{array}{cc}
    \mI_{k \times k} & \tilde{\mJ}^T \\
    \tilde{\mJ} & \mathbf{0}_{n \times n} \\
  \end{array}
\right)\in\mathbb{R}^{(k+n)\times(k+n)}\,.\nonumber \EE Additionally, we
define a new hidden variable vector
$\tilde{\vx}\triangleq\{x^{T},\vz^{T}\}^{T}\in\mathbb{R}^{(k+n)}$,
where $\vx\in\mathbb{R}^{k}$ is the solution vector and
$\vz\in\mathbb{R}^{n}$ is an auxiliary hidden vector, and a new
shift vector $\bar{\bar{\vh}}\triangleq\{\mathbf{0}_{k \times
  1}^{T},\vh^{T}\}^{T}\in\mathbb{R}^{(k+n)}$.

\begin{lemma}
Solving $\bar{\bar{x}} = \bar{\bar{J}}^{-1}\bar{\bar{h}}$ and taking the first $k$ entries
is identical to solving Eq. \ref{pi}.
\end{lemma}
\emph{Proof.}
Is given in \cite{ISIT2}.

 For applying our double-loop construction on the new system $(\bar{\bar{h}},\bar{\bar{J}})$ to obtain the
solution to Eq.~(\ref{pi}), we need to confirm that the matrix $\bar{\bar{J}}$ is positive definite. (See lemma 2).
To this end, we add a diagonal weighting $-\gamma I$ to the lower right block:
 \BE
 \hat{\mJ}\triangleq\left(
  \begin{array}{cc}
    \mI_{k \times k} & \tilde{\mJ}^T \\
    \tilde{\mJ} & -\gamma I \\
  \end{array}
\right)\in\mathbb{R}^{(k+n)\times(k+n)}\,.\nonumber \EE
Then we rescale $\hat{J}$ to make it unit diagonal (to deal with the negative sign of the lower right block we use a complex Gaussian notation as done in \cite{BibDB:MontanariEtAl}). It is clear that for a large enough $\gamma$ we are left with a walk-summable model, where the rescaled $\hat{J}$ is a hermitian positive definite matrix and $\rho(|\hat{J}-I|)<1$. Now it is possible to use the double-loop technique to compute Eq. \ref{pi}. Note that adding $-\gamma I$ to the lower right block of $\hat{J}$ is equivalent to adding $\gamma I$ into Eq. 7:
\BE x = (\tilde{J}^T\tilde{J} +  \gamma I)^{-1}\tilde{J}^{T}h\, \label{hatj} \EE
where $\gamma$ can be interpreted as a regularization parameter.

\chapter*{Part 2: Applications}
\addcontentsline{toc}{chapter}{Part 2: Applications}

\chapter[Peer-to-Peer Rating]{Rating Users and Data Items in Social Networks}
\label{PPNA}

We propose to utilize the distributed GaBP solver presented in Chapter~\ref{sec_GaBP}
to efficiently and distributively compute a solution
to a family of quadratic cost functions described below. By implementing our algorithm once, and choosing
the computed cost function dynamically on the run we allow a high
flexibility in the selection of the rating method deployed in the
Peer-to-Peer network.

We propose a unifying family of quadratic cost
functions to be used in Peer-to-Peer ratings. We show that our
approach is general since it captures many of the existing
algorithms in the fields of visual layout, collaborative filtering
and Peer-to-Peer rating, among them Koren spectral layout
algorithm, Katz method, Spatial ranking, Personalized PageRank and
Information Centrality. Beside of the theoretical interest in
finding common basis of algorithms that were not linked before,
we allow a single efficient implementation for computing those
various rating methods.

Using simulations over real social network topologies obtained
from various sources, including the MSN Messenger social network,
we demonstrate the applicability of our approach. We report
simulation results using networks of millions of nodes.

Whether you are browsing for a hotel, searching the web, or
looking for a recommendation on a local doctor, what your friends
like will bear great significance for you. This vision of virtual
social communities underlies the stellar success of a growing body
of recent web services, e.g., {\tt http://www.flickr.com}, {\tt
http://del.icio.us}, {\tt http://www.myspace.com}, and others.
However, while all of these services are centralized, the full
flexibility of social information-sharing can often be better
realized through direct sharing between peers.

This chapter presents a mechanism for sharing user {\em ratings\/}
(e.g., on movies, doctors, and vendors) in a social network. It
introduces distributed mechanisms for computing by the network
itself individual ratings that incorporate rating-information from
the network. Our approach utilizes message-passing algorithms from
the domain of Gaussian graphical models. In our system,
information remains in the network, and is never ``shipped'' to a
centralized server for any global computation. Our algorithms
provide each user in the network with an individualized rating per
object (e.g., per vendor). The end-result is a local rating per
user which minimizes her cost function from her own rating (if
exists) and, at the same time, benefits from incorporating ratings
from her network vicinity. Our rating converges quickly to an
approximate optimal value even in sizable networks.

Sharing views over a social network has many advantages. By taking
a peer-to-peer approach the user information is shared and stored
only within its community. Thus, there is no need for trusted
centralized authority, which could be biased by economic and/or
political forces. Likewise, a user can constrain information
pulling from its trusted social circle. This prevents spammers
from penetrating the system with false recommendations.

\section[Framework]{Our General Framework}\label{drating}
The social network is represented by a directed, weighted
communication graph $G=(V, E)$. Nodes $V=\{ 1, 2, ..., n\}$ are
users. Edges express social ties, where a non-negative edge weight
$w_{ij}$ indicates a measure of the mutual trust between the
endpoint nodes $i$ and $j$. Our goal is to compute an output
rating $\vx \in R^n$ to each data item (or node) where $x_i$ is
the output rating computed locally in node $i$. The vector $\vx$
is a solution that minimizes some cost function. Next, we propose
such a cost function, and show that many of the existing rating
algorithms conform to our proposed cost function.

We consider a single instance of the rating problem that concerns
an individual item (e.g., a movie or a user). In practice, the
system maintains ratings of many objects concurrently, but
presently, we do not discuss any correlations across items. A
straight forward generalization to our method for collaborative
filtering, to rank $m$ (possibly correlated) items, as done
in~\cite{koren-cf}.

In this chapter, we methodically derive the following quadratic cost
function, that quantifies the Peer-to-Peer rating problem:
\begin{equation}
\label{cost-function} \min E(x) \triangleq \sum_i w_{ii}(x_i -
y_i)^2 + \beta \sum_{i,j \in E} w_{ij} (x_i - x_j)^2,
\end{equation}
where $x_i$ is a desired rating computed locally by node $i$, $\vy$
is an optional prior rating, where $y_i$ is a local input to node
$i$ (in case there is no prior information, $\vy$ is a vector of
zeros).

We demonstrate the generality of the above cost function by
proving that many of the existing algorithms for visual layouts,
collaborative filtering and ranking of nodes in Peer-to-Peer
networks are special cases of our general framework:
\begin{enumerate}
\item{\bf Eigen-Projection method.} Setting $y_i = 1, w_{ii} = 1,
w_{ij} = 1$ we get the Eigen-Projection
method~\cite{eigen-projection} in a single dimension, an algorithm
for ranking network nodes for creating a intuitive visual layout.
\item{\bf Koren's spectral layout method.} Recently, Dell'Amico
proposed to use Koren's visual layout algorithm for ranking nodes
in a social network~\cite{matteo}. We will show that this ranking
method is a special case of our cost function, setting: $w_{ii} =
deg(i)$. \item{\bf Average computation.} By setting the prior
inputs $y_i$ to be the input of node $i$ and taking $\beta
\rightarrow \infty$ we get $x_i = 1/n\sum_i{y_i}$, the average
value of $\vy$. This construction, Consensus Propagation, was
proposed by Moallemi and Van-Roy in~\cite{CP}. \item{\bf
Peer-to-Peer Rating.} By generalizing the Consensus Propagation
algorithm above and supporting weighted graph edges, and finite
$\beta$ values we derive a new algorithm for Peer-to-Peer
rating~\cite{p2p-rating}. \item{\bf Spatial Ranking.} We propose a
generalization of the Katz method~\cite{katz-method}, for
computing a personalized ranking of peers in a social network. By
setting $y_i = 1, w_{ii} = 1$ and regarding the weights $w_{ij}$
as the probability of following a link of an absorbing
Markov-chain, we formulate the spatial ranking
method~\cite{p2p-rating} based on the work of~\cite{WS}.\item{\bf
Personalized PageRank.} We show how the PageRank and personalized
PageRank algorithms fits within our framework~\cite{PR,PR2}.
\item{\bf Information Centrality.} In the information centrality
node ranking method~\cite{CM}, the non-negative weighted graph
$G=(V,E)$ is considered as an electrical network, where edge
weights is taken to be the electrical conductance. We show that
this measure can be modelled using our cost function as well.
\end{enumerate}

Furthermore, we propose to utilize the Gaussian Belief Propagation
algorithm (GaBP) - an algorithm from the probabilistic graphical
models domain - for efficient and distributed computation of a
solution minimizing a single cost function drawn from our family
of quadratic cost functions. We explicitly show the relation
between the algorithm to our proposed cost function by deriving it
from the cost function.

Empirically, our algorithm demonstrates faster convergence
than the compared algorithms, including conjugate
gradient algorithms that were proposed in~\cite{matteo,koren-cf}
to be used in Peer-to-Peer environments. For comparative study of
those methods see~\cite{Allerton}.

\section[Quadratic Cost Functions]{Unifying Family of Quadratic Cost Functions}

We derive a family of unifying cost functions following the
intuitive explanation of Koren~\cite{koren-spectral}. His work
addresses graphic visualization of networks using spectral graph
properties. Recently, Dell'Amico proposed in~\cite{matteo} to use
Koren's visual layout algorithm for computing a distance metric that
enables ranking nodes in a social network. We further extend this
approach by finding a common base to many of the ranking
algorithms that are used in graph visualization, collaborative
filtering and in Peer-to-Peer rating. Beside of the theoretical
interest in finding a unifying framework to algorithms that were
not related before, we enable also a single efficient distributed
implementation that takes the specific cost function as input and
solves it.

Given a directed graph $G=(V,E)$ we would like to find an output
rating $\vx \in R^n$ to each item where $x_i$ is the output rating
computed locally in node $i$. $\vx$ can be expressed as the
solution for the following constraint minimization problem:
\begin{equation}
 \min_\vx E(\vx)
\triangleq \sum_{i,j \in E} w_{ij} (x_i - x_j)^2,
\end{equation}
Given
\begin{align*}
 \mathbf{Var(\vx)}= 1&, & \mathbf{Mean(\vx)} = 0.
\end{align*}
where $\mathbf{Var(x)} \triangleq \tfrac{1}{n}\sum_{i}(x_i - \mathbf{Mean(\vx)})^2\,,  \ \mathbf{Mean(x)} \triangleq \frac{1}{n}\sum_i x_i\,.$

The cost function $E(x)$ is combined of weighted sums of
interactions between neighbors. From the one hand, ''heavy`` edges
$w_{ij}$ force nodes to have a similar output rating. From the
other hand, the variance condition prevents a trivial solution of
all $x_i$ converging to a single value. In other words, we would
like the rating of an item to be scaled. The value of the variance
determines the scale of computed ratings, and is arbitrarily set
to one. Since the problem is invariant under translation, we add
also the mean constraint to force a unique solution. The mean is
arbitrarily set to zero.

In this chapter, we address visual layout computed for a single
dimension. The work of Koren and Dell'Amico is more general than
ours since it discusses rating in $k$ dimensions. A possible
future extension to this work is to extend our work to $k$
dimensions.

From the visual layout perspective, ``stronger'' edges $w_{ij}$
let neighboring nodes appear closer in the layout. The variance
condition forces a scaling on the drawing.

From the statistical mechanics perspective, the cost function
$E(x)$ is considered as the system energy that we would like to
minimize, the weighted sums are called ``attractive forces'' and
the variance constraint is a ``repulsive force''.

One of the most important observations we make is that using
Koren's framework, the chosen values of the variance and mean are
arbitrary and could be changed. This is because the variance
influences the scaling of the layout and the mean the translation
of the result. Without the loss of generality, in some of the
proofs we change the values of the mean and variance to reflect
easier mathematical derivation. However, normalization of rated
values can be always done at the end, if needed.
Following, we generalize Koren's method to support a larger
variety of cost functions. The main observation we have, is that
the variance condition is used only for scaling the rating,
without relating to the specific problem at hand. We propose to
add a second constraint which is local to each node:

\begin{equation}
\sum_i w_{ii}(x_i - y_i)^2 + \beta \sum_{i,j \in E} w_{ij} (x_i -
x_j)^2.
\end{equation}
Thus we allow higher flexibility, since we allow
$y_i$ can be regarded as {\it prior information} in the case of
Peer-to-Peer rating, where each node has some initial input it adds
to the computation. In other cases, $y_i$ can be some function
computed on $\vx$ like $y_i = \frac{1}{N}\sum_{i=1}^{N} x_i$ or a
function computed on the graph: $y_i = \sum_{N(i)}
\frac{w_{ij}}{deg(i)}$.\\

\begin{thm}
\label{PPNA-thm1}
The Eigen-Projection method is an instance of the
cost function~\ref{cost-function} when $w_{ij} = 1, w_{ii} = 1,
\beta = 1/2, y_i = 1$.
\end{thm}
\begin{proof} \label{PPNA-proof1} It is shown in~\cite{eigen-projection} that
the optimal solution to the cost function is $\vx = L^{-1}\ones$ where
$L$ is the graph Laplacian (as defined in \ref{def:Laplacian}). Substitute $\beta = 1/2, w_{ii} = 1,
w_{ij} = 1, y_i = 1$ in the cost function \eqref{cost-function} :
\[
\min_{x} E(x) \triangleq \sum_i 1(x_i - 1)^2 + 1/2 \sum_{i,j \in E}
(x_i - x_j)^2.
\]
The same cost function in linear algebra form ($\mathbf{1}$ is a vector of all ones):
\[
\min E(\vx) \triangleq \vx^T L \vx - 2 \vx \textbf{1} + n\,.
\]
Now we calculate the derivative and compare to zero and get
\[ \nabla_XE(\vx) = 2 \vx^T L - 2 \textbf{1}\,, \]
\[ \vx = L^{-1}\ones.  \]
\end{proof}
\begin{thm}
\label{PPNA-thm2} Koren's spectral layout algorithm/Del'Amicco
method in a single dimension, is an instance of the cost
function~\ref{cost-function} when $w_{ii} = 1, \beta = 1, y_i =
deg(i)$, where $deg(i) \triangleq \sum_{j \in N(i)} w_{ij}$, up to
a translation.
\end{thm}

\begin{proof}
 \label{PPNA-proof2} Using the notations of~\cite{matteo}
the cost function of Koren's spectral layout algorithm is:
\[ \min_x \sum_{i,j \in E} w_{ij} (x_i - x_j)^2, \]
 \begin{align*} \mbox{s.t. } &\ &\ \sum_i deg(i) x_i^2 = n & & \tfrac{1}{n}\sum_i deg(i) x_i = 0. \end{align*}

We compute the weighted cost function\[ \mathcal{L}(\vx,\beta,\gamma) = \sum_{ij} w_{ij} (x_i -
x_j)^2 - \beta (\sum_i  deg(i) x_i^2 - n) - \gamma \sum_i deg(i)
x_i\,.
\]
Substitute  the weights $\beta = 1, \gamma = 1/2$ we get:
\[ = \sum_{ij} w_{ij} (x_i -
x_j)^2 - \sum_i deg(i) (x_{i} - 1)^2,  \] Reverting to our cost
function formulation we get:
\[ = \sum_i deg(i) (x_i - 1)^2 + \sum_{ij} w_{ij} (x_i -
x_j)^2. \] In other words, we substitute $w_{ii} = deg(i), y_i = 1,
\beta = 1$ and we get Koren's formulation.
\end{proof}

It is interesting to note, that the optimal solution according to
Koren's work is $x_i = \sum_{j \in N(i)}\frac{w_{ij}x_j}{deg(i)}$
which is equivalent to the thin plate model image processing and
PDEs~\cite{LR}.

\subsection{Peer-to-Peer Rating}
\label{Rating} In~\cite{p2p-rating} we have proposed to generalize
the Consensus Propagation (CP) algorithm~\cite{CP} to solve the
general cost function (\ref{cost-function}).

The CP algorithm is a distributed algorithm for calculating the
network average, assuming each node has an initial value. We have
extended the CP algorithm in several ways. First, in the original
paper the authors propose to use a very large $\beta$ for
calculating the network average. As mentioned, large $\beta$
forces all the nodes to converge to the same value. We remove this
restriction by allowing a flexible selection of $\beta$ based on
the application needs. As $\beta$ becomes small the calculation is
done in a closer and closer vicinity of the node.

Second, we have extended the CP algorithm to support null value,
adapting it to omit the term $(y_i - x_i)^2$ when $y_i = \bot$,
i.e., when node $i$ has no initial rating. This construction is
reported in~\cite{p2p-rating} and not repeated here.

Third, we use non-uniform edge weights $w_{ij}$, which in our
settings represent mutual trust among nodes. This makes the rating
local to certain vicinities, since we believe there is no meaning
for getting a global rating in a very large network. This
extension allows also asymmetric links where the trust assigned
between neighbors is not symmetric. In that case we get an
approximation to the original problem.

Fourth, we introduce node weights, where node with higher weight
has an increasing linear contribution to the output of the
computation.

The Peer-to-Peer rating algorithm was reported in detail
in~\cite{p2p-rating}.

\begin{thm}
\label{PPNA-thm3}
 The Consensus propagation algorithm is an
instance of our cost function~\ref{cost-function} with $w_{ii} =
1, \beta \rightarrow \infty$.
\end{thm}

The proof is given in Chapter~\ref{CP}

\begin{thm}
\label{PPNA-thm4}
The Peer-to-Peer rating algorithm is an instance
of our cost function~\ref{cost-function}.
\end{thm}

The proof is given in~\cite{p2p-rating}.

\subsection{Spatial Ranking}
In~\cite{p2p-rating} we presented a new ranking method called
Spatial Ranking, based on the work of Jason K. Johnson
\etal~\cite{WS}. Recently, we found out that a related method was
proposed in 1953 in the social sciences field by Leo
Katz~\cite{katz-method}. The Spatial Ranking method described
below is a generalization of the Katz method, since it allows
weighted trust values between social network nodes (unlike the
Katz method which deals with binary relations). Furthermore, we
propose a new efficient implementation for a distributed
computation of the Spatial Ranking method.

In our method, each node ranks itself the list of all other nodes
based on its network topology and creates a personalized ranking
of its own.

We propose to model the network as a Markov chain with a
transition matrix $R$, and calculate the {\it fundamental matrix}
$P$, where the entry $P_{ij}$ is the expected number of times of a
random walk starting from node $i$ visits node $j$~\cite{MC}.

\begin{figure}[ht!]
\begin{center}
  \includegraphics[width=240pt]{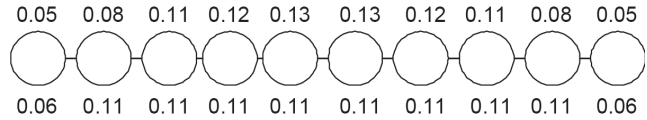}
  \caption{\it Example output of the Spatial ranking (on top) vs. PageRank (bottom) over a network
  of 10 nodes. In the Spatial ranking method node rank is biased towards the center, where in the
  PageRank method, non-leaf nodes have equal rank. This can be explained by the fact that
  the sum of self-returning random walks increases towards the center.}
  \label{PPNAfig1}
\end{center}
\end{figure}

We take the local value $P_{ii}$ of the fundamental matrix $P$,
computed in node $i$, as node $i$'s global importance.
Intuitively, the value signifies the weight of infinite number of
random walks that start and end in node $i$. Unlike the PageRank
ranking method, we explicitly bias the computation towards node
$i$, since we force the random walks to start from it. This
captures the local importance node $i$ has when we compute a
personalized rating of the network locally by node $i$. Figure \ref{PPNAfig1}
captures this bias using a simple network of 10 nodes.

\ignore{ In Section~\ref{extensions} we show how to compute the
full personal ranking of nodes (the full matrix $P$), where each
node has a personal ranking of all the other nodes.}

The fundamental matrix can be calculated by summing the
expectations of random walks of length one, two, three etc., $R +
R^2 + R^3 + \dots$. Assuming that the spectral radius $\varrho(R)
< 1$, we get $\sum_{l=1}^{\infty}R^l = (I - R)^{-1}$. Since $R$ is
stochastic, the inverse $(I-R)^{-1}$ does not exist. We therefore
slightly change the problem: we select a parameter $\alpha < 1$,
to initialize the matrix $J = I - \alpha R$ where $I$ is the
identity matrix. We know that $\varrho(\alpha R) < 1$ and thus
$\alpha R + \alpha^2R^2 + \alpha^3R^3 + \dots $ converges to $(I -
\alpha R)^{-1}$.

\begin{thm}
\label{PPNA-thm5}
The Spatial Ranking method is an instance of the cost
function~\ref{cost-function}, when $y_i = 0, w_{ii} = 1$, and
$w_{ij}$ are entries in an absorbing Markov chain $R$. \end{thm}

\begin{proof}\label{SR} \label{PPNA-proof5} We have shown that the
fundamental matrix is equal to $(I-R)^{-1}$. Assume that the edge
weights are probabilities of Markov-chain transitions (which means
the matrix is stochastic), substitute $\beta = \alpha, w_{ii}
= 1, y_i = 1$ in the cost function (~\ref{cost-function}):
\[
\min E(x) \triangleq \sum_i 1*(x_i - 1)^2 - \alpha \sum_{i,j \in
E} w_{ij}(x_i - x_j)^2.
\]
The same cost function in linear algebra form:
\[
\min E(x) \triangleq \vx^{T} I \vx - \alpha \vx^T R \vx - 2\vx.
\]
Now we calculate the derivative and compare to zero and get
\[ \vx = (I - \alpha R)^{-1}. \mbox{           }  \]
\end{proof}

We have shown that the Spatial Ranking algorithm fits well within
our unifying family of cost functions. Setting $\vy$ to a fixed
constant, means there is no individual prior input at the nodes,
thus we measure mainly the topological influence in ranking the
nodes. In case that we use $w_{ii} \ne 1$ we will get a biased
ranking, where nodes with higher weight have higher influence at
result of the computation.

\subsection{Personalized PageRank}
The PageRank algorithm is a fundamental algorithm in computing
node ranks in the network~\cite{PR}. The personalized PageRank
algorithm is a variant described in~\cite{PR2}. In a nutshell, the
Markov-chain transition probability matrix $\mM$ is constructed
out of the web links graph. A prior probability $\vx$ is taken to
weight the result. The personalized PageRank calculation can be
computed~\cite{PR2}:
\[ PR(x) = (1 - \alpha)(I - \alpha \mM)^{-1} \vx, \]
where $\alpha$ is a weighting constant which determines the speed
of convergence in trade-off with the accuracy of the solution and
$I$
is the identity matrix.

\begin{thm}
\label{PPNA-thm6} The Personalized PageRank algorithm can be expressed using
our cost function~\ref{cost-function}, up to a translation.
\end{thm}
\begin{proof}[proof sketch] \label{PPNA-proof6} \label{PP}
The proof is similar to the Spatial Ranking proof. There are two
differences: the first is that the prior distribution $\vx$ is set
in $\vy$ to weight the output towards the prior. Second, in the
the Personalized PageRank algorithm the result is multiplied by
the constant $(1-\alpha)$, which we omit in our cost function.
This computation can be done locally at each node after the
algorithm terminates, since $\alpha$ is a known fixed system
parameter.
\end{proof}
\subsection{Information Centrality} In the information
centrality (IC) node ranking method~\cite{CM}, the non-negative
weighted graph $G=(V,E)$ is considered as an electrical network,
where edge weights is taken to be the electrical conductance. A
vector of supply $b$ is given as input, and the question is to
compute the electrical potentials vector $p$. This is done by
computing the graph Laplacian and solving the set of linear
equations $Lp = b$. The IC method (a.k.a current flow betweenness 
centrality) is defined by:
\[ IC(i) = \frac{n-1}{\Sigma_{i \ne j}{p_{ij}(i) - p_{ij}(j)}}. \]
The motivation behind this definition is that a centrality of node
is measured by inverse proportion to the effective resistance
between a node to all other nodes. In case the effective
resistance is lower, there is a higher electrical current flow in
the network, thus making the node more ''socially influential''.

One can easily show that the IC method can be derived from our
cost function, by calculating $(L+J)^{-1}$ where $L$ is the graph
Laplacian and $J$ is a matrix of all ones. Note that the inverted
matrix is not sparse, unlike all the previous constructions.
Hence, a special construction which transforms this matrix into a
sparse matrix is needed. This topic is out of scope of this work.

\section{Experimental Results}
\label{experimental}
\begin{figure}[ht!]
\begin{minipage}[b]{0.5\linewidth} 
\centering
\includegraphics[width=6cm]{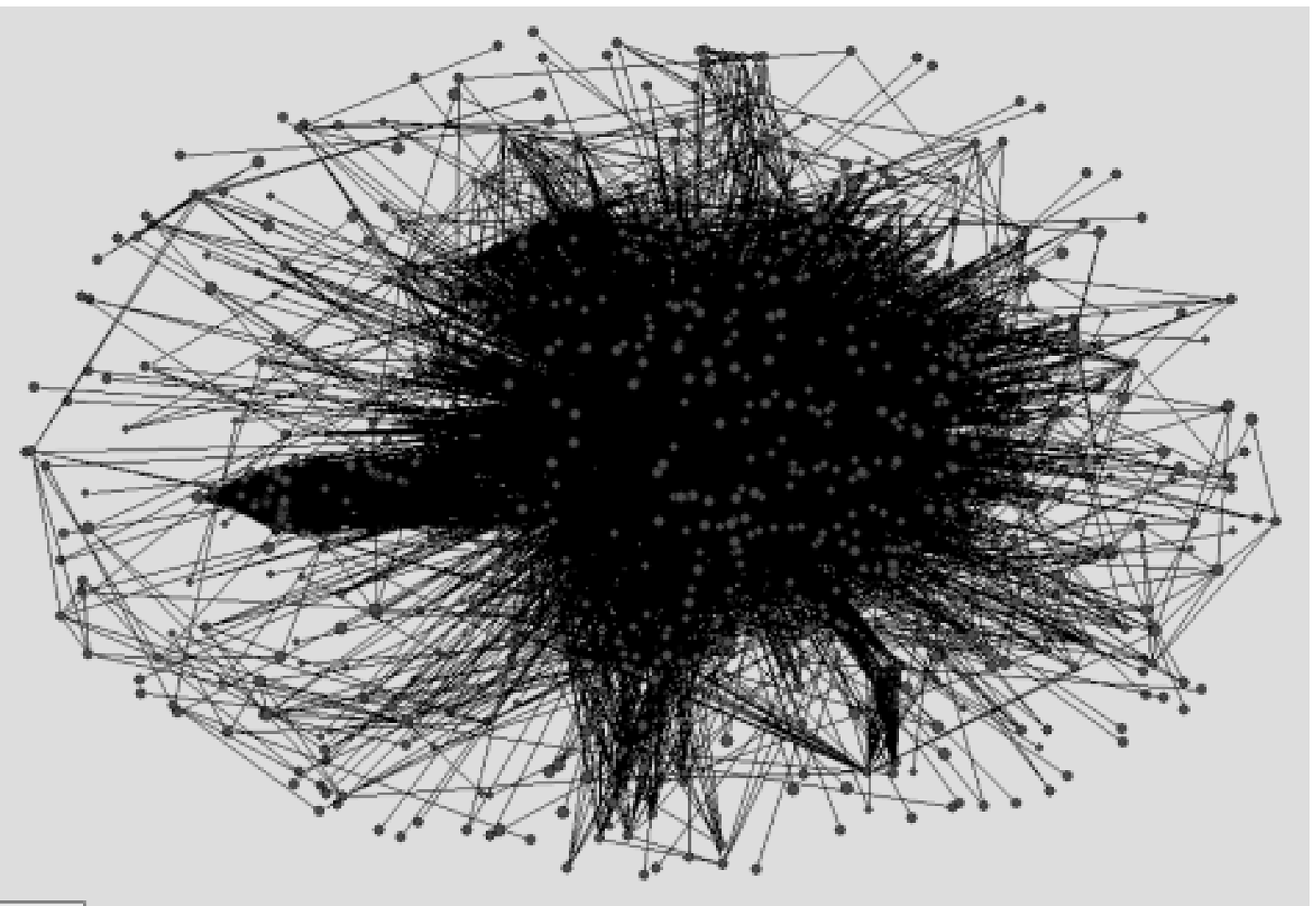}
\caption{Blog network topology}
\end{minipage}
\hspace{0.5cm} 
\begin{minipage}[b]{0.5\linewidth}
\centering
\includegraphics[width=6cm]{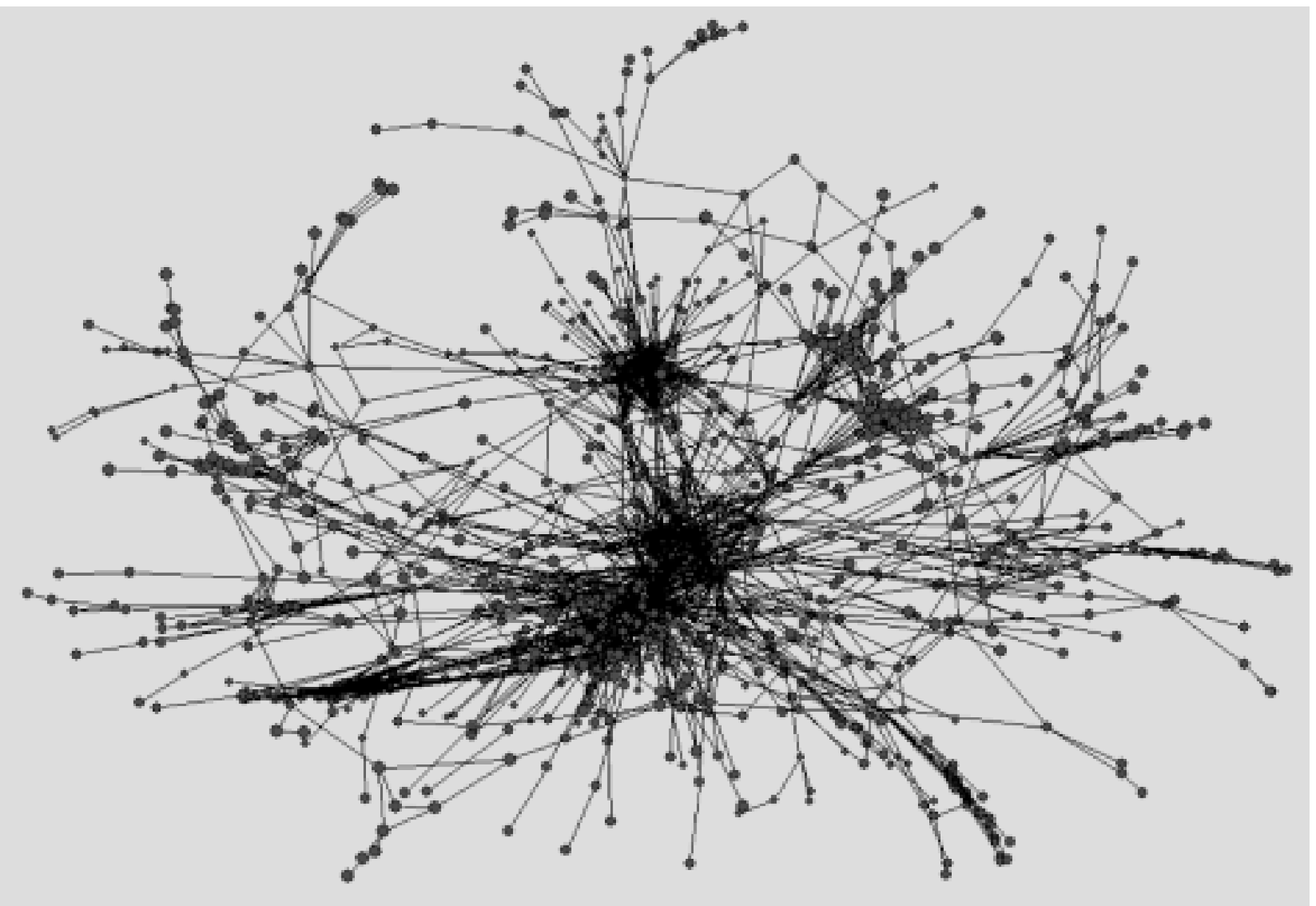}
\caption{DIMES Internet topology}
\end{minipage}
\end{figure}

\begin{figure}[h!]
\begin{minipage}[b]{0.5\linewidth} 
\centering
\includegraphics[width=6cm]{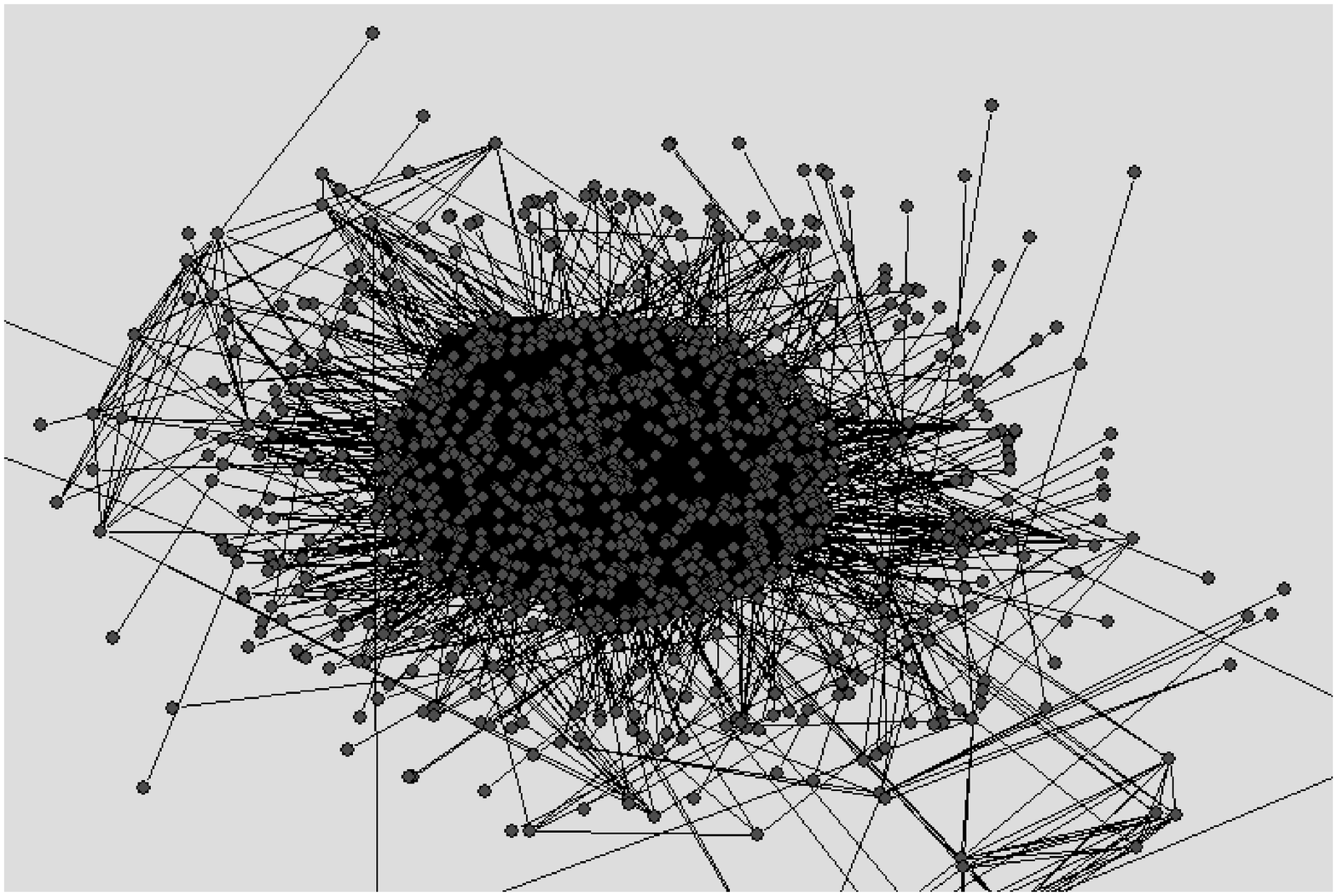}
\caption{US gov documents topology}
\end{minipage}
\hspace{0.5cm} 
\begin{minipage}[b]{0.5\linewidth}
\centering
\includegraphics[width=6cm]{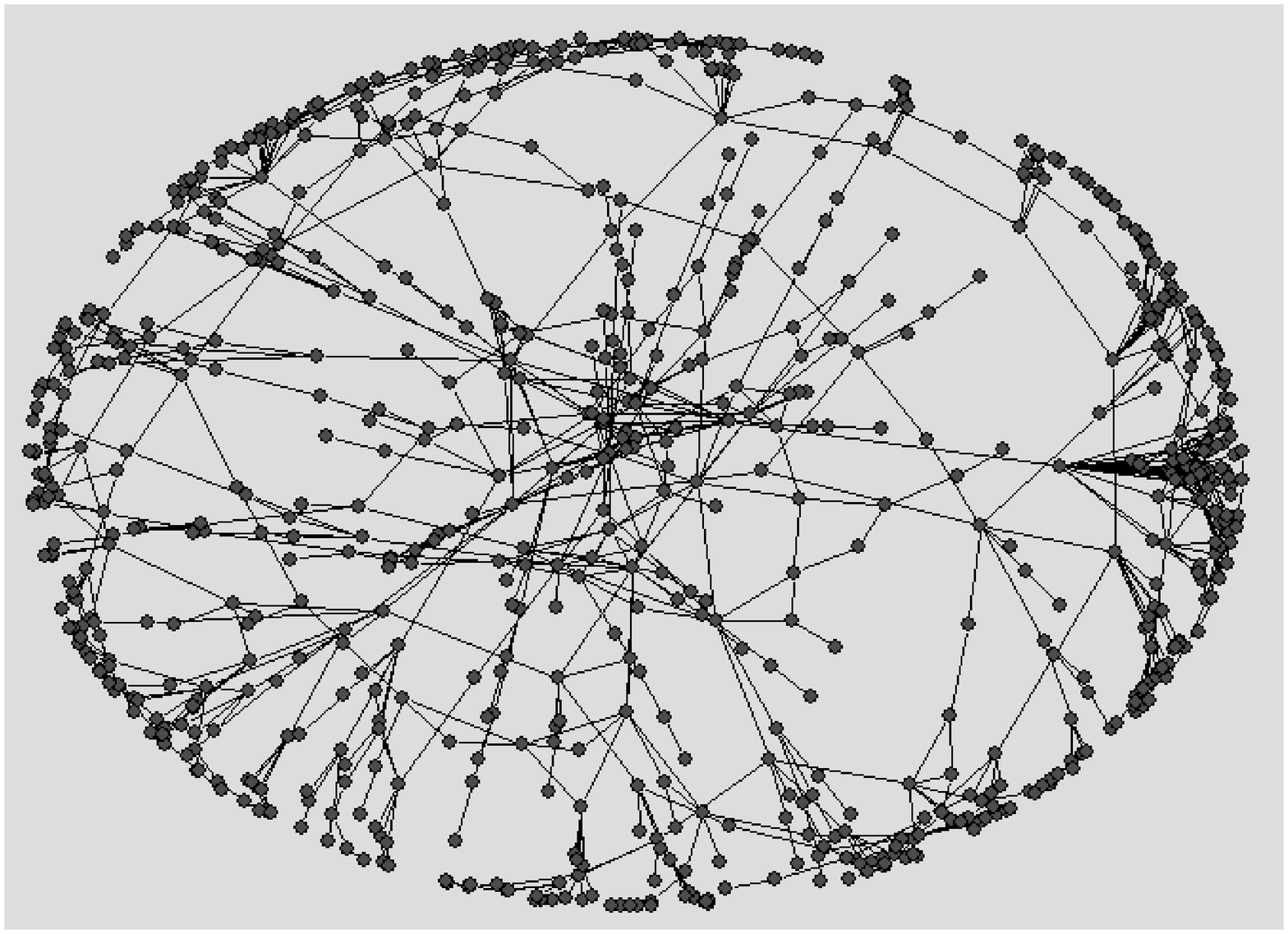}
\caption{MSN Messenger topology}
\end{minipage}
{\it Figures 2-5 illustrate subgraphs taken from the different
topologies plotted using the Pajek software~\cite{Pajek}. Despite
the different graph characteristics, the GaBP performed very well
on all tested topologies }
\end{figure}

\begin{table}
\begin{center}
\begin{tabular}{|c|c|c|c|}
  \hline
  Topology & Nodes & Edges & Data Source \\
  \hline
  MSN Messenger graph & ~1M & 9.4M & Microsoft  \\
  Blogs Web Crawl & 1.5M & 8M & IBM \\
  DIMES & 300K & 2.2M & DIMES Internet measurements \\
  US Government documents & 12M & 64M & Web research collection \\
  \hline
\end{tabular}
\caption{\mbox{              } Topologies used for
experimentation}
\end{center}
\end{table}

We have shown that various ranking methods can be computed by
solving a linear system of equations. We propose to use the GaBP algorithm, for
efficiently computing the ranking methods distributively by
the network nodes. Next, we bring simulation results which show
that for very large networks the GaBP algorithm performs remarkably well.

For evaluating the feasibility and efficiency of our rating
system, we used several types of large scale social networks
topologies:

\begin{enumerate}
\item{\bf MSN Live Messenger.} We used anonymized data obtained
from Windows Live Messenger that represents Messenger users' buddy
relationships. The Messenger network is rather large for
simulation (over two hundred million users), and so we cut
sub-graphs by starting at a random node, and taking a BFS cut of
about one million nodes. The diameter of the sampled graph is five
on average. \item{\bf Blog crawl data.} We used blog crawl data of
a network of about 1.5M bloggers, and 8M directed links to other
blogs. This data was provided thanks to Elad Yom-Tov from IBM
Research Labs, Haifa, Israel. \item{\bf DIMES Internet
measurements.} We used a snapshot of an Internet topology from
January 2007 captured by the DIMES project~\cite{DIMES}. The 300K
nodes are routers and the 2.2M edges are communication lines.
\item{\bf US gov document repository.} We used a crawl of 12M pdf
documents of US government, where the links are links within the
pdf documents pointing to other documents within the
repository~\cite{us_doc}.
\end{enumerate}

One of the interesting questions, is the practical convergence
speed of our rating methods. The results are given using the MSN
Messenger social network's topology, since all of the other
topologies tested obtained similar convergence results. We have
tested the Peer-to-Peer rating algorithm. We have drawn the input
ratings $y_i$ and edge weights $w_{ij}$ in uniformly at random in
the range $[0,1]$. We have repeated this experiment with different
initializations for the input rating and the edge weights and got
similar convergence results. We have tested other cost functions,
including PageRank and Spatial Ranking and got the same
convergence results.

Figure~\ref{conv-speed} shows the convergence speed of the
Peer-to-Peer rating algorithm. The x-axis represents round
numbers. The rounds are given only for reference, in practice
there is no need for the nodes to be synchronized in rounds as
shown in~\cite{RBP}. The y-axis represents the sum-total of change
in ratings relative to the previous round. We can see that the
node ratings converge very fast towards the optimal rating derived
from the cost function. After only five iterations, the total
change in nodes ratings is about $1$ (which means an average
change of $1 \times 10^{-6}$ per node).

\begin{figure}[ht!]
\begin{center}
\label{conv-speed}
  \includegraphics[width=350pt]{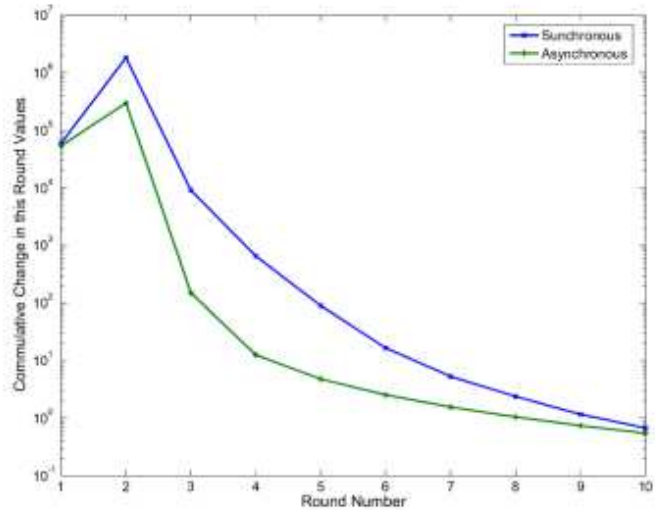}\\
  \caption{\it Convergence of rating over a social network of 1M nodes and 9.4M edges.
      Note, that using asynchronous rounds, the algorithm converges faster, as discussed in~\cite{RBP}}
\label{conv-speed}
\end{center}
\end{figure}

\subsection{Rating Benchmark}
\label{benchmark} For demonstrating the applicability of our
proposed cost functions, we have chosen to implement a
``benchmark'' that evaluates the effectiveness of the various cost
functions. Demonstrating this requires a quantitative measure
beyond mere speed and scalability. The benchmark approach we take
is as follows. First, we produce a ladder of ``social influence''
that is inferred from the network topology, and rank nodes by this
ladder, using the Spatial ranking cost function. Next, we test our
Peer-to-Peer rating method in the following settings. Some nodes
are initialized with rate values, while other nodes are
initialized with empty ratings. Influential nodes are given
different initial ratings than non-influential nodes. The expected
result is that the ratings of influential nodes should affect the
ratings of the rest of the network so long as they are not vastly
outnumbered by opposite ratings.

As a remark, we note that we can use the social ranks additionally
as trust indicators, giving higher trust values to edges which are
incident to high ranking nodes, and vice versa. This has the nice
effect of initially placing low trust on intruders, which by
assumption, cannot appear influential.

For performing our benchmark tests, we once again used simulation
over the $1$ million nodes sub-graph of the Messenger network.
Using the ranks produced by our spatial ranking, we selected the
seven highest ranking nodes and assigned them an initial rating
value $5$. We also selected seven of the lowest ranking nodes and
initialized them with rating value $1$. All other nodes started
with null input. The results of the rating system in this settings
are given in Figure 3. After about ten rounds, a majority of the
nodes converged to a rating very close to the one proposed by the
influential nodes. We ran a variety of similar tests and obtained
similar results in all cases where the influential nodes were not
totally outnumbered by opposite initial ratings; for brevity, we
report only one such test here.

The conclusion we draw from this test is that a combination of
applying our GaBP solver for computing first the rating of nodes,
and then using this rating for choosing influential nodes and
spreading their beliefs in the network has the desired effect of
fast and influential dissemination of the socially influential
nodes. This effect can have a lot of applications in practice,
including targeted commercials to certain zones in the social
network.

Quite importantly, our experiment shows that our framework provide
good protection against malicious infiltrators: Assuming that
intruders have low connectivity to the rest of the network, we
demonstrate that it is hard for them to influence the rating
values in the system. Furthermore, we note that this property will
be reinforced if the trust values on edges are reduced due to
their low ranks, and using users satisfaction feedback.

\begin{figure}[ht!]
\begin{center}
\label{Ranking}
  \includegraphics[width=240pt]{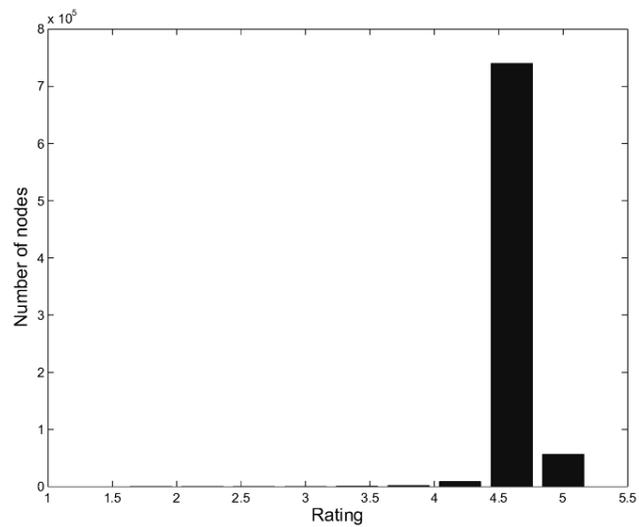}\\
  \caption{\it Final rating values in a network of 1M nodes. Initially, $7$ highest ranking nodes rate $5$ and $7$ lowest
  ranking nodes rate $1$.}
\end{center}
\end{figure}

\ignore{
\section{Conclusion} In this work we explore the connection
between different algorithms for collaborative filtering, visual
layouts and rating nodes and data items in Peer-to-Peer and social
networks. We connect those algorithms by introducing a family of
quadratic cost functions and proving analytically that other
methods can be reduced to our formulation. We propose a single
distributed solver which enables the efficient computation of
those different algorithms using different initializations of our
solver's input. Using experimental results over real social
network topologies of millions of nodes, we demonstrate the
applicability and attractiveness of our method. }

 \ignore{
\paragraph{Acknowledgements} We would like to thank Yaacov Fernandess
for his vision and support of the Nocturnal project, and Yair
Weiss for being a great research mentor in the field of Gaussian
Belief Propagation. We further like to thank Ciamak C. Moallemi ,
Benjamin Van-Roy, Erik Aurell, Danny Dolev and Jason K. Johnson
for interesting discussions regarding the GaBP algorithm. We thank
Elad Yom-Tov for supplying the Blogs crawl data and Udi Weinsberg
for his support in using the DIMES datasets. We also thank Matteo
Dell'Amico for pointing out the relation between visual layouts
and node ratings. We thank Ori Shental for clearly explaining the
GaBP algorithm in Section~\ref{gabp}. Finally, we thank the
anonymous reviewers for their helpful comments. } 

\chapter{Linear Detection}\label{sec_linear}
\label{linear_detection}
 Consider a discrete-time channel with a
real input vector $\vx=\{x_{1},\ldots,x_{K}\}^{T}$ and a corresponding
real output vector
$\vy=\{y_{1},\ldots,y_{K}\}^{T}=f\{\vx^{T}\}\in\mathbb{R}^{K}$.\footnote{An
extension to the complex domain is straightforward.} Here, the
function $f\{\cdot\}$ denotes the channel transformation. 
Specifically, CDMA\ multiuser detection problem is characterized by the following linear  channel

\[ \vy = \mS\vx +\vn\,, \]
where $\mS_{N \times K}$ is the CDMA chip sequence matrix, $\vx \in \{-1,1\}^K$ is the transmitted signal, $\vy \in \R^N$ is the observation vector and $\vn$ is a vector of AWGN noise. The multiuser detection problem is stated as follows. Given $\mS, \vy$ and knowledge about the noise, we would like to infer the most probable  $\vx$. This problem is NP-hard. 

The matched filter output is
\[ \mS^T\vy=\mR\vx+\mS^T\vn\,,
\] where $\mS^T\vn$ is a $K\times 1$ additive noise vector and
\mbox{${\mR_{\vk \times \vk}}=\mS^{T}\mS$} is a positive-definite symmetric matrix,
often known as the correlation matrix.

Typically, the binary constraint on $\vx$ is relaxed to $\vx \in [-1,1]$. This relaxation is called linear detection. In linear detection the decision rule is\BE\label{eq_gld}
\hat{\vx}=\Delta\{\vx^{\ast}\}=\Delta\{\mR^{-1}\mS^T\vy\}\,.\ \EE The vector $\vx^{\ast}$
is the solution (over $\mathbb{R}$) to $\mR\vx=\mS^T\vy$. Estimation is
completed by adjusting the (inverse) matrix-vector product to the
input alphabet, accomplished by using a
proper clipping function $\Delta\{\cdot\}$ (\eg, for binary
signaling $\Delta\{\cdot\}$ is the sign function).

Assuming linear channels with AWGN with variance $\sigma^{2}$ as the
ambient noise, the linear minimum mean-square error (MMSE) detector
can be described by using \mbox{$\mR+\sigma^{2}\mI_{K}$}, known
to be optimal when the input distribution $P_{\vx}$ is Gaussian.
In general, linear detection is suboptimal because of its
deterministic underlying mechanism (\ie, solving a given set of
linear equations), in contrast to other estimation schemes, such
as MAP or maximum likelihood, that emerge from an optimization
criteria.


The essence of detection theory is to estimate a hidden input to a
channel from empirically-observed outputs. An important class of
practical sub-optimal detectors is based on linear detection. This
class includes, for instance, the conventional single-user matched
filter, the decorrelator (also, called the zero-forcing
equalizer), the linear minimum mean-square error (LMMSE) detector,
and many other detectors with widespread
applicability~\cite{BibDB:BookVerdu,BibDB:BookProakis}. In general
terms, given a probabilistic estimation problem, linear detection
solves a deterministic system of linear equations derived from the
original problem, thereby providing a sub-optimal, but often
useful, estimate of the unknown input.

Applying the GaBP solver to linear detection, we establish a new
and explicit link between BP and linear detection. This link
strengthens the connection between message-passing inference and
estimation theory, previously seen in the context of optimal
maximum a-posteriori (MAP)
detection~\cite{BibDB:Kabashima,BibDB:ShentalITW} and several
sub-optimal nonlinear detection
techniques~\cite{BibDB:TanakaOkada} applied in the context of both
dense and sparse~\cite{BibDB:MontanariTse,BibDB:ConfWangGuo}
graphical models.

In the following experimental study, we examine the implementation
of a decorrelator detector in a noiseless synchronous CDMA system
with binary signaling and spreading codes based upon Gold
sequences of length $m=7$.\footnote{In this case, as long as the
system is not overloaded, \ie the number of active users $n$ is
not greater than the spreading code's length $m$, the decorrelator
detector yields optimal detection decisions.} Two system setups
are simulated,  corresponding to $n=3$ and $n=4$ users, resulting
in the cross-correlation matrices  \BE
\mR_{3} = \frac{1}{7}\left(%
\begin{array}{rrr}
  7 & -1 & 3 \\
  -1 & 7 & -5 \\
  3 & -5 & 7 \\
\end{array} \right), \EE and
\BE \mR_{4} = \frac{1}{7}\left(%
\begin{array}{rrrr}
  7 & -1 & 3 & 3\\
  -1 & 7 & 3 & -1\\
  3 & 3 & 7 & -1\\
  3 & -1 & -1 & 7\\
\end{array} \right), \EE respectively.\footnote{These particular correlation settings were taken from the simulation setup of Yener \etal~\cite{BibDB:YenerEtAl}.}

The decorrelator detector, a member of the family of linear
detectors, solves a system of linear equations,
\mbox{$\mR\vx=\mS^T\vy$}, thus the vector of decorrelator decisions is determined by
taking the signum of the vector $\mR^{-1}\mS^T\vy$. Note
that $\mR_{3}$ and $\mR_{4}$ are not strictly diagonally dominant,
but their spectral radius are less than unity, since
$\rho(|\mI_{3}-\mR_{3}|)=0.9008<1$ and
$\rho(|\mI_{4}-\mR_{4}|)=0.8747<1$, respectively. In all of the
experiments, we assumed the output vector was the all-ones vector.

Table~\ref{tab_1} compares the proposed GaBP algorithm with
standard iterative solution methods~\cite{BibDB:BookAxelsson}
(using random initial guesses), previously employed for CDMA
multiuser detectors (MUD). Specifically, MUD algorithms based on
the algorithms of Jacobi, Gauss-Seidel (GS) and (optimally
weighted) successive over-relaxation (SOR)\footnote{This moving
average improvement of Jacobi and GS algorithms is equivalent to
what is known in the BP literature as `damping'~\cite{Damping}.}
were investigated~\cite{grant99iterative,BibDB:TanRasmussen}. The
table lists the convergence rates for the two Gold code-based CDMA
settings. Convergence is identified and declared when the
differences in all the iterated values are less than $10^{-6}$. We
see that, in comparison with the previously proposed detectors
based upon the Jacobi and GS algorithms, the GaBP detectors
converge more rapidly for both $n=3$ and $n=4$. The serial
(asynchronous) GaBP algorithm achieves the best overall
convergence rate, surpassing even the SOR-based detector. 

\begin{table} \centerline{
\begin{tabular}{|c|r|r|}
  \hline
  \textbf{Algorithm}
  & Iterations $t$ ($\mR_{3}$) & Iterations $t$ ($\mR_{4}$) \\\hline\hline & & \\
  Jacobi & 111\comment{without dividing R_{3} with 7: 136} & 24\comment{50} \\\hline & & \\
  GS & 26 & 26\comment{32}\\\hline & & \\
  \textbf{Parallel GaBP} & \textbf{23} & \textbf{24}\\\hline & & \\
  Optimal SOR & 17 & 14\comment{20} \\\hline & & \\
  \textbf{Serial GaBP} & \textbf{16} & \textbf{13}\\
  \hline
\end{tabular}
}\vspace{0.5cm} \caption{Decorrelator for $K=3,4$-user, $N=7$ Gold
code CDMA. Total number of iterations required for convergence
(threshold $\epsilon=10^{-6}$) for GaBP-based solvers vs. standard
methods.}\label{tab_1}
\end{table}

\begin{figure}[t!]
\begin{minipage}[b]{0.5\linewidth}
\centering
    \includegraphics[width=\textwidth]{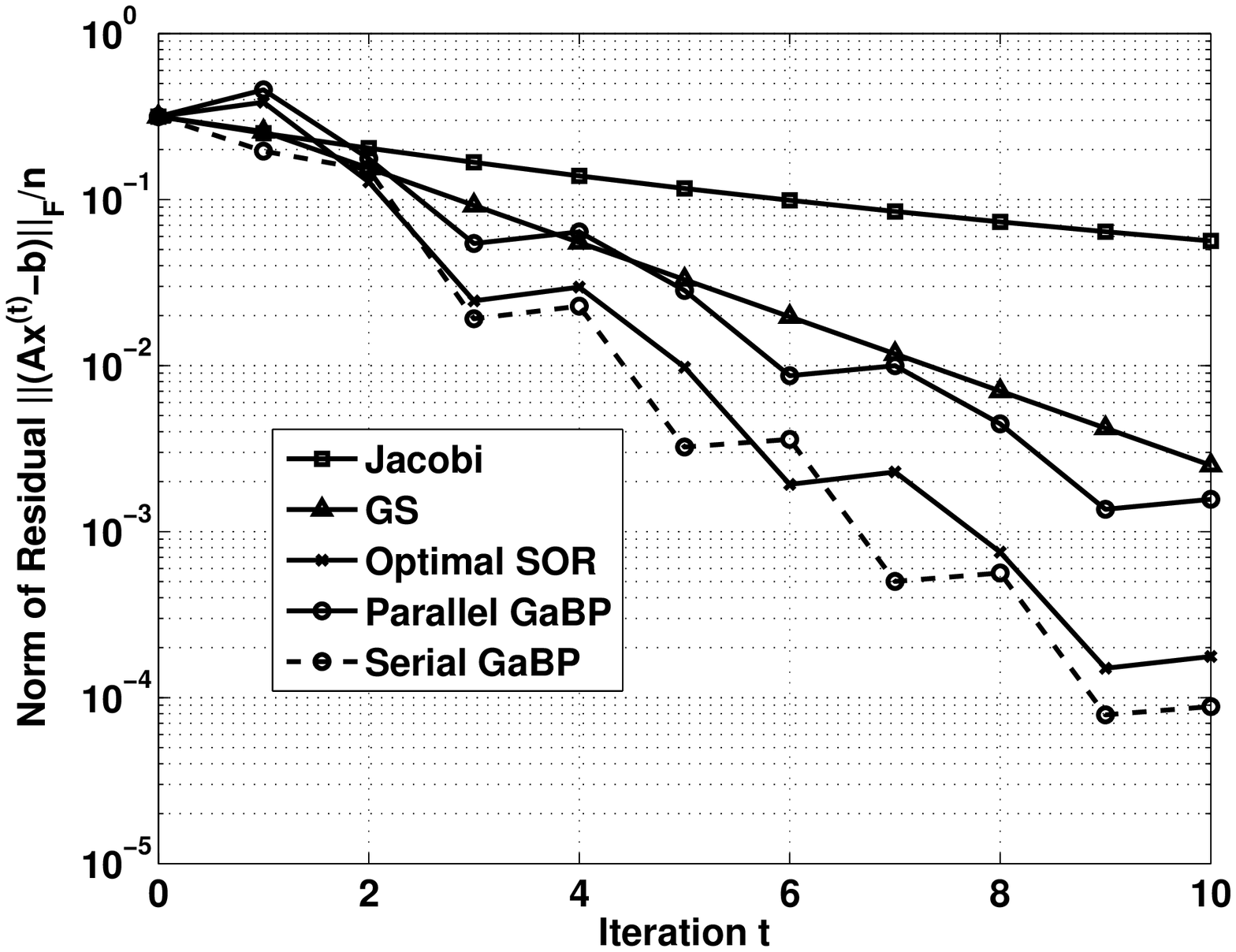}
 \label{fig_R3}
\end{minipage}
\begin{minipage}[b]{0.5\linewidth}
\centering
   \includegraphics[width=\textwidth]{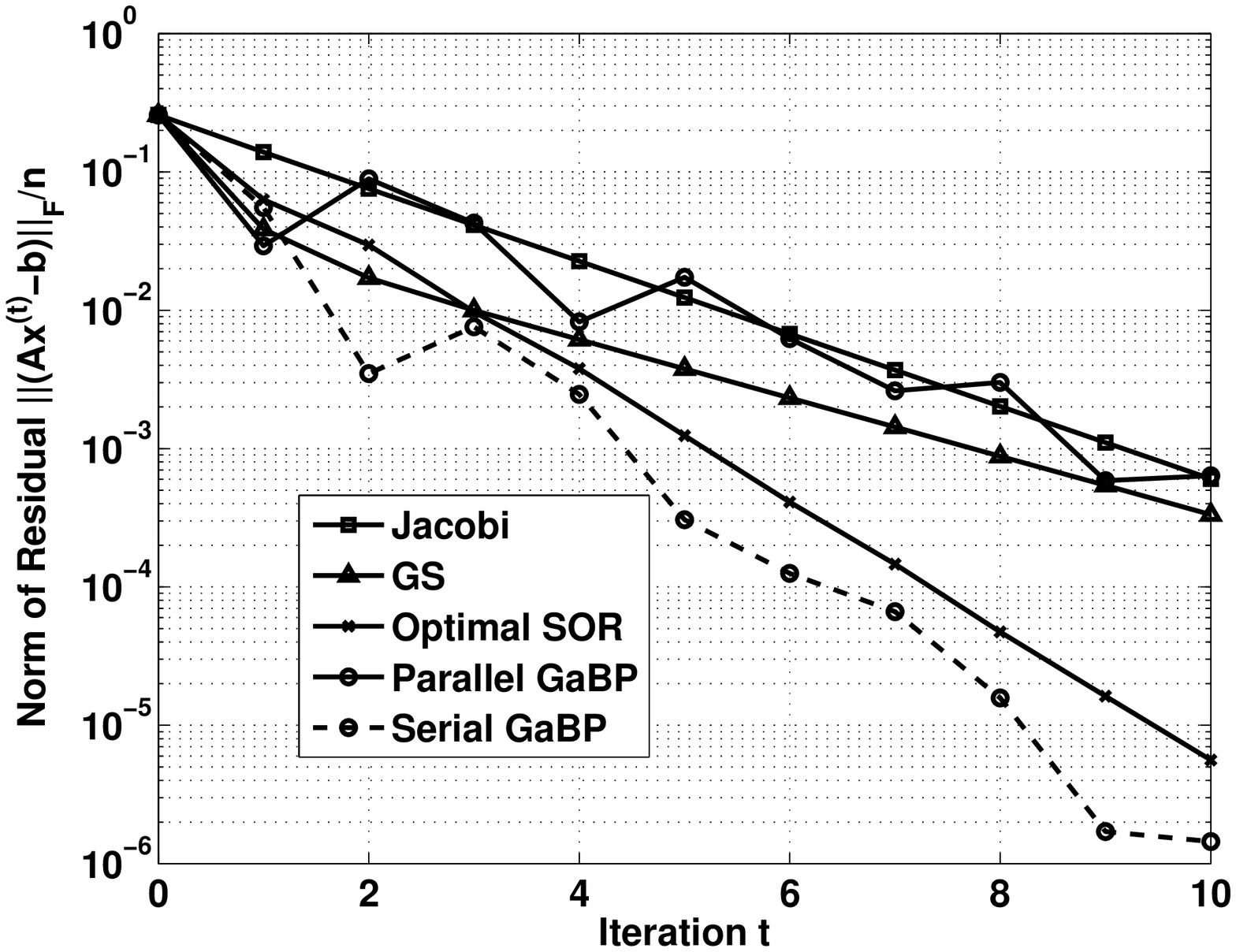}
 \label{fig_R4}
\end{minipage}
\caption{Convergence of the two gold CDMA matrices. To the left $\mR_3$, to the right, $\mR_4$. }
\end{figure}

\begin{table}[h!]
\centerline{
\begin{tabular}{|l|c|c|}
  \hline
  \textbf{Algorithm}
  & $\mR_{3}$ & $\mR_{4}$ \\
  \hline\hline & & \\
  Jacobi+Steffensen\footnotemark & 59\comment{51} & $-$ \\\hline & & \\
  \textbf{Parallel GaBP+Steffensen} & \textbf{13} & \textbf{13}\\\hline & & \\
  \textbf{Serial GaBP+Steffensen} & \textbf{9} & \textbf{7} \\
  \hline
\end{tabular}
}\vspace{0.5cm}\caption{Decorrelator for $K=3,4$-user, $N=7$ Gold
code CDMA. Total number of iterations required for convergence
(threshold $\epsilon=10^{-6}$) for Jacobi, parallel and serial
GaBP solvers accelerated by Steffensen iterations.}\label{tab_2}
\end{table}

Further speed-up of GaBP can be achieved by adapting known
acceleration techniques from linear algebra, such Aitken's method
and Steffensen's iterations, as explained in Section \ref{sec:accel}. 
 Table~\ref{tab_2} demonstrates the speed-up of GaBP
obtained by using these acceleration methods, in comparison with
that achieved by the similarly modified Jacobi
algorithm.\footnote{Application of Aitken and Steffensen's methods
for speeding-up the convergence of standard (non-BP) iterative
solution algorithms in the context of MUD was introduced by Leibig
\etal~\cite{LDF05}.} We remark that, although the convergence rate
is improved with these enhanced algorithms, the region of
convergence of the accelerated GaBP solver remains unchanged.

For the algorithms we examined, Figure \ref{fig_R3}.1 displays the Euclidean
distance between the tentative (intermediate) results and the
fixed-point solution as a function of the number of iterations. As
expected, all linear algorithms exhibit a logarithmic convergence
behavior. Note that GaBP converges faster on average, although
there are some fluctuations in the GaBP curves, in contrast to the
monotonicity of the other curves.

\begin{figure}[t!]
\begin{minipage}[b]{0.5\linewidth}
\centering
     \includegraphics[width=\textwidth]{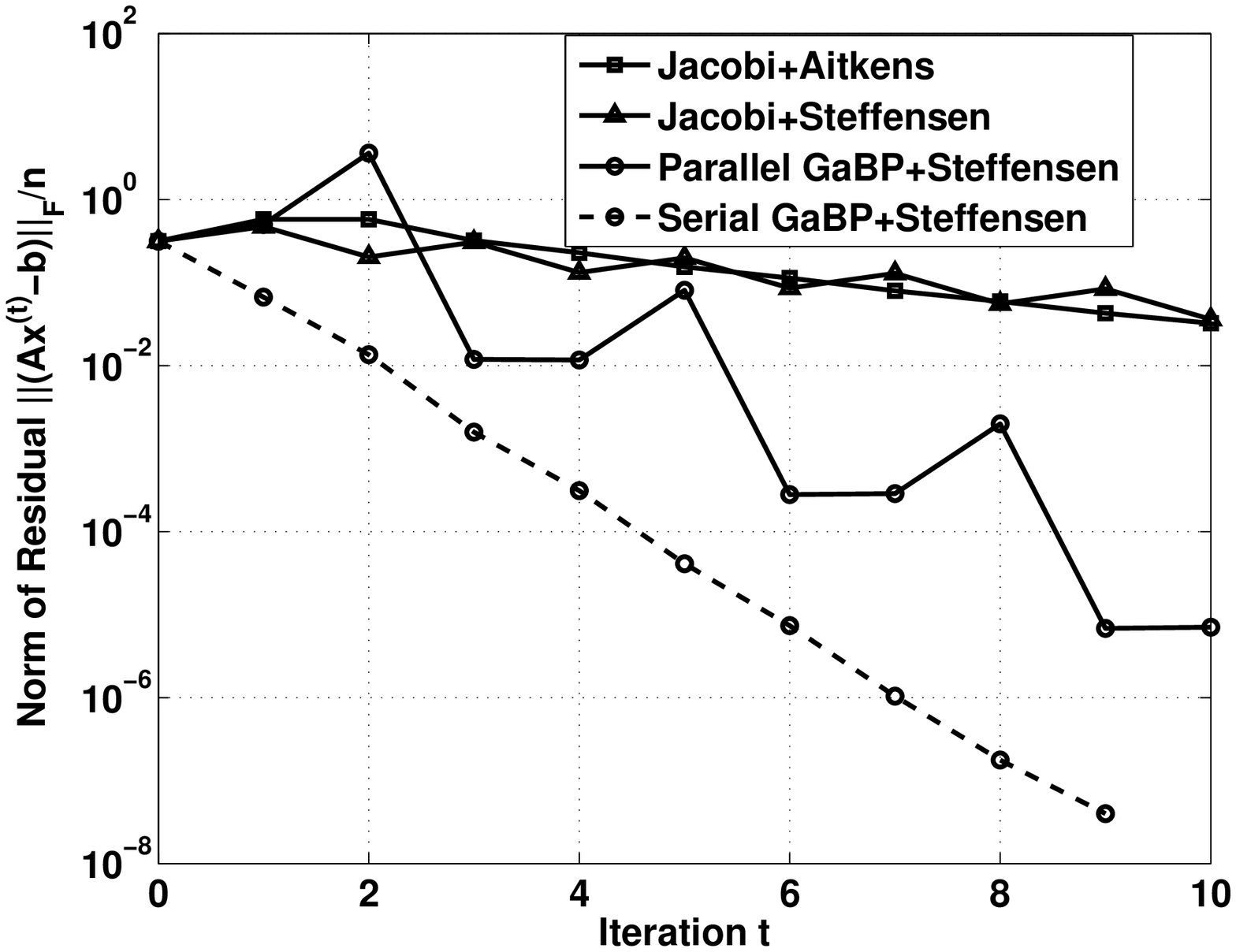}
  \label{fig_R3_accel}
\end{minipage}
\begin{minipage}[b]{0.5\linewidth}
\centering
\vspace{5mm}
   \includegraphics[width=\textwidth]{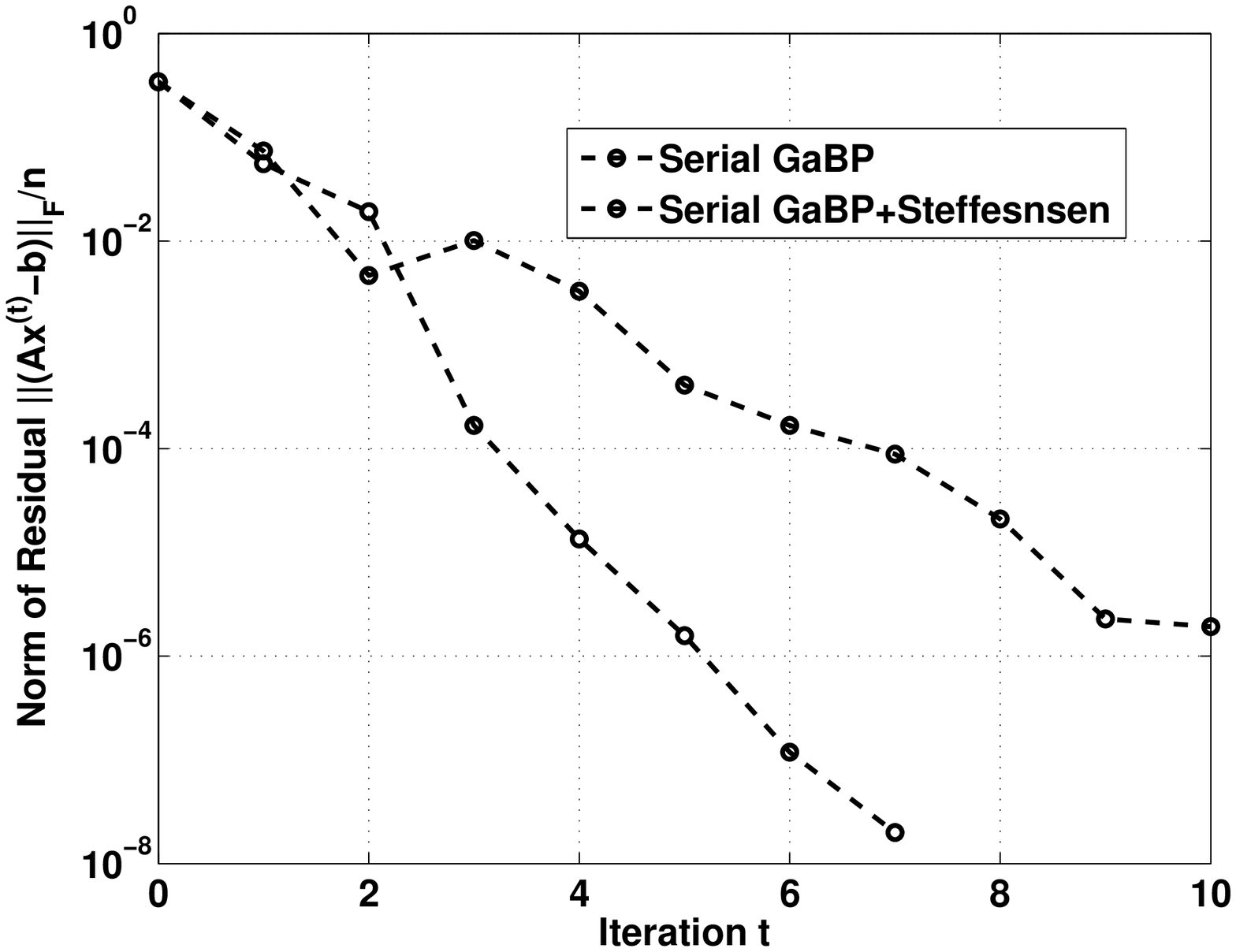}
  \label{fig_R4_accel}
\end{minipage}
 \caption{Convergence acceleration of the GaBP algorithm using Aitken and Steffensen methods.
  The left graph depicts a $3 \times 3$ gold CDMA matrix, the right graph $4 \times 4$ gold
  CDMA matrix. }
\end{figure}

\begin{figure}[h!]
\begin{center}
    \includegraphics[width=0.5\textwidth]{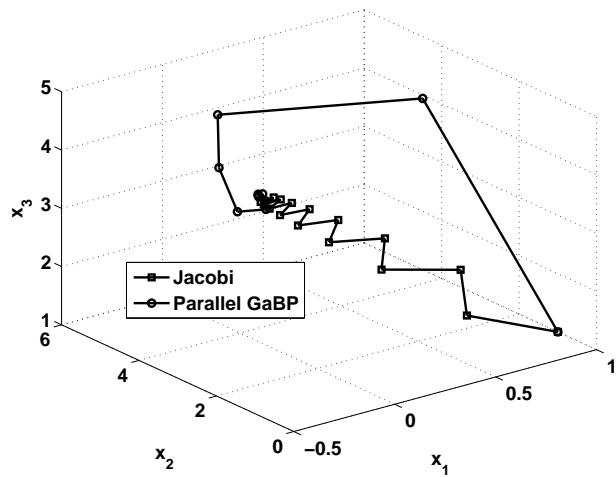}
  \caption{Convergence of the GaBP algorithm vs. Jacobi on a $3 \times 3$ gold CDMA matrix. Each dimension shows
  one coordinate of the solution. Jacobi converges in zigzags while GaBP has spiral convergence.}
\label{fig_R3_spiral}
\end{center}
\end{figure}

An interesting question concerns the origin of this convergence
speed-up associated with GaBP. Better understanding may be gained
by visualizing the iterations of the different methods for the
matrix $\mR_{3}$ case. The convergence contours are plotted in the
space of $\{x_{1},x_{2},x_{3}\}$ in Fig. \ref{fig_R3_spiral}. As expected, the
Jacobi algorithm converges in zigzags towards the fixed point
(this behavior is well-explained in Bertsekas and
Tsitsiklis~\cite{BibDB:BookBertsekasTsitsiklis}).\comment{SOR
algorithm averages the Jacobi algorithm convergence behavior thus
gains faster convergence speed. Aitken's speeds-up Jacobi's
convergence.} The fastest algorithm is serial GaBP. It is
interesting to note that GaBP convergence is in a spiral shape,
hinting that despite the overall convergence improvement,
performance improvement is not guaranteed in successive iteration
rounds. In this case the system was simulated with a specific
$\mR$ matrix for which Jacobi algorithm and other standard methods
did not even converge. Using Aitken's method, a further speed-up
in GaBP convergence was obtained.

Despite the fact that the examples considered correspond to small
multi-user systems, we believe that the results reflect the
typical behavior of the algorithms, and that similar qualitative
results would be observed in larger systems. In support of this
belief, we note, in passing, that GaBP was experimentally shown to
converge in a logarithmic number of iterations in the cases of
very large matrices both dense (with up to hundreds of thousands
of dimensions~\cite{SVM}) and sparse (with up to millions of
dimensions~\cite{p2p-rating}).

As a final remark on the linear detection example, we mention
that, in the case of a channel with Gaussian input signals,
for which linear detection is optimal, the proposed GaBP scheme reduces to the BP-based MUD scheme,
recently introduced by Montanari \etal \cite{BibDB:MontanariEtAl}, as shown in Chapter \ref{extended_mont}.
Montanari's BP scheme, assuming a Gaussian prior, has
been proven to converge to the MMSE (and optimal) solution for any
arbitrarily loaded, randomly-spread CDMA system (\ie, a system
where $\rho(|\mI_{n}-\mR|)\lesseqgtr1$).\footnote{For non-Gaussian
signaling, \eg with binary input alphabet, this BP-based detector
is conjectured to converge only in the large-system limit, as
$n,m\rightarrow\infty$~\cite{BibDB:MontanariEtAl}.}Thus
Gaussian-input additive white Gaussian noise CDMA is another
example for which the proposed GaBP solver converges to the MAP
decisions for any $m\times n$ random spreading matrix $\mS$,
regardless of the spectral radius.

\section[GaBP Extension]{Extending GaBP to support non-square matrices}
\label{chap:extended}
\label{improved} In the previous section, linear detection has
been explicitly linked to BP~\cite{Allerton}, using a Gaussian
belief propagation (GaBP) algorithm. This allows for an efficient iterative computation of the linear
detector~\cite{ISIT1}, circumventing the need
of, potentially cumbersome, direct matrix inversion (via, \eg,
Gaussian elimination). The derived iterative framework was
compared quantitatively with `classical' iterative methods for
solving systems of linear equations, such as those investigated in
the context of linear implementation of 
CDMA
demodulation~\cite{grant99iterative,BibDB:TanRasmussen,BibDB:YenerEtAl}.
GaBP is shown to yield faster convergence than these standard
methods. Another important work is the BP-based MUD, recently
derived and analyzed by Montanari \etal
~\cite{BibDB:MontanariEtAl} for Gaussian input symbols.

There are several drawbacks to the linear detection technique
presented earlier~\cite{Allerton}. First, the input matrix $\mR_{n
\times n} = \mS_{n \times k}^T\mS_{k \times n}$ (the chip
correlation matrix) needs to be computed prior to running the
algorithm. This computation requires $n^2k$ operations. In case
where the matrix $\mS$ is sparse~\cite{BibDB:ConfWangGuo}, the
matrix $\mR$ might not no longer be sparse. Second, GaBP uses
$2n^2$ memory to store the messages. For a large $n$ this could be
prohibitive.

In this section, we propose a new construction that addresses
those two drawbacks. In our improved construction, given a
non-rectangular CDMA matrix $\mS_{n \times k}$, we compute the
MMSE detector $\vx = (\mS^T\mS + \Psi)^{-1}\mS^Ty$, where $\Psi=\sigma^{-2}\mI$ is
the AWGN diagonal inverse covariance matrix. We utilize the GaBP algorithm
which is an efficient iterative distributed algorithm. The new
construction uses only $2nk$ memory for storing the messages. When
$k \ll n$ this represents significant saving relative to the
$2n^2$ in our previously proposed algorithm. Furthermore, we do
not explicitly compute $\mS^T\mS$, saving an extra $n^2k$
overhead.

In Chapter \ref{extended_mont} we show that Montanari's algorithm~\cite{BibDB:MontanariEtAl} is
an instance of GaBP. By showing this, we are able to prove
new convergence results for Montanari's algorithm. Montanari
proves that his method converges on normalized random-spreading
CDMA sequences, assuming Gaussian signaling. Using binary
signaling, he conjectures convergence to the large system limit.
Here, we extend Montanari's result, to show that his algorithm
converges also for non-random CDMA sequences when binary signaling
is used, under weaker conditions. Another advantage of our work is
that we allow different noise levels per bit transmitted.

\subsection{Distributed Iterative Computation of the MMSE Detector}
\label{sec_new_const} In this section, we efficiently extend the
applicability of the proposed GaBP-based solver for systems with
symmetric matrices~\cite{Allerton} to systems with any square
(\ie, also nonsymmetric) or rectangular matrix. We first construct
a new symmetric data matrix $\tilde{\mR}$ based on an arbitrary
(non-rectangular) matrix $\mS\in\mathbb{R}^{k\times n}$ \BE
\label{newR} \tilde{\mR}\triangleq\left(
  \begin{array}{cc}
    \mI_{k} & \mS^T \\
    \mS & -\Psi \\
  \end{array}
\right)\in\mathbb{R}^{(k+n)\times(k+n)}. \EE Additionally, we
define a new vector of variables
$\tilde{\vx}\triangleq\{\hat{\vx}^{T},\vz^{T}\}^{T}\in\mathbb{R}^{(k+n)\times1}$,
where $\hat{\vx}\in\mathbb{R}^{k\times1}$ is the (to be shown)
solution vector and $\vz\in\mathbb{R}^{n\times1}$ is an auxiliary
hidden vector, and a new observation vector
$\tilde{\vy}\triangleq\{\mathbf{0}^{T},\vy^{T}\}^{T}\in\mathbb{R}^{(k+n)\times1}$.

Now, we would like to show that solving the symmetric linear
system $\tilde{\mR}\tilde{\vx}=\tilde{\vy}$ and taking the first
$k$ entries of the corresponding solution vector $\tilde{\vx}$ is
equivalent to solving the original (not necessarily symmetric)
system $\mR\vx=\vy$. Note that in the new construction the matrix
$\tilde{\mR}$ is sparse again, and has only $2nk$ off-diagonal
nonzero elements. When running the GaBP algorithm we have only
$2nk$ messages, instead of $n^2$ in the previous construction.

Writing explicitly the symmetric linear system's equations, we get
\[     \hat{\vx}+\mS^T\vz=\mathbf{0}, \ \ \ \   \mS\hat{\vx}-\Psi \vz=\vy.     \]
Thus, \[ \hat{\vx}=\Psi^{-1}\mS^{T}(\vy-\mS\hat{\vx}), \] and
extracting $\hat{\vx}$ we have \[
\hat{\vx}=(\mS^{T}\mS+\Psi)^{-1}\mS^{T}\vy. \] Note, that when the
noise level is zero, $\Psi=0_{m \times m}$, we get the
Moore-Penrose pseudoinverse solution\[
\hat{\vx}=(\mS^{T}\mS)^{-1}\mS^{T}\vy=\mS^{\dag}\vy. \]

\subsection{Relation to Factor Graph} \label{sec_factor} In this
section we give an alternate proof of the correctness of our
construction. Given the inverse covariance matrix $\tilde{\mR}$
defined in (\ref{newR}), and the shift vector $ \tilde{\vx}$ we
can derive the matching self and edge potentials
\[ \psi_{ij}(x_{i},x_{j})\triangleq \exp(-x_{i}R_{ij}x_{j})\,, \]
\[ \phi_{i}(x_{i})\triangleq \exp(-1/2 x_{i}R_{ii}^2 x_{i} - x_i y_i)\,, \]
which is a factorization of the Gaussian system distribution
\[ p(\vx) \propto \prod_i \phi_i(x_i) \prod_{i,j} \psi_{ij}(x_i,
x_j) =  \prod_{i \le k} \phi_i(x_i) \prod_{i > k} \phi_i(x_i)
\prod_{i,j} \psi_{ij}(x_i, x_j) = \]
\[ = \prod_{i \le k} \overbrace{\exp( -\frac{1}{2} x_i^2 )}^{\mbox{ prior on x}} \prod_{i > k} \exp(-\frac{1}{2}
\Psi_i x_i^2 - x_i y_i) \prod_{i,j}
\exp(-x_{i}\overbrace{S_{ij}}^{ R_{ij}} x_{j}).
\]

\begin{figure}
\begin{center}
  \includegraphics[width=200pt]{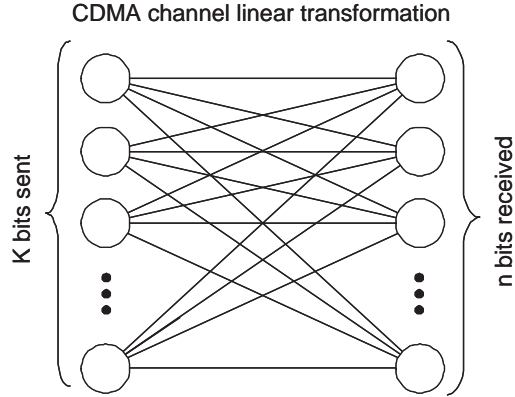}\\
  \caption{Factor graph describing the linear channel}\label{fig:factor}
\end{center}
\end{figure}

Next, we show the relation of our construction to a factor graph.
We will use a factor graph with $k$ nodes to the left (the bits
transmitted) and $n$ nodes to the right (the signal received),
shown in Fig \ref{fig:factor}. Using the definition
$\tilde{\vx}\triangleq\{\hat{\vx}^{T},\vz^{T}\}^{T}\in\mathbb{R}^{(k+n)\times1}$
the vector $\hat{\vx}$ represents the $k$ input bits and the
vector $\vz$ represents the signal received. Now we can write the
system probability as:
\[ p(\tilde{\vx}) \propto \int_{\hat{\vx}} \mathcal{N}(\hat{\vx};0,I) \mathcal{N}(\vz;S\hat{\vx}, \Psi) d\hat{\vx}\,. \]
It is known that the marginal distribution over $\vz$ is:
\[ = \mathcal{N}(\vz; 0, \mS^T\mS + \Psi). \]
The marginal distribution is Gaussian, with the following parameters:
\[ E(\vz|\hat{\vx}) = (\mS^T\mS + \Psi)^{-1}\mS^T\vy, \]
\[ Cov(\vz|\hat{\vx}) = (\mS^T\mS + \Psi)^{-1}. \]

 It is interesting to note that a similar
construction was used by Frey~\cite{frey99turbo} in his seminal
1999 work when discussing the factor analysis learning problem.
Comparison to Frey's work is found in Chapter ~\ref{sec:frey}.

\subsection{Convergence Analysis} \label{sec_conv}  
In this section we characterize the convergence properties of our linear detection method base on GaBP.
We know that if the matrix $\tilde{\mR}$ is strictly diagonally
dominant, then GaBP converges and the marginal means converge to
the true means~\cite[Claim 4]{BibDB:Weiss01Correctness}. Noting
that the matrix $\tilde{\mR}$ is symmetric, we can determine the
applicability of this condition by examining its columns.
As shown in Figure \ref{fig:CMDA_MAT} we see that in the first $k$ columns, we have the
$k$ CDMA sequences. We assume random-spreading binary CDMA
sequences where the total system power is normalized to one. In other words, the
absolute sum of each column is $1$. By adding $\epsilon$ to the main
diagonal, we insure that the first $k$ columns are diagonally dominant.
In the
next $n$ columns of the matrix $\tilde{\mR}$, we have the diagonal
covariance matrix $\Psi$ with different noise levels per bit in
the main diagonal, and zero elsewhere. The absolute sum of each
column of $S$ is $k/n$, thus when the noise level of each
bit satisfies $\Psi_i > k/n$, we have a convergence
guarantee. Note, that the convergence condition is a {\em
sufficient condition}. Based on Montanari's work, we also know
that in the large system limit, the algorithm converges for binary
signaling, even in the absence of noise.

\begin{figure}
\[ \left(
        \begin{array}{cccc|cccccc}
          1      & 0 & \cdots & 0   & 1/n & -1/n & -1/n & 1/n & \hdots & 1/n \\
          0      & 1 &        & 0   &  &  & &  &  &  \\
          \cdots &   & \ddots &    &  &  &  &  &  &  \\
          0      & 0 & \cdots & 1   & 1/n & -1/n & 1/n & -1/n & \hdots & -1/n \\ \hline
          1/n    & \cdots  &    & 1/n & \Psi_1 & 0 & 0 &  & \cdots & 0 \\
          -1/n   &   &        & -1/n& 0 & \Psi_2 &  &  &  &  \hdots \\
          -1/n   &   &        & 1/n &  &  &  &  &  &  \\
          1/n    &   &        & -1/n&  &  &  &  & &  \\
          \vdots   &   &      &\vdots&  &  &  & & \ddots & 0 \\
          1/n    &   &        & -1/n&   &  &  & \cdots & 0 & \Psi_n \\
        \end{array}
      \right) \]
\caption{An example randomly spreading CDMA sequences matrices using our new construction.
Using this illustration, it is easy to give a sufficient convergence proof to the GaBP algorithm.
Namely, when the above matrix is diagonally dominant.}
\label{fig:CMDA_MAT}
\end{figure}

In Chapter \ref{extended_mont}
we prove that Montanari's algorithm is an instance of our
algorithm, thus our convergence results apply to Montanari's
algorithm as well.

\section{Applying GaBP Convergence Fix}
\label{sec:fix-applied}
Next, we apply our novel double loop technique described in Chapter \ref{chap:conv_fix}, for forcing the
convergence of our linear detection algorithm using GaBP. We use the following setting:
given a random-spreading CDMA code\footnote{Randomly-spread CDMA\ code is a code where the matrix $\mS$ is initialized uniformly with the entries  $\pm \tfrac{1}{n}$.} with chip sequence length $n =
256$, and $k = 64$ users. We assume a diagonal AWGN with $\sigma^2 =
1$. Matlab code of our implementation is available on
\cite{MatlabGABP}.

Using the above settings, we have drawn at random random-spreading CDMA matrix.
Typically, the
sufficient convergence conditions for the GaBP algorithm do not
hold. For example, we have drawn at random a randomly-spread CDMA
matrix with $\rho(|I_{K}-{C^{N}}|) = 4.24$, where $C^N$ is a diagonally-normalized version of $(C+\sigma^2 I_K)$. Since $\rho(|I_K-{C^{N}}|) > 1$, the GaBP
algorithm for multiuser detection is not guaranteed to converge.

\begin{figure}[ht!]
\centering{
  \includegraphics[bb=83 180 504 488,scale=0.49,clip]{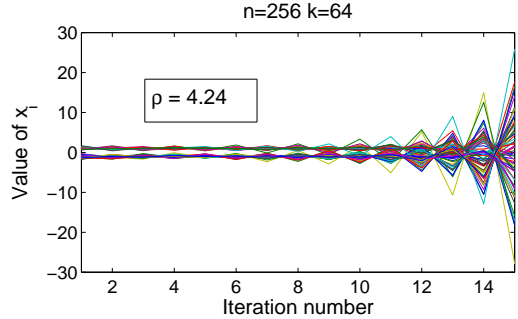}\\
  \caption{Divergence of the GaBP algorithm for the multiuser detection problem,
when $n=256, k=64$. }\label{fig:residual_gabp}
}
\end{figure}

Figure~\ref{fig:residual_gabp} shows that under the above settings,
the GaBP algorithm indeed diverged. The $x$-axis represent iteration
number, while the values of different $x_i$ are plotted using
different colors. This figure depicts well the fluctuating divergence
behavior.

\begin{figure}[h!]
\centering{
  \includegraphics[scale=0.44,clip]{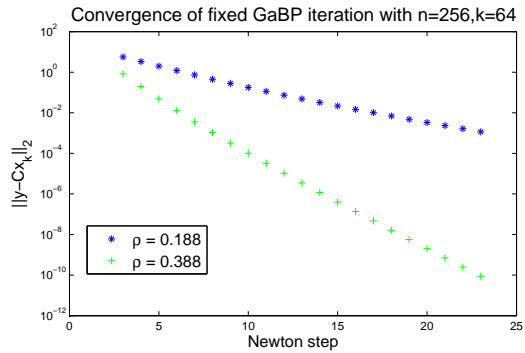}\\
  \caption{Convergence of the fixed GaBP iteration under the same settings
($n=256, k=64$).}\label{fig:newton}
}
\end{figure}

Next, we deployed our proposed construction and used a diagonal
loading to force convergence.  Figure~\ref{fig:newton} shows two
different possible diagonal loadings. The $x$-axis shows the Newton step
number, while the $y$-axis shows the residual. We experimented with two
options of diagonal loading.  In the first, we forced the matrix to be
diagonally-dominant (DD). In this case, the spectral radius $\rho =
0.188$. In the second case, the matrix was not DD, but the spectral
radius was $\rho = 0.388$. Clearly, the Newton method converges
faster when the spectral radius is larger. In both cases the inner iterations
converged
in five steps to an accuracy of $10^{-6}$.

The tradeoff between the amount of diagonal weighting to the total convergence speed is shown in Figures 3,4. A CDMA multiuser detection problem is shown ($k=128$, $n=256$). Convergence threshold for the inner and outer loops where $10^{-6}$ and $10^{-3}$. The $x$-axis present the amount of diagonal weighting normalized such that 1 is a diagonally-dominant matrix. $y$-axis represent the number of iterations. As expected, the outer loop number of iterations until convergence grows with $\gamma$. In contrast, the average number of inner loop iterations per Newton step (Figure 4)\ tends to decrease as $\gamma$ increases. The total number of iterations (inner $\times$ outer) represents the tradeoff between the inner and outer iterations and has a clear global~minima.

%

\begin{figure}[t!]
\begin{minipage}[b]{0.5\linewidth}
\centering
  \hspace{1.25in} 
  \includegraphics[scale=0.44,clip]{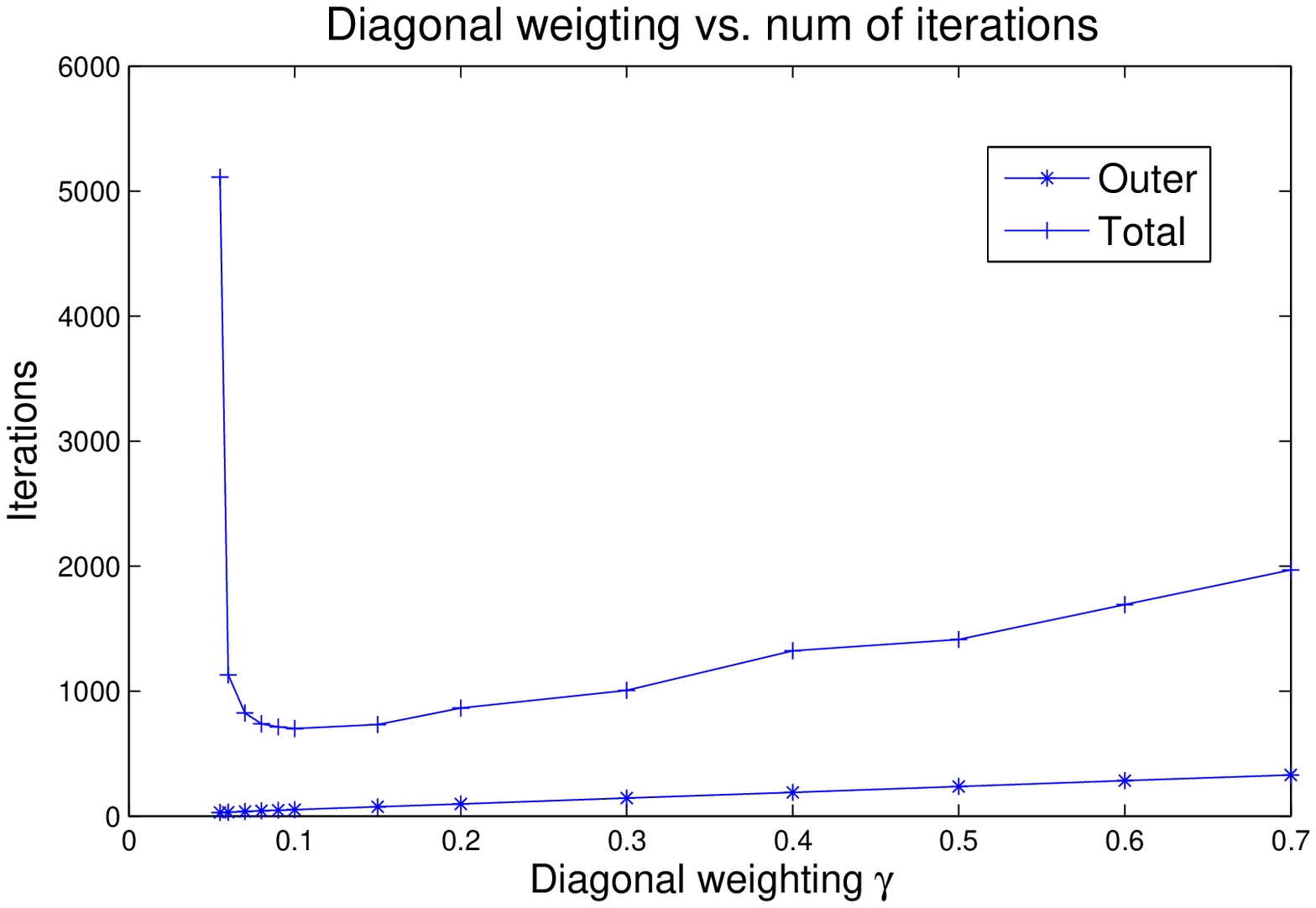}
   \caption{Effect of diagonal weighting on outer loop convergence speed. }\label{fig:outer}
  \label{figC}
\end{minipage}
\begin{minipage}[b]{0.5\linewidth}
\centering
  \hspace{1.25in} 
    \includegraphics[scale=0.44,clip]{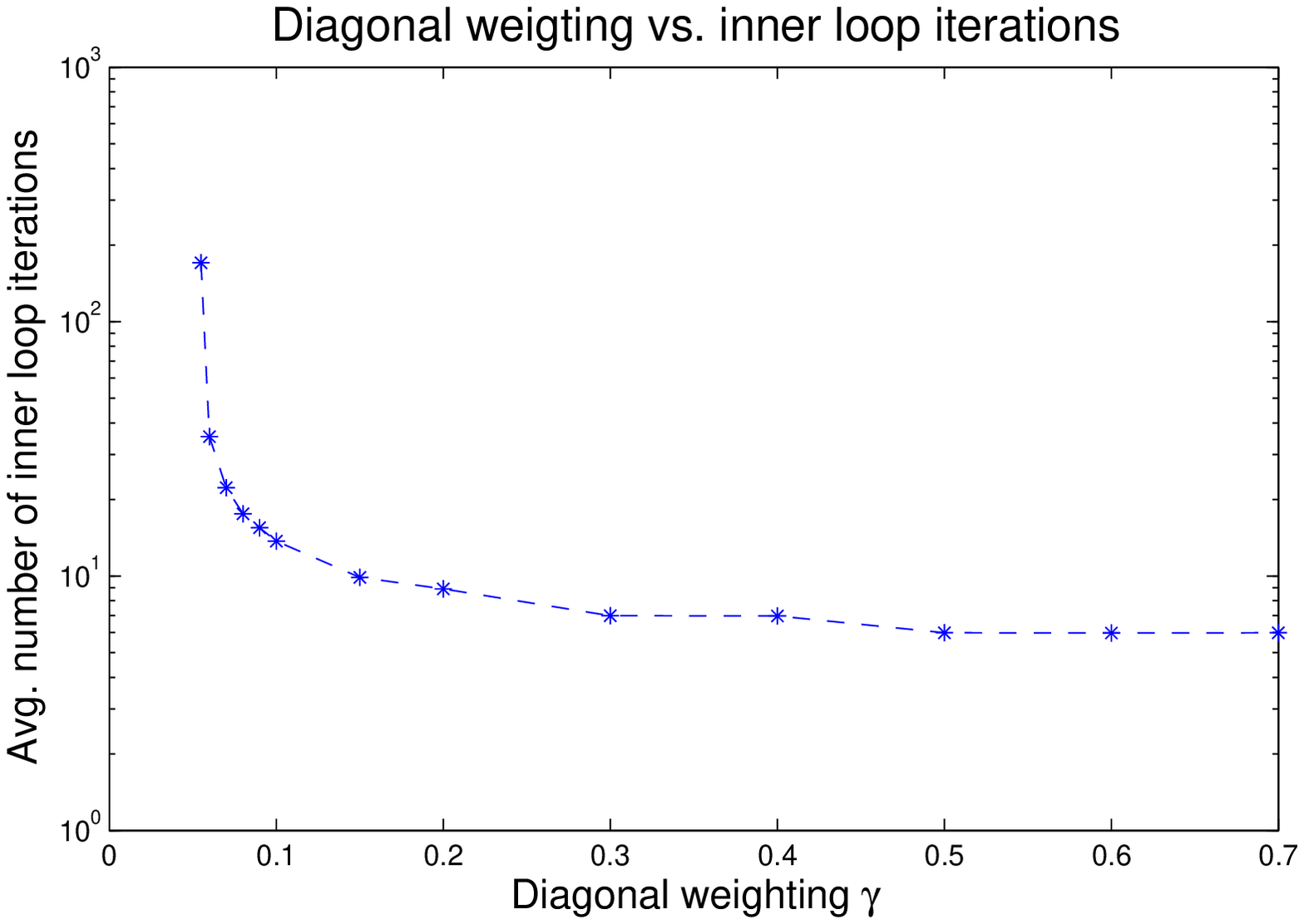}
   \caption{Effect of diagonal weighting on inner loop convergence speed.}\label{fig:inner}
  \label{figD}
\end{minipage}
\end{figure}

\chapter[SVR]{Support Vector Regression}\label{chap8}\label{chap:svm}
\label{SVM classifiers} \label{KRR}

In this chapter, we introduce a distributed support vector regression solver based on
the Gaussian Belief Propagation (GaBP) algorithm. We improve on
the original GaBP algorithm by reducing the communication load, as
represented by the number of messages sent in each optimization
iteration, from $n^2$ to $n$ aggregated messages, where $n$ is the
number of data points. Previously, it was known that the GaBP
algorithm is very efficient for sparse matrices. Using our novel
construction, we demonstrate that the algorithm exhibits very good
performance for dense matrices as well. We also show that the GaBP
algorithm can be used with kernels, thus making the algorithm more
powerful than previously possible.

Using extensive simulation we demonstrate the applicability of our
protocol vs. the state-of-the-art existing parallel SVM solvers.
Using a Linux cluster of up to a hundred machines and the IBM Blue
Gene supercomputer we managed to solve very large data sets up to
hundreds of thousands data point, using up to 1,024 CPUs working
in parallel. Our comparison shows that the proposed algorithm is
just as accurate as these solvers, while being significantly
faster.

We start by giving on overview of the related SVM problem.

\section[SVM Classification]{Classification Using Support Vector Machines} We begin
by formulating the SVM problem. Consider a training set:

\begin{equation}
\begin{array}{l}
D=\left\{\left(\mathbf{x}_i, y_i\right),\;\;\; i=1,\ldots,N,\;\;\;
  \mathbf{x}_i\in \R^m, \;\;\;y_i\in \left\{-1, 1\right\}\right\}.
\end{array}
\end{equation}
The goal of the SVM is to learn a mapping from $\mathbf{x}_i$ to
$y_i$ such that the error in mapping, as measured on a new
dataset, would be minimal. SVMs learn to find the linear weight
vector that separates the two classes so that
\begin{equation}
\begin{array}{l}
y_i \left( \mathbf{x_i} \cdot \mathbf{w} + b \right) \geq 1 \;\;
 for \;\; i = 1,\ldots,N.
\end{array}
\end{equation}

There may exist many hyperplanes that achieve such separation, but
SVMs find a weight vector $\mathbf{w}$ and a bias term $b$ that
maximize the margin $2 / \left\| \mathbf{w} \right\| $. Therefore,
the optimization problem that needs to be solved is
\begin{equation}
\min J_D(\mathbf{w}) = \frac{1}{2}\left\| \mathbf{w} \right\|,
\end{equation}
\begin{equation}
\mbox{subject to \;} y_i \left( \mathbf{x_i} \cdot \mathbf{w} + b
\right) \geq 1 \;\; for \;\; i = 1,\ldots,N.
\end{equation}

Points lying on the hyperplane $y_i \left( \mathbf{x_i} \cdot
\mathbf{w} + b \right) = 1 $ are called support vectors.

If the data cannot be separated using a linear separator, a slack
variable $\xi \geq 0$ is introduced and the constraint is relaxed
to:
\begin{equation}
y_i \left( \mathbf{x_i} \cdot \mathbf{w} + b \right) \geq 1 -
\xi_i \;\; for \;\; i = 1,\ldots,N.
\end{equation}

The optimization problem then becomes:
\begin{equation}
\min J_D(\mathbf{w}) = \tfrac{1}{2}\left\| \mathbf{w} \right\|^2_2 + C
\sum_{i=1}^N \xi_i,
\end{equation}
\begin{equation}
\mbox{subject to \;} y_i \left( \mathbf{x_i} \cdot \mathbf{w} + b
\right) \geq 1 \;\; for \;\; i = 1,\ldots,N,\newline
\end{equation}
\begin{equation}
\xi_i \geq 0  \;\; for \;\; i = 1,\ldots,N.
\end{equation}

The weights of the linear function can be found directly or by
converting the problem into its dual optimization problem, which
is usually easier to solve.

Using the notation of Vijayakumar and Wu~\cite{SVMSeq}, the dual
problem is thus:
\begin{equation}
\label{dual}
    \max \;\; L_D(h)=\sum_{i}h_i-\tfrac{1}{2}h^T D h\,, \\
\end{equation}
\begin{equation}
\label{cons1}
    \mbox{subject \; to \;\;} 0\leq h_i \leq C, \;\; i=1,...,N\,,\\
\end{equation}
\begin{equation}
\label{cons2}
    \; \; \; \;\; \;\;\; \Sigma_i h_i y_i = 0\,,
\end{equation}
where $D$ is a matrix such that $D_{ij} = y_i y_j K
\left(\mathbf{x}_i, \mathbf{x}_j \right) $ and $K \left(\cdot, \;
\cdot\right)$ is either an inner product of the samples or a
function of these samples. In the latter case, this function is
known as the kernel function, which can be any function that
complies with the Mercer conditions~\cite{SS2002}. For example,
these may be polynomial functions, radial-basis (Gaussian)
functions, or hyperbolic tangents. If the data is not separable,
$C$ is a tradeoff between maximizing the margin and reducing the
number of misclassifications.

        The classification of a new data point is then computed using the following equation:
\begin{equation}
         f\left( x \right) = \mathop{sign} \left( \sum_{i\in SV} h_i y_i K \left( x_i, x \right) + b \right)
\end{equation}

\section[KRR Problem]{Kernel Ridge Regression problem} Kernel Ridge Regression
(KRR) implements a regularized form of the least squares method
useful for both regression and classification. The non-linear
version of KRR is similar to  Support-Vector Machine (SVM)
problem. However, in the latter, special emphasis is given to
points close to the decision boundary, which is not provided by
the cost function used by KRR.

Given training data \[ \mathcal{D} = \{\vx_i, y_i\}_{i=1}^{l},
\mbox{   }\vx_i \in R^{d} \mbox{
  }, y_i \in R\,, \] the KRR algorithm determines the parameter
  vector $\vw \in R^d$ of a non-linear model (using the ``kernel trick''),
 via minimization of the following objective function:~\cite{SVM}:
\[ \min \lambda ||\vw||^2 + \sum_{i=1}^l (y_i - \vw^T\Phi(\vx_i))^2\,,\]
where $\lambda$ is a tradeoff parameter between the two terms of
the optimization function, and $\Phi(\dot)$ is a (possible
non-linear) mapping of the training patterns.

One can show that the dual form of this optimization problem is
given by:
\begin{equation}
\label{eq:SVR} \max W(\alpha) = \vy^T\alpha + \tfrac{1}{4}\lambda \alpha^T \mK
\alpha -
 \tfrac{1}{4} \alpha^T\alpha\,,
\end{equation}
where $\mK$ is a matrix whose $(i,j)$-th entry is the kernel
function $\mK_{i,j}= \Phi(\vx_i)^T \Phi(\vx_j)$.

The optimal solution to this optimization problem is: \[ \alpha =
2\lambda(\mK+\lambda \mI)^{-1}\vy
\]

The corresponding prediction function is given by:
\[ f(\vx) = \vw^T \Phi(\vx) = \vy^T(\mK+\lambda \mI)^{-1}\mK(\vx_i,\vx). \]

\section[Previous Approaches]{Previous Approaches for Solving Parallel SVMs}
        There are several main methods for finding a solution to an SVM problem
        on a single-node computer. (See \cite[Chapter 10]{SS2002}) for a taxonomy
        of such methods.) However, since solving an SVM is quadratic in time and
        cubic in memory, these methods encounter difficulty when scaling to datasets
        that have many examples and support vectors. The latter two are not synonymous.
        A large dataset with many repeated examples might be solved using sub-sampling
        approaches, while a highly non-separable dataset with many support vectors will
         require an altogether different solution strategy.
        The literature covers several attempts at solving SVMs in parallel,
        which allow for greater computational power and larger memory size. In Collobert et al.~\cite{CBB2002}
        the SVM solver is parallelized by training multiple SVMs, each on a subset
        of the training data, and aggregating the resulting classifiers into a single
        classifier. The training data is then redistributed to the classifiers according
         to their performance and the process is iterated until convergence is reached.
         The need to re-divide the data among the SVM classifiers implies that the data
         must be exchanged between nodes several times; this rules out the use of an
         approach where bandwidth is a concern.
        A more low-level approach is taken by Zanghirati et al.~\cite{ZZ2003}, where the quadratic
         optimization problem is divided into smaller quadratic programs, each of which is solved on a different node. The results
         are aggregated and the process is repeated until convergence. The performance
         of this method has a strong dependence on the caching architecture of the cluster.
        Graf et al.~\cite{GCBD2004} partition the data and solve an SVM for each partition. The support
        vectors from each pair of classifiers are then aggregated into a new training set
        for which an SVM is solved. The process continues until a single classifier remains.
         The aggregation process can be iterated, using the support vectors of the final
         classifier in the previous iteration to seed the new classifiers. One problem
         with this approach is that the data must be repeatedly shared between nodes,
         meaning that once again the goal of data distribution cannot be attained. The
         second problem, which might be more severe, is that the number of possible
         support vectors is restricted by the capacity of a single SVM solver.
         Yom Tov~\cite{SVMSeqY} proposed modifying the sequential algorithm
        developed in~\cite{SVMSeq} to batch mode. In this way, the complete
        kernel matrix is held in distributed memory and the Lagrange multipliers
         are computed iteratively. This method has the advantage that it can
         efficiently solve difficult SVM problems that have many
          support vectors to their solution. Based on that work, we show how an SVM solution can be obtained by adapting a Gaussian Belief
          Propagation algorithm to the solution of the algorithm proposed in~\cite{SVMSeq}.

Recently, Hazan \etal proposed an iterative algorithm for parallel
decomposition based on Fenchel Duality~\cite{Hazan}. Zanni \etal
propose a decomposition method for computing SVM in
parallel~\cite{Zanni}. We compare our running time results to both
systems in Section~\ref{exp_results}.

For our proposed solution, we take the exponent of dual SVM
formulation given in equation (\ref{dual}) and solve $\max
\exp(L_D( h))$. Since $\exp(L_D( h))$ is convex, the solution of
$\max \exp(L_D( h))$ is a global maximum that also satisfies $\max
L_D(h)$ since the matrix $D$ is symmetric and positive definite.
Now we can relate to the new problem formulation as a probability
density function, which is in itself Gaussian:
\[ p(h) \propto \exp(-\tfrac{1}{2}h^TDh + h^T\mathbf{1}), \] where $\mathbf{1}$ is a
vector of $(1,1,\cdots,1),$ and find the assignment of $\hat{h} =
\arg \max p(h)$. To solve the inference problem, namely computing the
marginals $\hat{h}$, we propose using the GaBP algorithm, which is
a distributed message passing algorithm. We take the computed
$\hat{h}$ as the Lagrange multiplier weights of the support
vectors of the original SVM data points and apply a threshold for
choosing data points with non-zero weight as support vectors.

Note that using this formulation we ignore the remaining
constraints~\eqref{cons1},~\eqref{cons2}. In other words we do not
solve the SVM problem, but the unconstrained kernel ridge regression problem \eqref{eq:SVR}.
Nevertheless, empirical results presented in
Chapter~\ref{emp_results} show that we achieve a very good
classification vs. state-of-the-art SVM solvers.

Finally, following \cite{SVMSeq}, we remove the explicit bias term
$b$ and instead add another dimension to the pattern vector
$\mathbf{x}_i$ such that $\mathbf{\hat{x_i}} = \left(x_1, x_2,
\ldots, x_N, \lambda \right)$, where $\lambda$ is a scalar
constant. The modified weight vector, which incorporates the bias
term, is written as $\mathbf{\hat{w}} = \left( w_1, w_2, \ldots,
w_N,b/\lambda \right)$. However, this modification causes a change
to the optimized margin. Vijayakumar and Wu \cite{SVMSeq} discuss
the effect of this modification and reach the conclusion that
``setting the augmenting term to zero (equivalent to neglecting
the bias term) in high dimensional kernels gives satisfactory
results on real world data''. We did not completely neglect the
bias term and in our experiments, which used the Radial Basis
Kernel, we set it to $1/N$, as proposed in \cite{SVMSeqY}.

\section{Our novel construction} \label{diagonal}
We propose to use the GaBP algorithm for solving the SVR problem \eqref{eq:SVR}. In order to
force the algorithm to converge, we artificially weight the main
diagonal of the kernel matrix $D$ to make it diagonally dominant.
Section~\ref{experimental} outlines our empirical results showing
that this modification did not significantly affect the error in
classifications on all tested data sets.

A partial justification for weighting the main diagonal is found
in~\cite{SVM}. In the 2-Norm soft margin formulation of the SVM
problem, the sum of squared slack variables is minimized:
\[ \min_{\xi,w,b} \|\mathbf{w} \|_2^2 + C \Sigma_i \mathbf{\xi}_i^2 \]
\[ \mbox{such that \ \ \      } y_i(\mathbf{w} \cdot \mathbf{x}_i + b) \ge 1 - \xi_i \]
The dual problem is derived:
\[ W(h) = \Sigma_{i} h_i - \frac{1}{2}\Sigma_{i,j} y_i y_j h_i h_j(\mathbf{x}_i \cdot \mathbf{x}_j +
                                                                      \tfrac{1}{C} \delta_{ij}), \]
where $\delta_{ij}$ is the Kronecker $\delta$ defined to be 1 when
$i=j$, and zero elsewhere. It is shown that the only change
relative to the 1-Norm soft margin SVM is the addition of $1/C$ to
the diagonal of the inner product matrix associated with the
training set. This has the effect of adding $1/C$ to the
eigenvalues, rendering the kernel matrix (and thus the GaBP
problem) better conditioned~\cite{SVM}.

One of the desired
properties of a large scale algorithm is that it should converge
in asynchronous settings as well as in synchronous settings. This
is because in a large-scale communication network, clocks are not
synchronized accurately and some nodes may be slower than others,
while some nodes experience longer communication delays.

\begin{corol}
Assuming one of the convergence conditions (Theorems~\ref{dd_theorem_weiss},~\ref{spectral_radius_thm}) holds, the GaBP algorithm convergence using serial (asynchronous) scheduling as well.
\end{corol}
\begin{proof}
The quadratic Min-Sum algorithm~\cite{MinSum} provides a convergence proof in the asynchronous
case. In Chapter \ref{MinSum} we show equivalence of both algorithms. Thus, assuming one of the convergence
conditions holds, the GaBP algorithm converges using serial scheduling as well.
\end{proof}


\section{Experimental Results}
\label{emp_results} \label{exp_results} \label{experimental} We
implemented our proposed algorithm using approximately 1,000 lines
of code in C. We implemented communication between the nodes using
the MPICH2 message passing interface.\footnote{\url{http://www-unix.mcs.anl.gov/mpi/mpich/}} Each node
was responsible for $d$ data points out of the total $n$ data
points in the dataset.

Our implementation used synchronous communication rounds because
of MPI limitations. In Section~\ref{discussion} we further
elaborate on this issue.

Each node was assigned several examples from the input file. Then,
the kernel matrix $D$ was computed by the nodes in a distributed
fashion, so that each node computed the rows of the kernel matrix
related to its assigned data points. After computing the relevant
parts of the matrix $D$, the nodes weighted the diagonal of the
matrix $D$, as discussed in Section~\ref{diagonal}. Then, several
rounds of communication between the nodes were executed. In each round,
using our optimization, a total of $n$ sums were calculated using
MPI\_Allreduce system call. Finally, each node output the solution
$x$, which was the mean of the input Gaussian that matched its own
data points. Each $x_i$ signified the weight of the data point $i$
for being chosen as a support vector.

To compare our algorithm performance, we used two algorithms:
Sequential SVM (SVMSeq) ~\cite{SVMSeq} and SVMlight
~\cite{SVMlight}. We used the SVMSeq implementation provided
within the IBM Parallel Machine Learning (PML) toolbox~\cite{IBM}.
The PML implements the same algorithm by Vijaykumar and Wu
\cite{SVMSeq} that our GaBP solver is based on, but the
implementation in through a master-slave architecture as described
in \cite{SVMSeqY}. SVMlight is a single computing node solver.

\begin{table}[t]
\begin{center}
\begin{tabular}{|l|l|l|l|c|c|c|}
\hline
Dataset                & Dimension & Train & Test & \multicolumn{3}{|c||}{\textsc{Error (\%)}} \\
\cline{5-7}
                         & & &                    & GaBP & Sequential    & SVMlight    \\
\hline
\texttt{Isolet}        & 617 & 6238 & 1559 & 7.06 & {\bf 5.84} & 49.97 \\
\texttt{Letter}        & 16 & 20000 &  & {\bf 2.06} & {\bf 2.06} & 2.3 \\
\texttt{Mushroom}      & 117 & 8124 & & 0.04 & 0.05 & {\bf 0.02}  \\
\texttt{Nursery}       & 25 & 12960 & & 4.16 & 5.29  & {\bf 0.02}  \\
\texttt{Pageblocks}   & 10 & 5473 & & 3.86 & 4.08  & {\bf 2.74}  \\
\texttt{Pen digits}    & 16 & 7494& 3498 & 1.66 & {\bf 1.37} & 1.57 \\
\texttt{Spambase}      & 57 & 4601 & & 16.3 & 16.5   & {\bf 6.57} \\
\hline
\end{tabular}
\caption{Error rates of the GaBP solver versus those of the
parallel sequential solver and SVMlight.} \label{err_run}
\end{center}
\end{table}

\begin{table}[t]
\begin{center}
\begin{tabular}{|l||c|c|}
\hline
Dataset                & \multicolumn{2}{|c|}{\textsc{Run times (sec)}}     \\
\cline{1-3}
                       & GaBP & Sequential     \\
\hline
\texttt{Isolet}        & 228     &  1328                  \\
\texttt{Letter}        & 468      &  601                    \\
\texttt{Mushroom}      & 226     &  176                   \\
\texttt{Nursery}       & 221     &  297                   \\
\texttt{Pageblocks}    & 26      &  37                    \\
\texttt{Pen digits}    & 45     &  155                   \\
\texttt{Spambase}      & 49      &  79                    \\
\hline
\end{tabular}
\caption{Running times (in seconds) of the GaBP solver (working in
a distributed environment) compared to that of the IBM parallel
solver.} \label{time_run}
\end{center}
\end{table}

Table \ref{err_run} describes the seven datasets we used to
compare the algorithms and the classification accuracy obtained.
These computations were done using five processing nodes (3.5GHz
Intel Pentium machines, running the Linux operating system) for
each of the parallel solvers. All datasets were taken from the UCI
repository~\cite{UCI}. We used medium-sized datasets so that
run-times using SVMlight would not be prohibitively high. All
algorithms were run with an RBF kernel. The parameters of the
algorithm (kernel width and misclassification cost) were optimized
using
 a line-search algorithm, as detailed in~\cite{RifkinK04}.

Note that SVMlight is a single node solver, which we use mainly as
a comparison for the accuracy in classification.

Using the Friedman test \cite{Demsar06}, we did not detect any
statistically significant difference between the performance of
the algorithms with regards
 to accuracy ($p<0.10^{-3}$).

\begin{figure}[t]
\begin{center}
  \includegraphics[width=340pt,clip,scale=0.35]{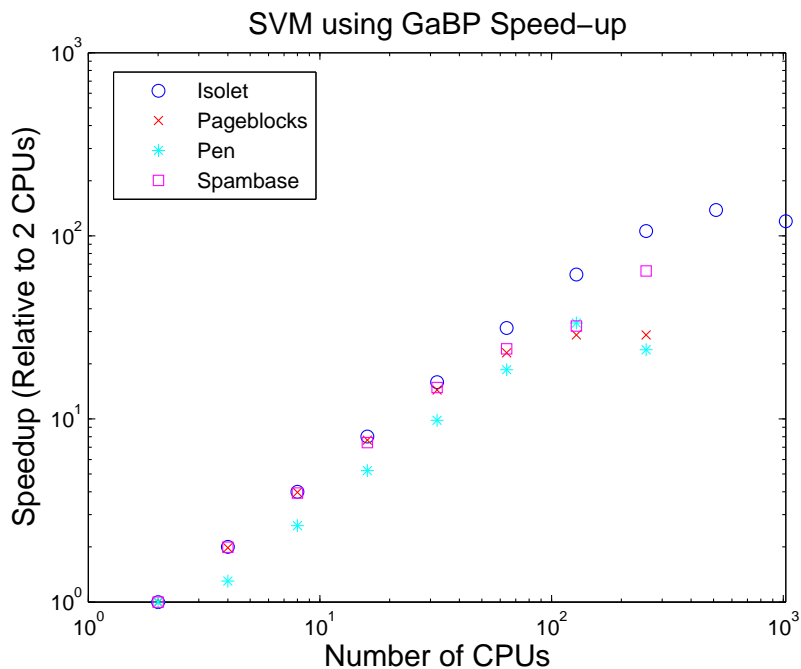}\\
  \caption{Speedup of the GaBP algorithm vs. 2 CPUS.}\label{speedup}
\end{center}
\end{figure}

Figure~\ref{speedup} shows the speedup results of the algorithm
when running the GaBP algorithm on a Blue Gene supercomputer. The
speedup with $N$ nodes is computed as the run time of the
algorithm on a single node, divided by the run time using $N$
nodes. Obviously, it is desirable to obtain linear speedup, i.e.,
doubling computational power halves the processing time, but this
is limited by the communication load and by parts of the algorithm
that cannot be parallelized. Since Blue Gene is currently limited
to 0.5 GB of memory at each node, most datasets could not be executed
on a single node. We therefore show speedup compared to two nodes.
As the figure shows, in most cases we get a linear speedup up to
256 CPUs, which means that the running time is linearly
proportional to one over the number of used CPUs. When using 512 -
1024 CPUs, the communication overhead reduces the efficiency of
the parallel computation. We identified this problem as an area
for future research into optimizing the performance for larger
scale grids.

We also tested the ability to build classifiers for larger
datasets. Table \ref{bg_times} shows the run times of the GaBP
algorithm using 1024 CPUs on two larger datasets, both from the
UCI repository. This demonstrates the ability of the algorithm to
process very large datasets in a reasonably short amount of time.
We compare our running time to state-of-the-art parallel
decomposition method by Zanni \etal ~\cite{Zanni} and Hazan \etal.
Using the MNIST dataset we where considerably slower by a factor
of two, but in the larger Covertype dataset we have a superior
performance. Note that the reported running times should be taken with a grain of salt, since
the machines used for experimentation are different. Zanni used 16
Pentium IV machines with 16Gb memory, Hazan used 10 Pentium IV
machines with 4Gb memory, while we used a larger number of weaker
Pentium III machines with 400Mb of memory. Furthermore, in the
Covertype dataset we used only 150,000 data points while Zanni and
Hazan used the full dataset which is twice larger.

\begin{table}[t]
\begin{center}
\begin{tabular}{|l|l|l|c|c|c|}
\hline
Dataset                   & Dim       & Num of examples    & Run time GaBP (sec) & Run time ~\cite{Zanni} (sec) & Run time~\cite{Hazan}\\
\hline
\texttt{Covertype}        & 54                      & 150,000/300,000        & {\bf 468}   & 24365 & 16742\\
\texttt{MNIST}            & 784                     & 60,000         & 756   & 359 &  {\bf 18}\\
\hline
\end{tabular}
\caption{Running times of the GaBP solver for large data sets
using 1024 CPUs on an IBM Blue Gene supercomputer. Running time
results are compared to two state-of-the-art solvers:
~\cite{Zanni} and~\cite{Hazan}.} \label{bg_times}
\end{center}
\end{table}

\section{Discussion}
\label{discussion}

In this chapter we demonstrated the application of the Gaussian
Belief Propagation to the solution of SVM problems. Our
experiments demonstrate the usefulness of this solver, being both
accurate and scalable.

We implemented our algorithm using a synchronous communication
model mainly because MPICH2 does not support asynchronous
communication. While synchronous communication is the mode of
choice for supercomputers such as Blue Gene, in many cases such as
heterogeneous grid environments, asynchronous communication will
be preferred. We believe that the next challenging goal will be to
implement the proposed algorithm in asynchronous settings, where
algorithm rounds will no longer be synchronized.

Our initial experiments with very large sparse kernel matrices
(millions of data points) show that asynchronous settings converge
faster. Recent work by Elidan \etal ~\cite{RBP} supports this claim by
showing that in many cases the BP algorithm converges faster in
asynchronous settings.

Another challenging task would involve scaling to data sets of
millions of data points. Currently the full kernel matrix is
computed by the nodes. While this is effective for problems with
many support vectors \cite{SVMSeqY}, it is not required in many
problems that are either easily separable or  where the
classification error is less important compared to the time
required to learn the mode. Thus, solvers scaling to much larger
datasets may have to diverge from the current strategy of
computing the full kernel matrix and instead sparsify the kernel
matrix as is commonly done in single node solvers.

Finally, it remains an open question whether SVMs can be solved
efficiently in Peer-to-Peer environments, where each node can
(efficiently) obtain data from only several close peers. Future
work will be required in order to verify how the GaBP algorithm
performs in such an environment, where only partial segments of
the kernel matrix can be computed by each node.

\chapter[Kalman Filter]{Distributed Computation of Kalman Filter}

\label{chap:Kalman}
In Chapter \ref{sec_linear} we show how to compute efficiently and
distributively the MMSE prediction for the multiuser detection
problem, using the Gaussian Belief Propagation (GaBP) algorithm.
The basic idea is to shift the problem from the linear algebra domain
into a probabilistic graphical model, solving an equivalent
inference problem using the efficient belief propagation inference
engine. 

In this chapter, we propose to extend the previous
construction, and show, that by performing the MMSE computation
twice on the matching inputs we are able to compute several
algorithms: Kalman filter, Gaussian information bottleneck and the Affine-scaling interior point method. We reduce the discrete Kalman filter
computation \cite{Kalman} to a matrix inversion problem and show
how to solve it using the GaBP algorithm. We show that Kalman
filter iteration that is composed from prediction and measurement
steps can be computed by two consecutive MMSE predictions. We explore the relation to Gaussian information bottleneck (GIB)
\cite{GIB} and show that Kalman filter is a special instance of
the GIB algorithm, when the weight parameter $\beta = 1$. To the
best of our knowledge, this is the first algorithmic link
between the information bottleneck framework and linear dynamical
systems. We discuss the connection to the Affine-scaling
interior-point method and show it is an instance of the Kalman
filter.

Besides of the theoretical interest of linking compression,
estimation and optimization together, our work is highly practical,
since it proposes a general framework for computing all of the
above tasks distributively in a computer network. This result can have many
applications in the fields of estimation, collaborative signal
processing, distributed resource allocation, etc.

A closely related work is \cite{BibDB:FactorGraph}.
In this work, Frey \etal focus on the belief propagation algorithm
(a.k.a sum-product algorithm) using factor graph topologies. They
show that the Kalman filter algorithm can be computed using belief
propagation over a factor graph. In this contribution we extend
their work in several directions. First, we extend the computation
to vector variables (relative to scalar variables in Frey's work).
Second, we use a different graphical model: an undirected
graphical model, which results in simpler update rules, where Frey
uses factor-graph with two types of messages: factor to variables
and variables to factors. Third and most important, we allow an
efficient distributed calculation of the Kalman filter steps,
where Frey's algorithm is centralized.

Another related work is \cite{KalmanKarmarkar}. In this work the
link between Kalman filter and linear programming is established.
In this thesis we propose a new and different construction that
ties the two algorithms.

\section{Kalman Filter}
\label{sec_kf}
 The Kalman filter is an efficient iterative
algorithm to estimate the state of a discrete-time controlled
process $x \in R^n$ that is governed by the linear stochastic
difference equation\footnote{We assume there is no
external input, namely $x_k = Ax_{k-1} + w_{k-1}$. However, our
approach can be easily extended to support external inputs.}: \BE
x_k = Ax_{k-1} + Bu_{k-1} + w_{k-1}\,, \EE \
with a measurement $z \in R^m$ that is $ z_k = Hx_k + v_k.$ The
random variables $w_k$ and $v_k$ that represent the process and
measurement AWGN noise (respectively). $p(w) \sim \mathcal{N}(0,
Q), p(v) \sim \mathcal{N}(0, R)$. We further assume that the
matrices $A,H,B,Q,R$ are given.\footnote{Another possible extension
is that the matrices $A,H,B,Q,R$ change in time, in this thesis we
assume they are fixed. However, our approach can be generalized to
this case as well.}

The discrete Kalman filter update equations are given
by~\cite{Kalman}:

The prediction step:
\begin{subequations}
\begin{eqnarray}
 \hat{x}^-_k &=& A\hat{x}_{k-1} + Bu_{k-1}, \label{hat_x_minus_k}\\
 P_k^- &=& AP_{k-1}A^T + Q. \label{p_k_minus}
\end{eqnarray}
\end{subequations}

The measurement step:
\begin{subequations}
\begin{eqnarray}
 K_k &=& P_k^-H^T(HP_k^-H^T + R)^{-1}, \label{kalman_gain} \\
\hat{x}_k &=& \hat{x}^-_k + K_k(z_k - H\hat{x}^-_k),
\label{hat_x_k} \\
 P_k &=& (I-K_kH)P_k^-. \label{p_k}
\end{eqnarray}
\end{subequations}
where $I$ is the identity matrix.

The algorithm operates in rounds. In round $k$ the estimates
$K_k,\hat{x}_k,P_k$ are computed, incorporating the (noisy)
measurement $z_k$ obtained in this round. The output of the
algorithm are the mean vector $\hat{x}_k$ and the covariance
matrix $P_k$.

\section{Our Construction} \label{new-cons} \label{kalman_solver}Our novel
contribution is a new efficient distributed algorithm for
computing the Kalman filter. We begin by showing that the Kalman
filter can be computed by inverting the following covariance
matrix: \BE E =
\left(%
\begin{array}{ccc}
  -P_{k-1} & A & 0 \\
  A^T & Q & H \\
  0 & H^T & R \\
\end{array}%
\right)\,, \label{mat_E} \EE and taking the lower right $1 \times 1$
block to be $P_k$.

The computation of $E^{-1}$ can be done efficiently using recent
advances in the field of Gaussian belief propagation
\cite{ISIT1,ISIT2}. The intuition for our approach, is that the
Kalman filter is composed of two steps. In the prediction step,
given $x_k$, we compute the MMSE prediction of $x_k^-$
\cite{BibDB:FactorGraph}. In the measurement step, we compute the
MMSE prediction of $x_{k+1}$ given $x_k^-$, the output of the
prediction step. Each MMSE computation can be done using the GaBP
algorithm \cite{ISIT2}. The basic idea is that given the joint
Gaussian distribution $p(\vx,\vy)$ with the covariance matrix $
C = \left(%
\begin{array}{cc}
  \Sigma_{xx} & \Sigma_{xy} \\
  \Sigma_{yx} & \Sigma_{yy} \\
\end{array}%
\right)$, we can compute the MMSE prediction \[ \hat{y} =
\argmax_{y}p(y|x) \propto \mathcal{N}(\mu_{y|x},
\Sigma_{y|x}^{-1})\,,
\] where \[ \mu_{y|x} = (\Sigma_{yy} -
\Sigma_{yx}\Sigma_{xx}^{-1}\Sigma_{xy})^{-1} \Sigma_{yx} \Sigma_{xx}^{-1}x\,, \]
\[ \Sigma_{y|x} = (\Sigma_{yy} -
\Sigma_{yx}\Sigma_{xx}^{-1}\Sigma_{xy})^{-1}\,. \] This in turn is
equivalent to computing the Schur complement of the lower right
block of the matrix $C$. In total, computing the MMSE prediction
in Gaussian graphical model boils down to a computation of a
matrix inverse. In \cite{ISIT1} we have shown that GaBP is an
efficient iterative algorithm for solving a system of linear
equations (or equivalently computing a matrix inverse). In
\cite{ISIT2} we have shown that for the specific case of linear
detection we can compute the MMSE estimator using the GaBP
algorithm. Next, we show that performing two consecutive
computations of the MMSE are equivalent to one iteration of the
Kalman filter.

\begin{thm} The lower right $1 \times 1$ block of the matrix inverse $E^{-1}$
(eq. \ref{mat_E}), computed by two MMSE iterations, is equivalent to the computation of $P_k$ done
by one iteration of the Kalman filter algorithm.\\
\end{thm}
\begin{proof}
We prove that inverting the matrix $E$ (eq. \ref{mat_E}) is
equivalent to one iteration of the Kalman filter for computing
$P_k$.

We start from the matrix $E$ and show that $P_k^-$ can be computed
in recursion using the Schur complement formula: \BE D - CA^{-1}B
\label{Schur} \EE applied to the $2 \times 2$ upper left submatrix
of E, where $D \triangleq Q, C \triangleq A^T, B \triangleq A, A
\triangleq P_{k-1}$, we get:
\[ P_{k}^- =
\overbrace{Q}^{D} \overbrace{+}^{-}
\overbrace{A^T}^{C} \overbrace{P_{k-1}}^{-A^{-1}}
\overbrace{A}^{B}. \]

Now we compute recursively the Schur complement of lower right $2
\times 2$ submatrix of the matrix $E$ using the matrix inversion
lemma: \[ A^{-1} + A^{-1}B (D-CA^{-1}B)^{-1} CA^{-1} \]
where $A^{-1}
\triangleq P_k^-, B \triangleq H^T, C \triangleq H, D \triangleq Q
.$
In total we get:
\small
\BE \overbrace{P_{k}^-}^{A^{-1}} +
\overbrace{P_{k}^-}^{A^{-1}}\overbrace{H^T}^{B} (\overbrace{R}^{D}
+ \overbrace{H}^{C}\overbrace{P_{k}^-}^{A^{-1}}\overbrace{H^T}^{B})^{-1}
\overbrace{H}^{C} \overbrace{P_{k}^-}^{A^{-1}} = \label{kalman_it}
\EE
\normalsize
\[ (I - \overbrace{P_k^-H^T(HP_k^-H^T + R)^{-1}}^{(\ref{kalman_gain})} H)P_k^-  = \overbrace{(I - K_kH)P_k^-}^{(\ref{p_k})} = P_k \]
\end{proof}

In Section \ref{kalman_solver} we explain how to utilize the above
observation to an efficient distributed iterative algorithm for
computing the Kalman filter.

\section{Gaussian Information Bottleneck}
\label{sec_GIB} Given the joint distribution of a source variable
X and another relevance variable Y, Information bottleneck (IB)
operates to compress X, while preserving information about
Y~\cite{InfoBottleneck,Slonim}, using the following variational
problem:

\[ \min_{p(t|x)} \mathcal{L} : \mathcal{L} \equiv I(X; T) - \beta I(T; Y ) \]
$T$ represents the compressed representation of $X$ via the
conditional distributions $p(t|x)$, while the information that $T$
maintains on $Y$ is captured by the distribution $p(y|t)$. $\beta
> 0 $ is a lagrange multiplier that weights the tradeoff between
minimizing the compression information and maximizing the relevant
information. As $\beta \rightarrow 0$ we are interested solely in
compression, but all relevant information about Y is lost $I(Y;T)
= 0$. When $\beta \rightarrow \infty$ we are focused on
preservation of relevant information, in this case T is simply the
distribution X and we obtain $I(T;Y) = I(X;Y)$. The interesting
cases are in between, when for finite values of $\beta$ we are
able to extract rather compressed representation of X while still
maintaining a significant fraction of the original information
about Y.

An iterative algorithm for solving the IB problem is given in
\cite{Slonim}:

\begin{subequations}
\label{IIB}
\begin{eqnarray}
P^{k+1}(t|x) =&\frac{P^k(t)}{Z^{k+1}(x,\beta)}  \nonumber 
 \exp(-\beta D_{KL}[p(y|x)||p^k(y|t)])\,, \label{IB_it1} \\
P^k(t)=& \int_x p(x)P^k(t|x)dx\,, \label{IB_it2} \\
P^k(y|t)=& \frac{1}{P^k(t)} \int_x P^k(t|x)p(x,y)dx\,\,, \label{IB_it3}
\end{eqnarray}
 \end{subequations}
where $Z^{k+1}$ is a normalization factor computed in round $k+1$.

The Gaussian information bottleneck (GIB) \cite{GIB} deals with
the special case where the underlying distributions are Gaussian.
In this case, the computed distribution $p(t)$ is Gaussian as
well, represented by a linear transformation $T_k = A_kX + \xi_k$
where $A_k$ is a joint covariance matrix of $X$ and $T$, $\xi_k \sim
\mathcal{N}(0, \Sigma_{\xi_k})$ is a multivariate Gaussian independent of X.
The outputs of the algorithm are the covariance matrices
representing the linear transformation T: $A_k, \Sigma_{\xi_k}$.

An iterative algorithm is derived by substituting Gaussian
distributions into (\ref{IIB}), resulting in the following
update rules: 
 \begin{subequations}
    \begin{eqnarray}
   \Sigma_{\xi+1}& = &(\beta\Sigma_{t_k|y} - (\beta - 1)\Sigma_{t_k}^{-1}) \label{GIB_xi}, \\
 A_{k+1} &=& \beta\Sigma_{\xi_k+1}\Sigma^{-1}_{t_k|y}A_k(I -
 \Sigma_{y|x}\Sigma_x^{-1}). \label{GIB_A}
    \end{eqnarray}
 \end{subequations}

\begin{table}[h!]
\begin{center}
 \caption{Summary of notations in the GIB
\cite{GIB} paper vs. Kalman filter \cite{Kalman}}
\begin{tabular}{|c|c|l|}
\hline \label{notations}
  GIB \cite{GIB} & Kalman \cite{Kalman} &  Kalman meaning \\
  \hline
  $\Sigma_x$ & $P_0$ & a-priori estimate error covariance \\
  $\Sigma_y$ & $Q$ & process AWGN noise\\
  $\Sigma_{t_k}$ & $R$ & measurement AWGN noise\\
  $\Sigma_{xy}$ & $A$ & process state transformation matrix\\
  $\Sigma_{yx}$ & $A^T$ & -"- \\
  $\Sigma_{xy}A$ & $H^T$ & measurement transformation matrix\\
  $A^T\Sigma_{yx}$ & $H$ & -"- \\
  $\Sigma_{\xi_k}$ & $P_k$ &  posterior error covariance in round k\\
  $\Sigma_{x|y_k}$ & $P_k^-$ & a-priori error covariance in round k\\ \hline
\end{tabular}
\end{center}
\end{table}

\begin{figure}
\begin{center}
  \includegraphics[scale=0.33]{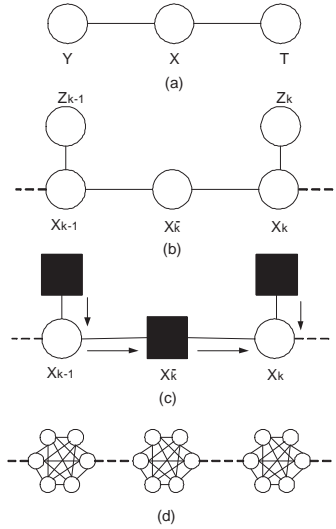}\\
  \caption{Comparison of the different graphical models used. (a) Gaussian Information Bottleneck \cite{GIB}
  (b) Kalman Filter (c) Frey's sum-product factor graph \cite{BibDB:FactorGraph} (d) Our new construction.}\label{graphical_model}
\end{center}
\end{figure}

Since the underlying graphical model of both algorithms (GIB and
Kalman filter) is Markovian with Gaussian probabilities, it is
interesting to ask what is the relation between them. In this work
we show, that the Kalman filter posterior error covariance
computation is a special case of the GIB algorithm when $\beta =
1$. Furthermore, we show how to compute GIB using the Kalman
filter when $\beta > 1$ (the case where $0 < \beta < 1$ is not
interesting since it gives a degenerate solution where $A_k \equiv
0$ \cite{GIB}.) Table \ref{notations} outlines the different
notations used by both algorithms.

\begin{thm}
The GIB algorithm when $\beta= 1$ is equivalent to the Kalman
filter algorithm.
\end{thm}
\begin{proof}
Looking at \cite[$\S 39$]{GIB}, when $\beta = 1$ we get
\[
\begin{array}{l}
  \Sigma_{\xi+1} = (\Sigma_{t_k|y}^{-1})^{-1} =
 \Sigma_{t_k|y} = 
 \overbrace{\Sigma_{t_k}- \Sigma_{t_ky} \Sigma_y^{-1} \Sigma_{yt_k}}^{\mbox{MMSE}} = 
 \overbrace{\Sigma_{t_k} + B^T
\Sigma_{y|t_k} B}^{\mbox{\cite[\S 38b]{GIB}}} = \\
 \Sigma_{t_k} + \overbrace{\Sigma_{t_k}^{-1} \Sigma_{t_ky} }^{\mbox{\cite[\S
34]{GIB}}}\Sigma_{y|t_k} \overbrace{\Sigma_{yt_k}
\Sigma_{t_k}^{-1}}^{\mbox{\cite[\S 34]{GIB}}} = 
  \overbrace{A^T \Sigma_x A + \Sigma_{\xi}}^{\mbox{\cite[\S 33]{GIB}}}
+ \overbrace{(A^T \Sigma_x A + \Sigma_{\xi})}^{\mbox{\cite[\S
33]{GIB}}} A^T \Sigma_{xy} \cdot \\
\cdot \Sigma_{y|t_k} \Sigma_{yx} A\overbrace{(A^T
\Sigma_xA + \Sigma_{\xi})^T}^{\mbox{\cite[\S 33]{GIB}}}  = 
  A^T \Sigma_x A + \Sigma_{\xi} + (A^T \Sigma_x A + \Sigma_{\xi}) A^T \Sigma_{xy} \cdot \\
\cdot \overbrace{(\Sigma_y + \Sigma_{yt_k} \Sigma_{t_k}^{-1}
\Sigma_{t_ky})}^{\mbox{MMSE}}
\Sigma_{yx} A (A^T \Sigma_x A + \Sigma_{\xi})^T  = 
A^T \Sigma_x A + \Sigma_{\xi}
+ (A^T \Sigma_x A + \Sigma_{\xi}) A^T \Sigma_{xy} \cdot \\
(\Sigma_y + \overbrace{A^T \Sigma_{yx}}^{\mbox{\cite[\S 5]{GIB}}}
\overbrace{(A\Sigma_xA^T + \Sigma_{\xi})}^{\mbox{(\cite[\S 5]{GIB}}}
\overbrace{\Sigma_{xy}A}^{\mbox{\cite[\S 5]{GIB}}}) \Sigma_{yx} A (A^T \Sigma_x A + \Sigma_{\xi})^T. \\
\end{array}
\]

Now we show this formulation is equivalent to the Kalman filter
with the following notations: \[  P_k^- \triangleq (A^T \Sigma_x A
+ \Sigma_{\xi}) \ \ , H \triangleq A^T \Sigma_{yx},\ \   R
\triangleq \Sigma_y, P_{k-1} \triangleq \Sigma_x,
 Q \triangleq \Sigma_{\xi}. \]
 Substituting we get:
\[
\begin{array}{l}
 \overbrace{(A^T \Sigma_x A + \Sigma_{\xi})}^{P_k^-} +
\overbrace{(A^T \Sigma_x A + \Sigma_{\xi})}^{P_k^-}
\overbrace{A^T \Sigma_{xy}}^{H^T} 
  \cdot (\overbrace{\Sigma_y}^{R} + \overbrace{A^T
\Sigma_{yx}}^{H}\overbrace{(A^T \Sigma_x A +
\Sigma_{\xi})}^{P_k^-}\overbrace{\Sigma_{xy}A}^{H^T}) \overbrace{\Sigma_{yx}A}^{H}\overbrace{(A^T \Sigma_x A +
\Sigma_{\xi})}^{P_k^-}. \\
\end{array}
\]
Which is equivalent to (\ref{kalman_it}). Now we can apply Theorem
1 and get the desired result.
\end{proof}
\begin{thm}
The GIB algorithm when $\beta > 1$ can be computed by a modified
Kalman filter iteration.
\end{thm}
\begin{proof}
In the case where $\beta > 1$, the MAP covariance matrix as
computed by the GIB algorithm is: \BE \Sigma_{\xi_{k+1}} = \beta
\Sigma_{t_k|y} + (1 - \beta) \Sigma_{t_k}
\label{weighted-GIB-kalman} \EE This is a weighted average of two
covariance matrices. $\Sigma_{t_k}$ is computed at the first phase
of the algorithm (equivalent to the prediction phase in Kalman
literature), and $\Sigma_{t_k|y}$ is computed in the second phase
of the algorithm (measurement phase). At the end of the Kalman
iteration, we simply compute the weighted average of the two
matrices to get (\ref{weighted-GIB-kalman}). Finally, we compute
$A_{k+1}$ using (eq. \ref{GIB_A}) by substituting the modified
$\Sigma_{\xi_{k+1}}$.
\end{proof}

\begin{figure}
\begin{center}
  \includegraphics[scale=0.35]{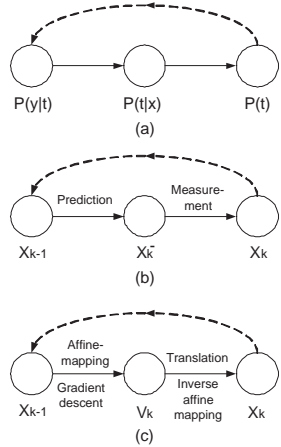}\\
  \caption{Comparison of the schematic operation of the different algorithms. (a) iterative information bottleneck operation (b) Kalman filter operation (c) Affine-scaling operation.}\label{iterative-model}
\end{center}
\end{figure}

There are some differences between the GIB algorithm and Kalman
filter computation. First, the Kalman filter has input
observations $z_k$ in each round. Note that the observations do
not affect the posterior error covariance computation $P_k$ (eq.
\ref{p_k}), but affect the posterior mean $\hat{x}_k$ (eq.
\ref{hat_x_k}).  Second, Kalman filter computes both posterior
mean $\hat{x_k}$ and error covariance $P_k$. The covariance
$\Sigma_{\xi_k}$ computed by the GIB algorithm was shown to be
identical to $P_k$ when $\beta = 1$. The GIB algorithm does not
compute the posterior mean, but computes an additional covariance
$A_k$ (eq. \ref{GIB_A}), which is assumed known in the Kalman
filter.

From the information theoretic perspective, our work extends the
ideas presented in \cite{PredictiveIB}. Predictive information is
defined to be the mutual information between the past and the
future of a time serias. In that sense, by using Theorem 2, Kalman
filter can be thought of as a prediction of the future, which from
the one hand compresses the information about past, and from the
other hand maintains information about the present.

The origins of similarity between the GIB algorithm and Kalman
filter are rooted in the IB iterative algorithm: For computing
(\ref{IB_it1}), we need to compute (\ref{IB_it2},\ref{IB_it3}) in
recursion, and vice~versa.

\section{Relation to the Affine-Scaling Algorithm}
\label{sec_as} One of the most efficient interior point methods
used for linear programming is the Affine-scaling algorithm
\cite{Affine-scaling}. It is known that the Kalman filter is
linked to the Affine-scaling algorithm \cite{KalmanKarmarkar}. In
this work we give an alternate proof, based on different
construction, which shows that Affine-scaling is an instance of
Kalman filter, which is an instance of GIB. This link between
estimation and optimization allows for numerous applications.
Furthermore, by providing a single distribute efficient
implementation of the GIB algorithm, we are able to solve numerous
problems in communication networks.

The linear programming problem in its canonical form is given by:
 \begin{subequations}
    \begin{eqnarray}
   \mbox{minimize} & \vc^T\vx\,, \label{can-linear} \\
    \mbox{subject to} & A\vx = \vb , \ \ \ \ \vx \ge 0\,, \label{can-constraints}
    \end{eqnarray}
 \end{subequations}
where $A \in \mathbb{R}^{n \times p}$ with $\mathbf{rank}\{ A \} =
p < n$. We assume the problem is solvable with an optimal $\vx^*$.
We also assume that the problem is strictly feasible, in other
words there exists $\vx \in \mathbb{R}^n$ that satisfies $A\vx =
\vb$ and $\vx > 0$.

The Affine-scaling algorithm \cite{Affine-scaling} is summarized
below. Assume $\vx_0$ is an interior feasible point to
(\ref{can-constraints}). Let $D = diag(\vx_0)$. The Affine-scaling
is an iterative algorithm which computes a new feasible point that
minimizes the cost function (\ref{can-linear}): \BE \vx_1 = \vx_0
- \frac{\alpha}{\gamma}D^2 \vr\,, \label{af_11}  \EE where $0 <
\alpha < 1$ is the step size, $\vr$ is the step direction.
\begin{subequations}
    \begin{eqnarray}
   \vr &=& (\vc - A^T\vw)\,, \label{af_9b} \\
   \vw &=& (AD^2A^T)^{-1}AD^2\vc\,, \label{af_9c} \\
   \gamma &=& \max_i(\ve_i P D \vc)\,. \label{af_9d}
    \end{eqnarray}
 \end{subequations}
where $\ve_i$ is the $i^{th}$ unit vector and $P$ is a projection matrix
 given by: \BE P = I - DA^T(AD^2A^T)^{-1}AD\,. \label{af_10} \EE
The algorithm continues in rounds and is guaranteed to find an
optimal solution in at most $n$ rounds. In a nutshell, in each
iteration, the Affine-scaling algorithm first performs an
Affine-scaling with respect to the current solution point $\vx_i$
and obtains the direction of descent by projecting the gradient of
the transformed cost function on the null space of the constraints
set. The new solution is obtained by translating the current
solution along the direction found and then mapping the result
back into the original space \cite{KalmanKarmarkar}. This has
interesting analogy for the two phases of the Kalman filter.

\begin{thm}
The Affine-scaling algorithm iteration is an instance of the
Kalman filter algorithm iteration.
\end{thm}
\begin{proof}
We start by expanding the Affine-scaling update rule:\\
 \[ \begin{array}{c}
 \vx_1 = \overbrace{\vx_0 -
\frac{\alpha}{\gamma}D^2\vr}^{(\ref{af_11})} = \vx_0 -
\frac{\alpha}{\underbrace{\max_i \ve_i P D
\vc}_{(\ref{af_9d})}}D^2\vr =
= \vx_0 - \frac{\alpha}{\max_i \ve_i \underbrace{(I -
DA^T(AD^2A^T)AD)}_{(\ref{af_10})} D \vc}D^2\vr =\\ %
= \vx_0 - \frac{\alpha D^2 \overbrace{(\vc - A^T
\vw)}^{(\ref{af_9b})}}{\max_i \ve_i (I - DA^T(AD^2A^T)^{-1}AD) D \vc} = 
\vx_0 - \frac{\alpha D^2 (\vc - A^T
\overbrace{(AD^2A^T)^{-1}AD^2\vc}^{(\ref{af_9c})})}{\max_i \ve_i (I
- DA^T(AD^2A^T)AD)^{-1} D \vc} =\\
=\vx_0 - \frac{\alpha D (I - DA^T
(AD^2A^T)^{-1}AD )D \vc}{\max_i \ve_i (I - DA^T(AD^2A^T)^{-1}AD) D
\vc}\,.
\\
\end{array}
\]
Looking at the numerator and using the Schur complement formula
(\ref{Schur}) with the following notations: $A \triangleq
(AD^2A^T)^{-1}, B \triangleq AD, C \triangleq DA^T, D \triangleq
I$ we get the following matrix:
$ \left(%
\begin{array}{cc}
  AD^2A^T & AD \\
  DA^T & I \\
\end{array}%
\right) $. Again, the upper left block is a Schur complement $A
\triangleq 0, B \triangleq AD, C \triangleq DA^T, D \triangleq I$
of the following matrix:
$ \left(%
\begin{array}{cc}
  0 & AD \\
  DA^T & I \\
\end{array}%
\right) $. In total we get a $3 \times 3$ block matrix of the
form:
$ \left(%
\begin{array}{ccc}
  0 & AD & 0 \\
  DA^T & I & AD \\
  0 & DA^T & I \\
\end{array}%
\right)$.\\
Note that the divisor is a scalar that affects the scaling of the
step size.

Using Theorem 1, we get a computation of Kalman filter with the
following parameters: $A,H \triangleq AD, Q \triangleq I, R
\triangleq I, P_0 \triangleq 0$. This has an
interesting interpretation in the context of Kalman filter: both
prediction and measurement transformation are identical and equal
$AD$. The noise variance of both transformations are Gaussian
variables with prior $\sim \mathcal{N}(0, I)$.
\end{proof}

We have shown how to express the Kalman filter,
Gaussian information bottleneck and Affine-scaling algorithms as a
two step MMSE computation. Each step involves inverting a $2
\times 2$ block matrix. The MMSE computation can be done
efficiently and distributively using the Gaussian belief
propagation algorithm. 

\chapter{Linear Programming}
\label{chap:LP}
In recent years, considerable attention has been dedicated to the
relation between belief propagation message passing and linear
programming schemes. This relation is natural since the maximum
a-posteriori (MAP) inference problem can be translated into
integer linear programming (ILP)~\cite{LPWeiss}.

Weiss \etal~\cite{LPWeiss} approximate the solution to the ILP
problem by relaxing it to a LP problem using convex variational
methods. In~\cite{LPWeiss2}, tree-reweighted belief propagation
(BP) is used to find the global minimum of a convex approximation
to the free energy. Both of these works apply discrete forms of
BP. Globerson \etal~\cite{Globerson,Globerson2} assume convexity
of the problem and modify the BP update rules using
dual-coordinate ascent algorithm. Hazan \etal~\cite{Hazan}
describe an algorithm for solving a general convex free energy
minimization. In both cases the algorithm is guaranteed to
converge to the global minimum as the problem is tailored to be
convex.

This chapter takes a different path. Unlike most of the
previous work, which uses gradient-descent methods, we show how to
use interior-point methods, which are shown to have strong
advantages over gradient and steepest descent methods. (For a
comparative study see \cite[$\S 9.5$,p. 496]{BV04}.) The main
benefit of using interior point methods is their rapid
convergence, which is quadratic once we are close enough to the
optimal solution. Their main drawback is that they require heavier
computational effort for forming and inverting the Hessian matrix
needed for computing the Newton step. To overcome this, we propose
the use of Gaussian BP (GaBP)~\cite{ISIT1,Allerton}, which is a
variant of BP applicable when the underlying distribution is
Gaussian. Using GaBP, we are able to reduce the time associated
with the Hessian inversion task, from $O(n^{2.5})$ to
$O(np{\log(\epsilon)}/{\log(\gamma)})$ at the worst case, where
$p<n$ is the size of the constraint matrix $\mA$, $\epsilon$ is
the desired accuracy, and $1/2 < \gamma < 1$ is a parameter
characterizing the matrix $\mA$. This computational savings is
accomplished by exploiting the sparsity of the Hessian matrix.

An additional benefit of our GaBP-based approach is that the
polynomial-complexity LP solver can be implemented in a
distributed manner, enabling efficient solution of large-scale
problems.


\section{Standard Linear Programming}
\label{linear-programming} Consider the standard linear program
\begin{subequations}
\begin{eqnarray}
 \mbox{minimize}_{\vx} &  {\vc^T\vx }\,, \label{can-linear} \\
\mbox{subject to} & \mA\vx = \vb,\ \ \ \ \vx\ge 0\,,
\end{eqnarray}
\end{subequations}
where $\mA \in \mathbb{R}^{n \times p}$ with
$\mathbf{rank}\{\mA\} = p < n$. We assume the problem is solvable
with an optimal $\vx^*$ assignment. We also assume that the
problem is strictly feasible, or in other words there exists $\vx
\in \mathbb{R}^n$ that satisfies $\mA\vx = \vb$ and $\vx > 0$.

Using the log-barrier method \cite[$\S 11.2$]{BV04}, one gets
\begin{subequations}
\begin{eqnarray}
 \mbox{minimize}_{\vx, \mu} & \vc^T\vx - \mu \Sigma_{k=1}^{n}
\log x_k\,, \label{log-barrier}\\
 \mbox{subject to} & \mA\vx = \vb.
 \label{log-barrier-constraints}
\end{eqnarray}
\end{subequations}

 This is an approximation to the original problem
 (\ref{can-linear}). The quality of the approximation improves as
 the parameter $\mu \rightarrow 0$.

\begin{table*}[htb!]
\begin{center}
\normalsize \caption{The Newton algorithm \cite[$\S 9.5.2$]{BV04}
.}
\begin{tabular}{lcl}
  \hline
  Given & &  feasible starting point $\vx_0$ and tolerance $\epsilon > 0$, $k = 1$  \\
  \hline
  Repeat & 1 & Compute the Newton step and decrement \\
  & & $\Delta \vx = f''(\vx)^{-1} f'(\vx), \ \ \ \ \ \lambda^2 = f'(\vx)^T
  \Delta \vx$ \\
   & 2 & Stopping criterion. quit if $ \lambda^2/2 \le \epsilon$\\
   & 3 & Line search. Choose step size t by
backtracking line search. \\
  & 4 & Update. $\vx_k := \vx_{k-1} + t\Delta \vx, \ \ \ k = k + 1$ \\
  \hline
\label{newton-method}
\end{tabular}
\end{center}
\end{table*}

Now we would like to use the Newton method for
solving the log-barrier constrained objective
function~(\ref{log-barrier}), described in Table
\ref{newton-method}. Suppose that we have an initial feasible
point $\vx_0$ for the canonical linear program~(\ref{can-linear}).
We approximate the objective function~(\ref{log-barrier}) around
the current point $\tilde{\vx}$ using a second-order Taylor
expansion \BE \label{taylor} f(\tilde{\vx} + \Delta \vx) \simeq
f(\tilde{\vx}) + f'(\tilde{\vx}) \Delta \vx + 1/2 \Delta \vx^T
f''(\tilde{\vx}) \Delta \vx. \EE Finding the optimal search
direction $\Delta \vx$ yields the computation of the gradient and
compare it to zero \BE \frac{\partial f}{\partial \Delta \vx} =
f'(\tilde{\vx}) + f''(\tilde{\vx}) \Delta \vx = 0,
\label{taylor-gradient} \EE \BE \Delta \vx = -
f''(\tilde{\vx})^{-1} f'(\tilde{\vx}). \label{newton-step} \EE

Denoting the current point $\tilde{\vx}\triangleq(\vx, \mu, \vy)$
and the Newton step $\Delta \vx \triangleq (\vx, \vy, \mu)$, we
compute the gradient
\begin{subequations}
\begin{eqnarray*}
 f'(\vx,\mu,\vy) \equiv ({\partial f(\vx,
\mu, \vy) }/{ \partial \vx}, {\partial f(\vx, \mu, \vy) }/{
\partial \mu},
{\partial f(\vx, \mu, \vy) }/{ \partial \vy})
\end{eqnarray*}
\end{subequations}

The Lagrangian is \BE \mathcal{L}( \vx, \mu, \vy) = \vc^T\vx  -
\mu \Sigma_k \log x_k + \vy^T(\vb - \mA\vx), \EE

\BE \frac{
\partial \mathcal{L}(\vx,\mu,\vy)}{
\partial \vx} = \vc - \mu \mX^{-1} \mathbf{1} - \vy^T\mA = 0
\label{log-bar-der-x-1}, \EE \BE \frac{
\partial^2 \mathcal{L}(\vx,\mu,\vy)}{
\partial \vx}= \mu \mX^{-2} \label{log-bar-der-x-2}, \EE where $\mX\triangleq\textrm{diag}(\vx)$ and $\mathbf{1}$ is the all-one column vector.
Substituting~(\ref{log-bar-der-x-1})-(\ref{log-bar-der-x-2}) into
(\ref{taylor-gradient}), we get \BE \vc - \mu \mX^{-1} \mathbf{1}
- \vy^T\mA + \mu \mX^{-2} \vx = 0\,, \EE \BE \vc - \mu \mX^{-1}
\mathbf{1}  + \vx \mu \mX^{-2} = \vy^T \mA\,
\label{first-order-condition1}, \EE

\BE \frac{
\partial \mathcal{L}(\vx,\mu,\vy)}{ \partial \vy } = \mA\vx = 0\,\label{log-bar-der-y-1}. \EE
Now multiplying (\ref{first-order-condition1}) by $\mA\mX^2$, and
using (\ref{log-bar-der-y-1}) to eliminate $\vx$ we get \BE
\mA\mX^2\mA^T\vy = \mA\mX^2\vc - \mu\mA\mX \mathbf{1}\,. \EE These
normal equations can be recognized as generated from the linear
least-squares problem \BE \min_\vy || \mX\mA^T\vy - \mX\vc -
\mu\mA\mX \mathbf{1} ||_2^2 \label{primal-ls}. \EE Solving for
$\vy$ we can compute the Newton direction $\vx$, taking a step
towards the boundary and compose one iteration of the Newton
algorithm. Next, we will explain how to shift the deterministic LP
problem to the probabilistic domain and solve it distributively
using GaBP.

\section[From LP to Inference]{From LP to Probabilistic Inference} \label{lp-to-gm}
We start from the
least-squares problem~(\ref{primal-ls}), changing notations to \BE
\min_{\vy} ||\mF \vy - \vg ||^2_2\,, \label{ls-fg} \EE where $\mF
\triangleq \mX\mA^T, \vg \triangleq \mX\vc  + \mu \mA\mX
\mathbf{1}$. Now we define a multivariate Gaussian \BE
p(\hat{\vx}) \triangleq p(\vx,\vy) \propto \exp(-1/2(\mF \vy -
\vg)^T\mI(\mF \vy- \vg))\,. \label{gaussian-mult} \EE It is clear
that $\hat{\vy}$, the minimizing solution of (\ref{ls-fg}), is the
MAP estimator of the conditional probability

\BE \label{pi} \hat{\vy} = \argmax_{\vy} p(\vy|\vx) = \nonumber \mathcal{N}((\mF^T\mF)^{-1} \mF^T \vg , (\mF^T\mF)^{-1})\,. \EE


As shown in Chapter  \ref{linear_detection}, the pseudo-inverse solution can be computed efficiently and distributively
by using the GaBP algorithm.

\ignore{ Schur complement of the following covariance matrix \BE
\mC = \left(%
\begin{array}{cc}
  -\mI & \mF \\
  \mF^T & \mathbf{0} \\
\end{array}%
\right) \label{C_cov} \EE }

The formulation~(\ref{gaussian-mult}) allows us to shift the
least-squares problem from an algebraic to a probabilistic domain.
Instead of solving a deterministic vector-matrix linear equation,
we now solve an inference problem in a graphical model describing
a certain Gaussian distribution function. We define the joint covariance matrix \BE \mC \triangleq \left(%
\begin{array}{cc}
  -\mI & \mF \\
  \mF^T & \mathbf{0} \\
\end{array}%
\right)\,, \label{C_cov} \EE and the shift vector $\vb
\triangleq\{\mathbf{0}^{T},\vg^{T}\}^{T}\in\mathbb{R}^{(p+n)\times
1}.$

Given the covariance matrix $\mC$ and the shift vector $\vb$, one
can write explicitly the Gaussian density function, $p(\hat{\vx})$
, and its corresponding graph $\mathcal{G}$ with edge potentials
(`compatibility functions') $\psi_{ij}$ and self-potentials
(`evidence') $\phi_{i}$. These graph potentials are determined
according to the following pairwise factorization of the Gaussian
distribution $p(\vx) \propto
\prod_{i=1}^{n}\phi_{i}(x_{i})\prod_{\{i,j\}}\psi_{ij}(x_{i},x_{j}),$
        resulting in $ \psi_{ij}(x_{i},x_{j})\triangleq \exp(-x_{i}C_{ij}x_{j}),$ and
        $ \phi_{i}(x_{i}) \triangleq
        \exp\big(b_{i}x_{i}-C_{ii}x_{i}^{2}/2\big).$
        The set of edges $\{i,j\}$ corresponds to the set of
        non-zero entries in $\mC$ (\ref{C_cov}). Hence, we would like to calculate the
marginal densities, which must also be Gaussian,
\BE p(x_{i})\sim\mathcal{N}(\mu_{i}=\{\mC^{-1}\vg\}_{i},P_{i}^{-1}=\{\mC^{-1}\}_{ii}),
\nonumber \EE
\BE \forall i > p, \nonumber \EE where $\mu_{i}$ and $P_{i}$ are the
marginal mean and inverse variance (a.k.a. precision),
respectively. Recall that in the GaBP algorithm, the inferred
mean $\mu_{i}$ is identical to the desired solution $\hat{y}$ of
(\ref{pi}). \comment{The move to the probabilistic domain calls
for the utilization of BP as an efficient inference engine. GaBP
is a special case of continuous BP where the underlying
distribution is Gaussian. In~\cite{Allerton,ISIT1} we show how to
derive the GaBP update rules by substituting Gaussian
distributions in the continuous BP equations. The output of this
derivation is update rules that are computed locally by each
node.}

\section[Extended Construction]{Extending the Construction to the Primal-Dual Method}
\label{primal-dual} In the previous section we have shown how to
compute one iteration of the Newton method using GaBP. In this
section we extend the technique for computing the primal-dual
method. This construction is attractive, since the extended
technique has the same computation overhead.

The dual problem (\cite{GHare}) conforming to (\ref{can-linear})
can be computed using the Lagrangian \BE \mathcal{L}(\vx,\vy,\vz)
= \vc^T\vx + \vy^T(\vb - \mA\vx) - \vz^T\vx , \ \ \  \vz \ge 0,
\nonumber \EE
\begin{subequations}
\begin{eqnarray}
\label{lagrange1}
  g(\vy, \vz) = \inf_{x} \mathcal{\mathcal{L}}(\vx, \vy,
\vz), \\
  \mbox{subject to} \ \ \ \ \mA\vx = \vb, \vx \ge 0.
\end{eqnarray}
\end{subequations}
while
 \BE \frac{\partial \mathcal{L}(\vx,\vy,\vz) }{\partial \vx} =
\vc - \mA^T\vy - \vz = 0 \label{lagrange2}. \EE

Substituting (\ref{lagrange2}) into (\ref{lagrange1}) we get
\begin{subequations}
\begin{eqnarray*}
  \mbox{maximize}_{\vy}  & \vb^T \vy \label{dual2} & \\
  \mbox{subject to} & \mA^T\vy + \vz = \vc , & \vz \ge 0.
\end{eqnarray*}
\end{subequations}

Primal optimality is obtained using (\ref{log-bar-der-x-1}) \cite{GHare}
\BE \vy^T \mA = \vc - \mu \mX^{-1}\mathbf{1}. \label{opt-primal} \EE
Substituting (\ref{opt-primal}) in (\ref{dual2}) we get the connection
between the primal and dual
\[ \mu \mX^{-1} \mathbf{1}= \vz. \]
In total, we have a primal-dual system (again we assume that the
solution is strictly feasible, namely $\vx > 0, \vz > 0$)
\begin{subequations}
\begin{eqnarray*}
  \mA\vx = \vb, & \vx > 0, \\
  \mA^T \vy + \vz = \vc, & \vz > 0, \\
  \mX\vz = \mu \mathbf{1}.  &  \\
\end{eqnarray*}
\end{subequations}
The solution $[\vx(\mu), \vy(\mu), \vz(\mu)]$ of these equations
constitutes the central path of solutions to the logarithmic
barrier method \cite[11.2.2]{BV04}. Applying the Newton method to
this system of equations we get 
\small
\BE \label{newton-3-3-matrix}
\left(%
\begin{array}{ccc}
  0 & \mA^T & I \\
  \mA & 0 & 0 \\
  \mZ & 0 & \mX \\
\end{array}%
\right)
\left(%
\begin{array}{c}
  \Delta \vx \\
  \Delta \vy \\
  \Delta \vz \\
\end{array}%
\right) =
\left(%
\begin{array}{c}
  \vb - \mA\vx \\
  \vc - \mA^T\vy - \vz \\
  \mu \mathbf{1} - \mX\vz \\
\end{array}%
\right). \EE 
\normalsize
The solution can be computed explicitly by

\BE
\begin{array}{cl}
  \Delta \vy = & (\mA\mZ^{-1}\mX\mA^T)^{-1}  (\mA\mZ^{-1}\mX(\vc - \mu \mX^{-1}\mathbf{1} -\mA^T\vy) + \vb - \mA\vx), \\
  \Delta \vx =& \mX\mZ^{-1}(\mA^T \Delta \vy + \mu \mX^{-1}\mathbf{1} = \vc
+ \mA^T\vy), \\
  \Delta \vz = & -\mA^T\Delta \vy + \vc - \mA^T\vy - \vz.  \\
\end{array}
\nonumber \EE The main computational overhead in this method is the
computation of $(\mA\mZ^{-1}\mX\mA^T)^{-1}$, which is derived from
the Newton step in (\ref{newton-step}).

 Now we would like to use GaBP for computing the solution.
 We make the following simple change to (\ref{newton-3-3-matrix})
 to make it symmetric: since $\vz > 0$, we can multiply the third
 row by $\mZ^{-1}$ and get a modified symmetric system
\small
\[ \label{newton-3-3-matrix-mod}
\left(%
\begin{array}{ccc}
  0 & \mA^T & I \\
  \mA & 0 & 0 \\
  I & 0 & \mZ^{-1}\mX \\
\end{array}%
\right)
\left(%
\begin{array}{c}
  \Delta \vx \\
  \Delta \vy \\
  \Delta \vz \\
\end{array}%
\right) =
\left(%
\begin{array}{c}
  \vb - \mA\vx \\
  \vc - \mA^T\vy - \vz \\
  \mu \mZ^{-1} \mathbf{1} - \mX \\
\end{array}%
\right). \]
\normalsize
Defining $ \tilde{\mA} \triangleq \left(%
\begin{array}{ccc}
  0 & \mA^T & I \\
  \mA & 0 & 0 \\
  I & 0 & \mZ^{-1}\mX \\
\end{array}%
\right),$ and $\tilde{\vb} \triangleq \left(%
\begin{array}{c}
  \vb - \mA\vx \\
  \vc - \mA^T\vy - \vz \\
  \mu \mZ^{-1} \mathbf{1} - \mX \\
\end{array}%
\right). $ one can use the GaBP algorithm.

In general, by looking at (\ref{taylor-gradient}) we see that the
solution of each Newton step involves inverting the Hessian matrix
$f''(\vx)$. The state-of-the-art approach in practical
implementations of the Newton step is first computing the Hessian
inverse $f''(\vx)^{-1}$ by using a (sparse) decomposition method
like (sparse) Cholesky decomposition, and then multiplying the
result by $f'(\vx)$. In our approach, the GaBP algorithm computes
directly the result $\Delta \vx$, without computing the full
matrix inverse. Furthermore, if the GaBP algorithm converges, the
computation of $\Delta \vx$ is guaranteed to be accurate.

\subsection{Applications to Interior-Point Methods}
We would like to compare the running time of our proposed method
to the Newton interior-point method, utilizing our new convergence
results of the previous section. As a reference we take the
Karmarkar algorithm~\cite{Karmarkar} which is known to be an
instance of the Newton method \cite{Karmarkar-newton}. Its running
time is composed of $n$ rounds, where on each round one Newton
step is computed. The cost of computing one Newton step on a dense
Hessian matrix is $O(n^{2.5})$, so the total running time is
$O(n^{3.5})$.

Using our approach, the total number of Newton iterations, $n$,
remains the same as in the Karmarkar algorithm. However, we
exploit the special structure of the Hessian matrix, which is both
symmetric and sparse. Assuming that the size of the constraint
matrix $\mA$ is $n \times p, \ \ \ p < n$, each iteration of GaBP
for computing a single Newton step takes $O(np)$, and based on the convergence analysis in Section \ref{chap:conv_rate}, for a desired accuracy $\epsilon||\vb||_{\infty}$
we need to iterate for $r = \ceil{{
\log(\epsilon)}/{\log(\gamma)}}$ rounds, where $\gamma$ is defined
in (\ref{gamma}). The total computational burden for a single
Newton step is $O(np{ \log(\epsilon)}/{\log(\gamma)})$. There are
at most $n$ rounds, hence in total we get $O(n^2p{
\log(\epsilon)}/{\log(\gamma)})$. 

%

\begin{figure}
\begin{center}
  \includegraphics[scale=0.45, bb=49 234 554 585]{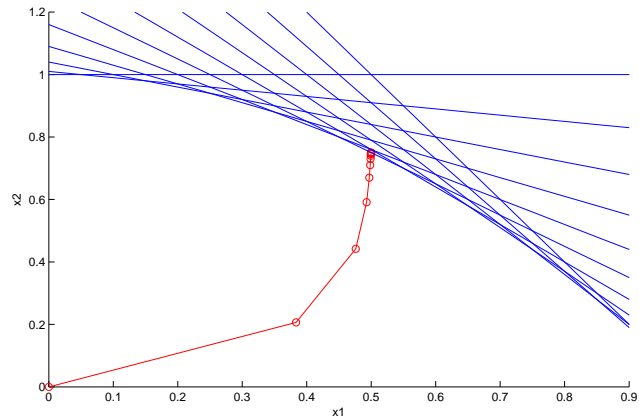}\\
  \caption{A simple example of using GaBP for solving linear programming with two variables and eleven constraints. Each red circle shows one iteration of the Newton method.}\label{toy-example}
\end{center}
\end{figure}

\section[NUM]{Case Study: Network Utility\ Maximization}

We consider a network that supports a set of flows,
each of which has a nonnegative flow rate, and an associated
utility function.
Each flow passes over a route, which is a subset of the edges of
the network. Each edge has a given capacity, which is the maximum
total traffic (the sum of the flow rates through it) it can support.
The network utility maximization (NUM) problem is to choose the
flow rates to maximize the total utility, while respecting the
edge capacity constraints \cite{Sri:04,Ber:98}.
We consider the case where all utility functions are concave, in which
case the NUM problem is a convex optimization problem.

A standard technique for solving NUM problems is based on
dual decomposition \cite{DaW:60,Shor:85}.
This approach yields fully decentralized algorithms, that can scale
to very large networks.
Dual decomposition was first applied to the NUM problem
in \cite{Kelly:97},
and has led to an extensive body of research on distributed algorithms
for network optimization \cite{Low:99,CLCD:07,PalChiang:06}
and new ways to interpret existing network protocols \cite{Low:03}.

Recent work by Zymnis \etal~\cite{NUM}, presents a specialized primal-dual interior-point
method for the NUM problem. Each Newton step is computed using the
preconditioned conjugate gradient method (PCG). This proposed method had a significant
performance improvement over the dual decomposition approach, especially when the
network is congested. Furthermore, the method can handle utility functions that
are not strictly concave. The main drawback of the primal-dual method is that it is centralized,
while the dual decomposition methods are easily distributed.

Next, we compare the performance of the GaBP solver proposed in this chapter, to the truncated Newton method and dual decomposition approaches. We provide the first comparison of performance of
the GaBP algorithm vs. the PCG method. The PCG method is a state-of-the-art method used extensively in
large-scale optimization applications. Examples include $\ell_1$-regularized
logistic regression \cite{KKB:07}, gate sizing \cite{joshi:eml}, and slack allocation
\cite{joshi:eml2}.
Empirically, the GaBP algorithm is immune to numerical
problems with typically occur in the PCG method, while demonstrating a faster convergence. The only previous work comparing the performance of GaBP vs. PCG we are aware of is \cite{Sthesis}, which used a small example of $25$ nodes,
and the work of \cite{BibDB:Weiss01Correctness} which used a grid of $25 \times 25$ nodes.

We believe that our approach is general and not limited to the NUM problem. It could potentially
be used for the solution of other large scale distributed optimization problems.

\subsection{NUM\ Problem Formulation}
\label{s-probform}

There are $n$ flows in a network, each of which is associated with
a fixed route, \ie, some subset of $m$ links.
Each flow has a nonnegative
\emph{rate}, which we denote $f_1, \ldots, f_n$.
With the flow $j$ we associate a utility function $U_j:\reals \rightarrow
\reals$, which is concave and twice differentiable, with $\dom U_j
\subseteq \reals_+$.
The utility derived by a flow rate $f_j$ is given by $U_j(f_j)$.
The total utility associated with all the flows is then
$U(f)=U_1(f_1)+\cdots +U_n(f_n)$.

The total traffic on a link in the network
is the sum of the rates of all flows that utilize that link.
We can express the link traffic compactly using the
\emph{routing} or \emph{link-route} matrix $R \in\reals^{m\times n}$,
defined as
\[
R_{ij} = \left\{\begin{array}{ll}
1 & \mbox{flow $j$'s route passes over link $i$}\\
0 & \mbox{otherwise}.
\end{array}\right.
\]
Each link in the network has a (positive)
\emph{capacity} $c_1, \ldots, c_m$.
The traffic on a link cannot exceed its capacity, \ie, we have
$Rf \leq c$, where $\leq$ is used for componentwise inequality.

The NUM problem is to choose the
rates to maximize total utility, subject to the link capacity
and the nonnegativity constraints:
\BEQ\label{e-rate_control}
\begin{array}{ll}
\mbox{maximize} & U(f)\\
\mbox{subject to} & Rf \leq c, \quad f \geq 0,
\end{array}
\EEQ
with variable $f \in \reals^n$.
This is a convex optimization problem and can be solved by a
variety of methods. We say that
$f$ is {\em primal feasible} if it satisfies $Rf\leq c$, $f\geq 0$.

The dual of problem (\ref{e-rate_control}) is
\BEQ\label{e-rate_control_dual}
\begin{array}{ll}
\mbox{minimize} & \lambda^Tc + \sum_{j=1}^n (-U_j)^*(-r_j^T\lambda)\\
\mbox{subject to} & \lambda \geq 0,
\end{array}
\EEQ
where $\lambda \in \reals^m_+$ is the dual variable associated with
the capacity constraint of problem (\ref{e-rate_control}),
$r_j$ is the $j$th column of $R$ and $(-U_j)^*$ is the
conjugate of the negative $j$th utility function \cite[\S 3.3]{BoV:04},
\[
(-U_j)^*(a) = \sup_{x \geq 0} (ax+U_j(x)).
\]
We say that $\lambda$ is {\em dual feasible} if it satisfies
$\lambda \geq 0$ and $\lambda \in \cap_{j=1}^n \dom (-U_j)^*$.

\subsection{Previous Work}
\label{s-prev}
In this section we give a brief overview of the dual-decomposition
method and the primal-dual interior point method proposed in~\cite{NUM}.
%

%


Dual decomposition \cite{DaW:60,Shor:85,Kelly:97,Low:99}
is a projected (sub)gradient algorithm
for solving problem (\ref{e-rate_control_dual}), in the case
when all utility functions are strictly concave.
We start with any positive $\lambda$, and repeatedly
carry out the update
\begin{eqnarray*}
f_j &:=& \argmax_{x\geq 0}
\left(U_j(x)- x (r_j^T\lambda) \right), \quad j=1, \ldots, n,\\
\lambda &:=& \left(\lambda-\alpha\left(c-Rf\right)\right)_+,
\end{eqnarray*}
where $\alpha>0$ is the step size, and $x_+$ denotes the entrywise
nonnegative part of the vector $x$.
It can be shown
that for small enough $\alpha$, $f$ and $\lambda$ will converge
to $f^\star$ and $\lambda^\star$, respectively, provided all $U_j$ are
differentiable and strictly concave.
The term $s=c-Rf$ appearing in the update is the \emph{slack} in
the link capacity constraints (and can have negative entries during
the algorithm execution).
It can be shown that the slack is exactly the gradient
of the dual objective function.

Dual decomposition is a distributed algorithm. Each flow is
updated based on information obtained from the links it
passes over, and each link dual variable is updated based
only on the flows that pass over it.


The primal-dual interior-point method is based on using a Newton step,
applied to a suitably modified form of the optimality conditions.
The modification is parameterized by a parameter $t$, which is adjusted
during the algorithm based on progress, as measured by the actual
duality gap (if it is available) or a surrogate duality gap (when the
actual duality gap is not available).

We first describe the search direction.
We modify the complementary slackness conditions to obtain
the modified optimality conditions
\begin{eqnarray*}
-\nabla U(f) + R^T \lambda-\mu &=& 0 \\
\diag(\lambda)s &=& (1/t)\ones \\
\diag(\mu)f &=& (1/t)\ones\,,
\end{eqnarray*}
where $t>0$ is a parameter that sets the accuracy of the approximation.
(As $t \rightarrow \infty$, we recover the optimality conditions for
the NUM problem.)
Here we implicitly assume that $f,s,\lambda,\mu > 0$.
The modified optimality conditions can be compactly written
as $r_t(f,\lambda,\mu) = 0$, where
\[
r_t(f,\lambda,\mu) = \left[\begin{array}{c}
-\nabla U(f) + R^T \lambda - \mu \\
\diag(\lambda)s - (1/t)\ones\\
\diag(\mu)f -(1/t)\ones
\end{array}\right].
\]

The primal-dual search direction is the
Newton step for solving the nonlinear equations
$r_t(f,\lambda,\mu) = 0$. If $y = (f,\lambda,\mu)$
denotes the current point, the Newton step
$\Delta y = (\Delta f,\Delta \lambda,\Delta \mu)$
is characterized by the linear equations
\[
r_t(y+\Delta y) \approx r_t(y)+ r'_t(y)\Delta y = 0\,,
\]
which, written out in more detail, are
\small
\begin{equation}\label{e-primdualstep}
\left[\begin{array}{ccc}
-\nabla^2U(f)     &   R^T  & -I   \\
-\diag(\lambda)R &  \diag(s) & 0  \\
\diag(\mu) & 0 & \diag(f)
\end{array}\right]
\left[\begin{array}{c}
\Delta f \\ \Delta \lambda \\ \Delta \mu
\end{array}\right]
= -r_t(f,\lambda,\mu)\,.
\end{equation}
\normalsize

During the algorithm, the parameter $t$ is increased,
as the primal and dual variables approach optimality.
When we have easy access to a dual feasible point during the algorithm,
we can make use of the exact duality gap $\eta$ to set the value of $t$;
in other cases, we can use the surrogate duality gap $\hat \eta$.

The primal-dual interior point algorithm is given in \cite[\S 11.7]{BoV:04},
\cite{Wri:97}.

The most expensive part of computing the primal-dual
search direction is solving equation (\ref{e-primdualstep}).
For problems of modest size,
\ie, with $m$ and $n$ no more than $10^4$,
it can be solved using direct methods such as a
sparse Cholesky decomposition.

For larger problem instances \cite{NUM} proposes to solve (\ref{e-primdualstep})
\emph{approximately}, using a preconditioned conjugate gradient (PCG)
algorithm \cite[\S 6.6]{Dem:97}, \cite[chap. 2]{CTK:95}, \cite[chap. 5]{NoW:99}.
When an iterative method is used to approximately solve
a Newton system, the algorithm
is referred to as an {\em inexact}, \emph{iterative},
or {\em approximate} Newton method
(see \cite[chap. 6]{CTK:95} and its references).
When an iterative method is used inside a primal-dual
interior-point method, the overall algorithm is called a \emph{truncated-Newton primal-dual
interior-point method}. For details of the PCG algorithm,
we refer the reader to the references cited above.
Each iteration requires multiplication of the matrix by a vector,
and a few vector inner products.
\subsection{Experimental Results}
\label{s-exp-res}

In our first example we look at the performance of our method
on a small network. 
The utility functions are all logarithmic, \ie,
$U_j(f_j) = \log f_j$.
There are $n=10^3$ flows, and $m=2\cdot 10^3$ links.
The elements of $R$ are chosen randomly and independently,
so that the average route length is $10$ links.
The link capacities $c_i$ are chosen independently from a
uniform distribution on $[0.1,1]$. For this particular example, there are about $10^4$
nonzero elements in $R$ ($0.5\%$ density).

We compare three different algorithms for solving the NUM problem:
The dual-decomposition method, a truncated Newton method via PCG and a customized Newton method via the GaBP solver.
Out of the examined algorithms, the Newton method is centralized, while the dual-decomposition and GaBP solver are distributed algorithms. The source code of our Matlab simulation is available on \cite{MatlabGABP}.
\begin{figure}[ht!]
\centering{
  \includegraphics[width=250pt,clip]{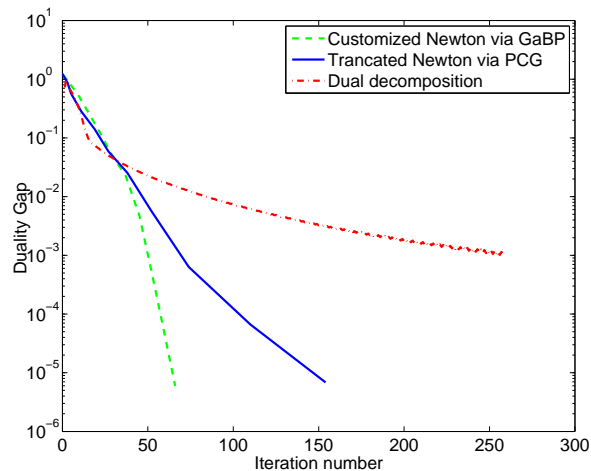}\\
  \caption{Convergence rate using the small settings.}\label{small_fig}
}
\end{figure}

Figure \ref{small_fig} depicts the solution quality, where the X-axis represents the number of algorithm
iterations, and the Y-axis is the surrogate duality gap (using a logarithmic scale).
As clearly shown, the GaBP algorithm has a comparable performance to the sparse Cholesky decomposition,
while it is a distributed algorithm. The dual decomposition method has much slower convergence.

Our second example is too large to be solved using
the primal-dual interior-point method with direct search direction
computation, but is readily handled
by the truncated-Newton primal-dual algorithm using PCG,
the dual decomposition method and the customized Newton method via GaBP.
The utility functions are all logarithmic: $U_j(f_j) = \log f_j$.
There are $n=10^4$ flows, and $m=2\cdot 10^4$ links.
The elements of $R$ and $c$ are chosen as for the small example.
For dual decomposition, we initialized all $\lambda_i$ as $1$.
For the interior-point method, we initialized all $\lambda_i$ and $\mu_i$
as $1$. We initialize all $f_j$ as $\gamma$, where we choose $\gamma$
so that $Rf \leq 0.9 c$.

\begin{figure}[ht!]
\centering{
  \includegraphics[width=250pt,clip]{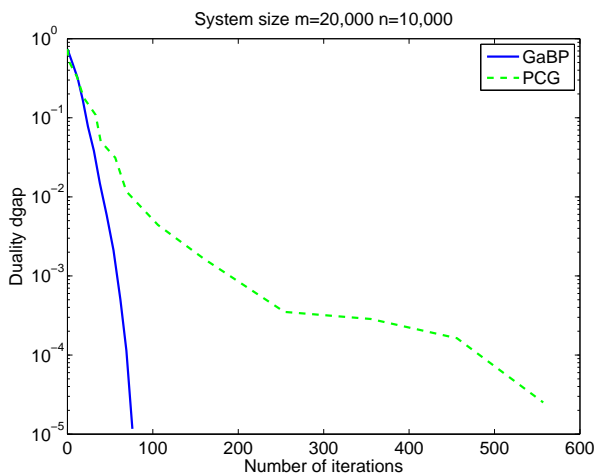}\\
  \caption{Convergence rate in the larger settings.}\label{larger_fig}
}
\end{figure}

Our experimental results shows, that
as the system size grows larger, the GaBP solver has favorable performance.
Figure \ref{larger_fig} plots the duality gap of both algorithms, vs. the number of iterations performed.

\begin{figure}[ht!]
\centering{
  \includegraphics[width=250pt,clip]{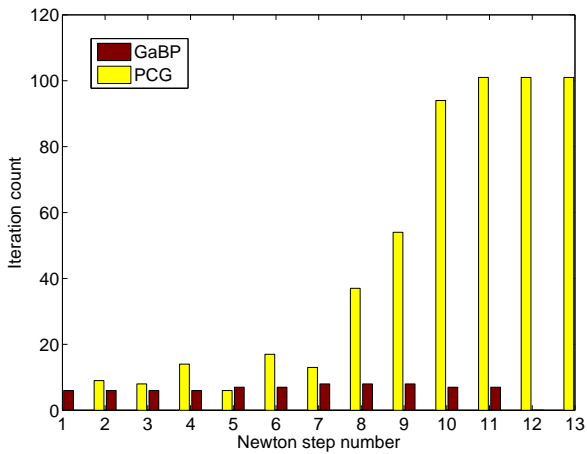}\\
  \caption{Iteration count per Newton step.}\label{gabp_vs_pcg}
}
\end{figure}

Figure \ref{gabp_vs_pcg} shows that in terms of Newton steps, both methods had comparable performance. The Newton method via the GaBP algorithm converged in 11 steps, to an accuracy of $10^{-4}$ where the truncated Newton method implemented via PCG converged in 13 steps to the same accuracy. However, when examining the iteration count in each Newton step (the Y-axis) we see that the GaBP remained constant, while the PCG iterations significantly increase as we are getting closer to the optimal point.

We have experimented with larger settings, up to $n=10^5$ flows, and $m=2\cdot 10^5$ links.
The GaBP algorithm converged in 11 Newton steps with 7-9 inner iteration in each Newton step.
The PCG method converged in 16 Newton steps with an average of 45 inner iterations.

Overall, we have observed three types of numerical problems with
the PCG method. First, the PCG Matlab implementation runs into numerical problems and failed to compute
the search direction. Second, the line search failed, which means that no progress is possible in the computed direction without violating the problem constraints. Third, when getting close to the optimal solution, the number of PCG iterations significantly increases.

The numerical problems of the PCG algorithm are well known, see of example \cite{PCG1,PCG2}. In contrary, the GaBP algorithm did not suffer from the above numerical problems.

Furthermore, the PCG is harder to distribute, since in each PCG iteration a vector dot product and a matrix product
are performed. Those operations are global, unlike the GaBP which exploits the sparseness of the input matrix.


We believe that the NUM problem serves as a case study for demonstrating the superior performance of the GaBP algorithm
in solving sparse systems of linear equations. Since the problem of solving a system of linear equations is a fundamental problem in computer science and engineering, we envision many other applications for our proposed method.

%
%
%

\chapter{Relation to Other Algorithms}\label{chap5}\label{chap:other_algos}
In this chapter we discuss the relation between the GaBP algorithm
and several other algorithms, and show that all of them are
instances of the GaBP algorithm. Thus, a
single efficient implementation of GaBP can efficiently
compute different cost functions without the need of deriving new
update rules. Furthermore, it is easier to find sufficient
conditions for convergence in our settings.

\section[Montanari's MUD]{Montanari's Linear Detection Algorithm}
\label{extended_mont}

In this section we show that the Montanari's algorithm~\cite{BibDB:MontanariEtAl}, which is
 an iterative algorithm for linear detection, is an instance of Gaussian BP.
 A reader that is not familiar with Montanari's algorithm or linear detection is referred to Chapter \ref{linear_detection}. Our improved algorithm for linear detection described in Section~\ref{improved} is more general. First, we allow different noise level for each received bit,
unlike his work that uses a single fixed noise for the whole
system. In practice, the bits are transmitted using different
frequencies, thus suffering from different noise levels. Second,
the update rules in his paper are fitted only to the
randomly-spreading CDMA codes, where the matrix $A$ contains only
values which are drawn uniformly from $\{-1,1\}$. Assuming binary
signalling, he conjectures convergence to the large system limit.
Our new convergence proof holds for any CDMA matrices provided
that the absolute sum of the chip sequences is one, under weaker
conditions on the noise level. Third, we propose
in~\cite{Allerton} an efficient broadcast version for saving
messages in a broadcast supporting network.

The probability distribution of the factor graph used by Montanari
is:
\[ d\mu_y^{N,K} = \frac{1}{Z_y^{N,K}} \prod_{a=1}^{N}
\exp(-\frac{1}{2} \sigma^2 \omega_a^2 + j y_a \omega_a)
\prod_{i=1}^{K} \exp(-\frac{1}{2}x_i^2)  \cdot \prod_{i,a} \exp(
-\frac{j}{\sqrt{N}}s_{ai}\omega_a x_i) d\omega \]

Extracting the self and edge potentials from the above probability
distribution:
\[ \psi_{ii}(\vx_i) \triangleq \exp(-\frac{1}{2}\vx_i^2) \propto \mathcal{N}(\vx; 0, 1) \]
\[ \psi_{aa}(\omega_a) \triangleq \exp(-\frac{1}{2}\sigma^2 \omega^2_a +
j\vy_a \omega_a) \propto \mathcal{N}(\omega_a ; j \vy_a, \sigma^2)
\]
\[ \psi_{ia}(\vx_i, \omega_a) \triangleq \exp(
-\frac{j}{\sqrt{N}}s_{ai}\omega_a \vx_i) \propto \mathcal{N}(\vx;
\frac{j}{\sqrt{N}}s_{ai}, 0) \]

For convenience, Table~\ref{tab:montanari_vs_gabp} provides a translation between the
notations used in ~\cite{Allerton} and that
used by Montanari {\it et al.} in ~\cite{BibDB:MontanariEtAl}:

\begin{table}[h!]
\begin{center}
\caption{Summary of notations}
\label{tab:montanari_vs_gabp}
\begin{tabular}{|c|c|c|}
  \hline
  GaBP~\cite{Allerton} & Montanari {\it el al.}~\cite{BibDB:MontanariEtAl} & Description\\
  \hline
  $P_{ij}$ & $\lambda_{i \rightarrow a}^{(t+1)}$ & precision msg from left to right \\
  & $\hat{\lambda}_{a \rightarrow i}^{(t+1)}$ & precision msg from right to left \\
  $\mu_{ij}$ & $\gamma_{i \rightarrow a}^{(t+1)}$ & mean msg from left to right \\
  & $\hat{\gamma}_{a \rightarrow i}^{(t+1)}$ & mean msg from right to left  \\
  $\mu_{ii}$ & $y_i$ & prior mean of left node\\
  & $0$ & prior mean of right node\\
  $P_{ii}$ & $1$ & prior precision of left node\\
  $\Psi_{i}$ & $\sigma^2$ & prior precision of right node \\
  $\mu_i$ & $\frac{G_i}{L_i}$ & posterior mean of node \\
  $P_{i}$ & $L_i$ & posterior precision of node \\
  $A_{ij}$ & $\frac{-j s_{ia}}{\sqrt{N}}$ & covariance \\
  $A_{ji}$ & $\frac{-j s_{ai}}{\sqrt{N}}$ & covariance \\
      & $j$ & $j=\sqrt{-1}$ \\
  \hline
\end{tabular}
\label{tab_mont1}
\end{center}
\end{table}

\begin{thm}
Montanari's update rules are special case of the GaBP algorithm.
\end{thm}

\begin{proof}
Now we derive Montanari's update rules. We start with the
precision message from left to right:
\[ \overbrace{\lambda_{i \rightarrow a}^{(t+1)}}^{P_{ij}} = 1 + \frac{1}{N} \Sigma_{b
\ne a} \frac{s_{ib}^2}{\hat{\lambda}^{(t)}_{b \rightarrow i}} =
 \overbrace{1}^{P_{ii}} + \Sigma_{b \ne a} \overbrace{\frac{1}{N}
\frac{s_{ib}^2}{\hat{\lambda}^{(t)}_{b \rightarrow i}}}^{P_{ki}} \]
\[ = \overbrace{1}^{P_{ii}} - \Sigma_{b \ne a} \overbrace{\frac{- j s_{ib}}{
\sqrt{N}}}^{-A_{ij}}
 \overbrace{\frac{1}{\hat{\lambda}^{(t)}_{b \rightarrow i}}}^{(P_{j \backslash i})^{-1}} \overbrace{\frac{- j s_{ib}}{ \sqrt{N}}}^{A_{ji}} . \]
By looking at Table \ref{tab_mont1}, it is easy to verify that this precision
update rule is equivalent to \ref{eq_prec}.

Using the same logic we get the precision message from right to
left:
\[ \overbrace{\hat{\lambda}_{i \rightarrow a}^{(t+1)}}^{P_{ji}} = \overbrace{\sigma^2}^{P_{ii}} + \overbrace{\frac{1}{N}
\Sigma_{k \ne i} \frac{s_{ka}^2}{\lambda^{(t)}_{k \rightarrow
a}}}^{- A_{ij}^2 P_{j \backslash i}^{-1}} .
\]

The mean message from left to right is given by
\[
\gamma_{i \rightarrow a}^{(t+1)} = \frac{1}{N} \Sigma_{b \ne a}
\frac{s_{ib}}{\lambda_{b \rightarrow i}^{(t)}}
\hat{\gamma}^{(t)}_{b \rightarrow i} =
 \overbrace{0}^{\mu_{ii}} -
 \Sigma_{b \ne a} \overbrace{\frac{- j s_{ib}}{
\sqrt{N}}}^{-A_{ij}} \overbrace{\frac{1}{\hat{\lambda}_{b
\rightarrow i}^{(t)}}}^{P_{j \backslash i}^{-1}}
\overbrace{\hat{\gamma}^{(t)}_{b \rightarrow i}}^{\mu_{j
\backslash i}} .
\]
The same calculation is done for the mean from right to left:
\[
\hat{\gamma}_{i \rightarrow a}^{(t+1)} = y_a - \frac{1}{N}
\Sigma_{k \ne i} \frac{s_{ka}}{\lambda_{k \rightarrow a}^{(t)}}
\gamma^{(t)}_{k \rightarrow a}.
\]
Finally, the left nodes calculated the precision and mean by
\[
G_{i}^{(t+1)} = \frac{1}{\sqrt{N}} \Sigma_{b}
\frac{s_{ib}}{\lambda_{b \rightarrow i}^{(t)}}
\hat{\gamma}^{(t)}_{b \rightarrow i} \mbox{  ,  }  J_{i} =
G_{i}^{-1}.
\]
\[ L_i^{(t+1)} =1 + \frac{1}{N} \Sigma_{b
} \frac{s_{ib}^2}{\lambda^{(t)}_{b \rightarrow i}} \mbox{   , }
 \mu_i = L_i G_i^{-1} . \]
\end{proof}

The key difference between the two constructions is that Montanari
uses a directed factor graph while we use an undirected graphical
model. As a consequence, our construction provides additional
convergence results and simpler update rules.

\section[Frey's IPP Algorithm]{Frey's Local Probability Propagation Algorithm} \label{sec:frey}
Frey's local probability propagation~\cite{LPP} was published in 1998,
before the work of Weiss on Gaussian Belief propagation. That is
why it is interesting to find the relations between those two
works. Frey's work deals with the factor analysis learning
problem. The factor analyzer network is a two layer densely
connected network that models bottom layer sensory inputs as a
linear combination of top layer factors plus independent gaussian
sensor noise.

In the factor analysis model, the underlying distribution is
Gaussian. There are $n$ sensors that sense information generated from $k$ factors. The prior distribution of the sensors is \begin{align*} p(z) =
\mathcal{N}(z;0,I) &\,,& p(x|z) = \mathcal{N}(z; Sx, \Psi)\,. \end{align*}

The matrix $S$ defines the linear relation between the factors and the sensors.
Given $S$, the observation $y$ and the noise level $\Psi$, the goal is to infer the most probable factor states $x$.
It is
shown that maximum likelihood estimation of $\mS$ and $\Psi$
performs factor analysis.

The marginal distribution over $\vx$ is:
\[ p(x) \propto \int_{\vx} \mathcal{N}(\vx;0,I) \mathcal{N}(\vz;\mS\vx, \Psi) d\vx = \mathcal{N}(\vz; 0, \mS^T\mS + \Psi). \]
This distribution is Gaussian as well, with the following
parameters:
\[ E(\vz|\vx) = (\mS^T\mS + \Psi)^{-1}\mS^T\vy, \]
\[ Cov(\vz|\vx) = (\mS^T\mS + \Psi)^{-1}. \]

\begin{table}
\caption{Notations of GaBP (Weiss) and Frey's algorithm}
\begin{center}
\begin{tabular}{|c|c|l|}
  \hline
  ~\cite{Allerton} & Frey & comments\\
  \hline
  $P_{ij}^{-1} $ & $-\phi_{nk}^{(i)}$ & precision message from left $i$ to right $a$ \\
  & $\nu_{kn}^{(i)} $ & precision message from right $a$ to right $i$ \\
  $\mu_{ij}$ & $\mu_{nk}^{(i)}$ & mean message from left $i$ to right $a$ \\
  & $\eta_{kn}^{(i)}$ & mean message from right $a$ to left $i$ \\
  $\mu_{ii}$ & $x_n$ & prior mean of left node $i$\\
  & $0$ & prior mean of right node $a$\\
  $P_{ii}$ & $1$ & prior precision of left node $i$\\
  & $\psi_n $ & prior precision of right node $i$\\
  $\mu_i$ & $\upsilon_{k}^{(i)} $ & posterior mean of node $i$\\
  $P_{i}$ & $\hat{z}_k^{(i)}$ & posterior precision of node $i$ \\
  $A_{ij}$ & $\lambda_{nk}$ & covariance of left node $i$ and right node $a$\\
  $A_{ji}$ & $1$ & covariance of right node $a$ and left node $i$ \\
  \hline
\end{tabular}
\end{center}
\end{table}

\begin{thm}
Frey's iterative propagability propagation is an instance of GaBP.
\end{thm}

\begin{proof}
We start by showing that Frey's update rules are equivalent to the
GaBP update rules. From right to left:
\[ \overbrace{\phi_{nk}^{(i)}}^{-P_{ij}^{-1}} = \frac{\Psi_n + \Sigma_k
\lambda_{nk}^2\upsilon_{kn}^{(i-1)}}{\lambda^2_{nk}} -
\upsilon_{kn}^{(i-1)} = \frac{ \overbrace{\Psi_n + \Sigma_{k \ne
j} \upsilon_{jn}}^{P_{i \backslash
j}}}{\underbrace{\lambda^2_{nk}}_{A_{ij}^2}}
, \]
\[ \overbrace{\mu_{nk}^{(i)}}^{\mu_{ij}} = \frac{x_n - \Sigma_k \lambda_{nk}
\eta_{kn}^{(i-1)} }{ \lambda_{nk}} + \eta_{kn}^{(i-1)} =
\frac{\overbrace{x_n}^{\mu_{ii} P_{ii}} - \overbrace{\Sigma_{j \ne
k} \eta_{kn}^{(i-1)}}^{\Sigma_{j \ne k} -P_{ki} \mu_{ki}}
}{\underbrace{\lambda_{nk}}_{A_{ij}}}.
\]
And from left to right:
\[ \overbrace{\eta_{kn}^{(i)}}^{P_{ji}^{-1}} =
 1/(1/(1/(1+ \Sigma_n 1/\psi_{nk}^{(i)} - 1/\psi_{nk}^{(i)}))) =
\overbrace{1}^{A_{ji}^2}/(\overbrace{1}^{P_{ii}} +
\overbrace{\Sigma_{j \ne k} 1/\psi_{nk}^{(i)}}^{\Sigma_{j \ne k}
-P_{ij}}), \]

\[ \nu_{kn}^{(i)} = \eta_{kn}^{(i)}(\Sigma_n
\mu_{nk}^{(i)}/\psi_{nk}^{(i)} - \mu_{nk}^{(i)}/\psi_{nk}^{(i)}) =
\eta_{kn}^{(i)}(\Sigma_{j \ne k}\mu_{nk}^{(i)}/\psi_{nk}^{(i)} ) =
\]
\[
\overbrace{1}^{A_{ij}}\overbrace{\eta_{kn}^{(i)}}^{P_{ij}^{-1}}(
\overbrace{0}^{\mu_{ii} P_{ii} } + \Sigma_{j \ne
k}\overbrace{\mu_{nk}^{(i)}/\psi_{nk}^{(i)}}^{\mu_{ji} P_{ji}} ).
\]
It is interesting to note, that in Frey's model the graph is
directed. That is why the edges $A_{ji} = 1$ in the update rules
from right to left.
\end{proof}

\section[Consensus Propagation]{Moallami and Van-Roy's Consensus Propagation}\label{CP} The consensus propagation (CP)~\cite{CP} is a
distributed algorithm for calculating the average value in the
network. Given an adjacency graph matrix $Q$ and a vector 
input values $y$, the goal is to compute the average value $\bar{y}$. 

It is clear that the CP algorithm is related to the BP algorithm. However the relation to GaBP was not given.  In this section we
derive the update rules used in CP from the GaBP algorithm
(Weiss).

The self potentials of the nodes are
\[ \psi_{ii}(x_i) \propto  \exp(-(x_i - y_i)^2) \]
The edge potentials are:
\[ \psi_{ij}(x_i,x_j) \propto \exp(-\beta Q_{ij}(x_i - x_j)^2) \]
In total, the probability distribution $p(x)$
\[ p(x) \propto \Pi_i \exp(-(x_i - y_i))^2 \Pi_{e \in E} \exp(-\beta Q_{ij}(x_i - x_j)^2) \]
\[ = \exp(\Sigma_i(-(x_i - y_i))^2 -\beta \Sigma_{e \in E}Q_{ij}(x_i -
x_j)^2) \] We want to find an assignment $x^* = max_x p(x)$. It is
shown in~\cite{CP} that this assignment conforms to the mean value
in the network: $\overline{y} = \frac{1}{n} \Sigma_i y_i$ when
$\beta$ is very large.

For simplicity of notations, we list the different notations in
Table~\ref{tab:weiss_vs_cp}.
\begin{table}
\caption{Notations of GaBP (Weiss) and Consensus Propagation}
\label{tab:weiss_vs_cp}
\begin{center}
\begin{tabular}{|c|c|l|}
  \hline
  Weiss & CP & comments\\
  \hline
  $P_0$ & $\tilde{{\cal K}}$ & precision of $\psi_{ii}(x_i) \prod_{x_k \in N(x_i) \backslash x_j} m_{ki}(x_i)$\\
  $\mu_0$ & $\mu({\cal K})$ & mean of $\psi_{ii}(x_i) \prod_{x_k \in N(x_i) \backslash x_j} m_{ki}(x_i)$\\
  $P_{ij}$ & ${\cal F}_{ij}({\cal K})$ & precision message from $i$ to $j$ \\
  $\mu_{ij}$ & ${\cal G}_{ij}({\cal K})$ & mean message from $i$ to $j$ \\
  $\mu_{ii}$ & $y_i$ & prior mean of node $i$\\
  $P_{ii}$ & $1$ & prior precision of node $i$\\
  $\mu_i$ & $\overline{y}$ & posterior mean of node $i$ (identical for all nodes) \\
  $P_{i}$ & $P_{ii}$ & posterior precision of node $i$ \\
  $b$ & $Q_{ji}$ & covariance of nodes $i$ and $j$\\
  $b'$ & $Q_{ij} $& covariance of nodes $i$ and $j$ \\
  $a$ & $0$ & variance of node $i$ in the pairwise covariance matrix $V_{ij}$ \\
  $c$ & $0$ & variance of node $j$ in the pairwise covariance matrix $V_{ij}$ \\
  \hline
\end{tabular}
\end{center}
\end{table}

\begin{thm}
The consensus propagation algorithm is an instance of the Gaussian
belief propagation algorithm.
\end{thm}

\begin{proof}
We prove this theorem by substituting the self and edge potentials
used in the CP paper in the belief propagation update rules,
deriving the update rules of the CP algorithm.
\[ m_{ij}(x_j) \propto \int_{x_i}  \overbrace{\exp( -(x_i - \mu_{ii})^2)}^{\psi_{ii}(x_i)} \overbrace{\exp(-\beta Q_{ik}(x_i - x_j)^2)}^{\psi_{ij}(x_i,x_j)} \overbrace{\exp(-\Sigma_{k} P_{ki}(x_i -
\mu_{ki})^2)dx_i }^{\prod m_{ki}(x_i)} =
\]

\[ \int_{x_i} \exp(-x_i^2 + 2 x_i
\mu_{ii} - \mu_{ii}^2 - \beta Q_{ik}x_i^2 + 2 \beta Q_{ij}x_i x_j -
\beta Q_{ij} x_j^2 -\Sigma_{k}P_{ki} x_i^2 + \Sigma_{k} P_{ki}
\mu_{ki} - \Sigma_{k} P_{ki}^2 \mu_{ki}^2) dx_i =
\]
\begin{equation}
 \exp( -\mu_{ii}^2 - \beta Q_{ij} x_j^2 + \Sigma_k {P_{ij}^2 m_{ij}^2})
\int_{x_i} \overbrace{\exp((1 + \beta Q_{ij} + \Sigma_k P_{ki} ) x_i ^
2}^{ax^2} + \overbrace{2 (\beta Q_{ij} x_j + P_{ki} \mu_{ki} +
\mu_{ii} ) x_i }^{bx} dx_i \label{eq5}
\end{equation}
Now we use the following integration rule:
\[ \int_x \exp(-(ax^2 + bx)) dx = \sqrt{\pi/a}
\exp(\frac{b^2}{4a}) \] We compute the integral:
\[ \int_{x_i} \overbrace{\exp(1 + \beta Q_{ij} + \Sigma_k P_{ki} ) x_i
^ 2}^{ax^2} + \overbrace{2(\beta Q_{ij} x_j + P_{ki} \mu_{ki} +
\mu_{ii} ) x_i }^{bx} dx_i = \]

\[ = \sqrt{\pi/a} \exp( \frac{\overbrace{\cancel{4} (\beta Q_{ij} x_j
+ P_{ki} \mu_{ki} + \mu_{ii} )^2}^{b^2}}{\underbrace{\cancel{4}(1
+ \beta Q_{ki} + \sum_k P_{kl})}_{4a}}) \propto  \exp( \frac{\beta
^2Q_{ij}^2 x_j ^2 + 2 \beta Q_{ij} (P_{ki} \mu_{ki} + \mu_{ii})
x_j + (\mu_{ki} + \mu_{ii})^2}{1 + \beta Q_{ki} + \Sigma_k
P_{ki}})
\]
Substituting the result back into (\ref{eq5}) we get:
\begin{equation}
m_{ij}(x_j) \propto  \exp( -\mu_{ii}^2 - \beta Q_{ij} x_j^2 +
\Sigma_k {P_{ij}^2 \mu_{ij}^2}) \exp( \frac{\beta ^2Q_{ij}^2 x_j
^2 + 2 \beta Q_{ij} (P_{ki} \mu_{ki} + \mu_{ii}) x_j + (\mu_{ki} +
\mu_{ii})^2}{1 + \beta Q_{ki} + \Sigma_k P_{ki}}) \label{eq6}
\end{equation}
For computing the precision, we take all the terms of $x_j^2$ from
(\ref{eq6}) and we get:
\[ P_{ij}(x_j)  \propto \exp (-\beta Q_{ij}  +  \frac{ \beta^2 Q_{ij}^2
}{1+ \beta Q_{ij} + \Sigma P_{ki} }) = \]
\[ \frac{-\beta Q_{ij} (1+ \beta Q_{ij} + \Sigma P_{ki}) + \beta^2 Q_{ij}^2
}{1+ \beta Q_{ij} + \Sigma P_{ki} }
  = \]
\[ \frac{-\beta Q_{ij} - \cancel{\beta^2 Q_{ij}^2} - \beta Q_{ij} \Sigma P_{ki} + \cancel{\beta^2
Q_{ij}^2}}{1+ \beta Q_{ij} + \Sigma P_{ki} }
  = \]
\[ = -\frac{1 + \Sigma P_{ki}}{\beta Q_{ij} + 1 + \Sigma P_{ki}} \]
The same is done for computing the mean, taking all the terms of
$x_j$ from equation (\ref{eq6}):
\[ \mu_{ij}(x_j) \propto \frac{ \beta Q_{ij} (\mu_{ii} + \Sigma  P_{ki} \mu_{ki})}{1+ \beta Q_{ij} + \Sigma P_{ki}
} = \frac{\mu_{ii} + \Sigma  P_{ki} \mu_{ki}}{ 1 + \frac{1 +
\Sigma P_{ki}}{ \beta Q_{ij}}} \]
\end{proof}

\section[Quadratic Min-Sum]{Quadratic Min-Sum Message Passing Algorithm  \mbox{          }}
\label{MinSum}
The quadratic Min-Sum message passing algorithm was initially presented in \cite{MinSum}.
It is a variant of the max-product algorithm, with underlying Gaussian distributions.
The quadratic Min-Sum algorithm is an iterative algorithm for solving a quadratic cost function. Not surprisingly, as we have shown in Section \ref{MaxProductRule} that the Max-Product and the Sum-Product algorithms are identical when the underlying distributions are Gaussians. In this chapter, we show that the quadratic Min-Sum algorithm
is identical to the GaBP algorithm, although it was derived differently.

In \cite{MinSum} the authors discuss the application for solving linear system of equations
using the Min-Sum algorithm. Our work \cite{Allerton} was done in parallel to their work, and
both papers appeared in the 45th Allerton 2007 conference.
\begin{thm}
The Quadratic Min-Sum algorithm is an instance of the GaBP algorithm.
\end{thm}
\begin{proof}
We start in the quadratic parameter updates:
\[ \gamma_{ij} = \frac{1}{1 - \Sigma_{u \in N(i) \backslash j}\Gamma^2_{ui} \gamma_{ui}} =
\overbrace{(\overbrace{1}^{A_{ii}} - \Sigma_{u \in N(i) \backslash j}\overbrace{\Gamma_{ui}}^{A_{ui}} \overbrace{\gamma_{ui}}^{P_{ui}^{-1}} \overbrace{\Gamma_{iu}}^{A_{iu}})^{-1}}^{P_{i \backslash j}^{-1}}
\]
Which is equivalent to \ref{eq_prec}.
Regarding the mean parameters,
\[ z_{ij} = \frac{\Gamma_{ij} }{1 - \Sigma_{u \in N(i) \backslash j}\Gamma^2_{ui} \gamma_{ui} } (h_i - \Sigma_{u \in N(i) \backslash j} z_{ui}) = \overbrace{ \overbrace{\Gamma_{ij}}^{A_{ij}} \overbrace{\gamma_{ij}}^{(P_{i \backslash j})^{-1}} (\overbrace{h_i}^{b_i} - \Sigma_{u \in N(i) \backslash j} z_{ui})}^{\mu_{i \backslash j}} \]
Which is equivalent to \ref{eq_mean}.
\end{proof}

For simplicity of notations, we list the different notations in
Table~\ref{tab:minsum_vs_gabp}.
\begin{table}[h!]
\caption{Notations of Min-Sum \cite{MinSum} vs. GaBP}
\label{tab:minsum_vs_gabp}
\begin{center}
\begin{tabular}{|c|c|l|}
  \hline
  Min-Sum~\cite{MinSum} & GaBP~\cite{Allerton} & comments\\
  \hline
  $\gamma_{ij}^{(t+1)}$ & $P_{i \backslash j}^{-1} $ & quadratic parameters / product rule precision from $i$ to $j$ \\
  $z_{ij}^{(t+1)}$ & $\mu_{i \backslash j}$ & linear parameters / product rule mean rom $i$ to $j$ \\
  $h_i$ & $b_i$ & prior mean of node $i$\\
  $A_{ii}$ & $1$ & prior precision of node $i$\\
  $x_i$ & $x_i$ & posterior mean of node $i$ \\
  $-$ & $P_{i}$ & posterior precision of node $i$ \\
  $\Gamma_{ij}$ & $A_{ij}$ & covariance of nodes $i$ and $j$ \\
  \hline
\end{tabular}
\end{center}
\end{table}
As shown in~Table~\ref{tab:minsum_vs_gabp}, the Min-Sum algorithm assumes the covariance matrix $\Gamma$ is first normalized s.t. the main
diagonal entries (the variances) are all one.
The messages sent in the Min-Sum algorithm are called linear parameters (which are equivalent to the mean messages in GaBP) and quadratic parameters (which are equivalent to variances). The difference between the algorithm is that in the GaBP algorithm, a node computes the product rule and the integral, and sends the result to its neighbor. In the Min-Sum algorithm, a node computes the product rule, sends the intermediate result, and
the receiving node computes the integral. In other words, the same computation is performed but on different
locations. In the Min-Sum algorithm terminology, the messages are linear and quadratic parameters vs.
Gaussians in our terminology.

\chapter{Appendices}

\section[Weiss vs. Johnson]{Equivalence of Weiss and Johnson Formulations}
One of the confusing aspects of learning GaBP is that each paper is using its own model as well as its own notations. For completeness, 
we show equivalence of notations in Weiss' vs. Johnsons papers. In~\cite{JMLR} it is shown that one of the possible ways of
converting the information form to pairwise potentials form is
when the inverse covariance matrices used are of the type:
\[ V_{ij} = \left(
\begin{array}{cc}
  0 & J_{ij} \\
  J_{ji} & 0 \\
\end{array}%
\right)\,, \] where the terms $J_{ij} = J_{ji}$ are the entries of
the inverse covariance matrix $J$ in the information form. First we list
the equivalent notations in the two papers:\\
\begin{center}
\begin{tabular}{|c|c|l|}
  \hline
  Weiss & Johnson & comments\\
  \hline
  $P_0$ & $\hat{J}_{i \backslash j}$ & precision of $\psi_{ii}(x_i)
\prod_{x_k \in N(x_i) \backslash x_j} m_{ki}(x_i)$\\
  $\mu_0$ & $\hat{h}_{i \backslash j}$ & mean of $\psi_{ii}(x_i)
\prod_{x_k \in N(x_i) \backslash x_j} m_{ki}(x_i)$\\
  $P_{ij}$ & $\Delta J_{i \rightarrow j}$ & precision message from $i$ to $j$ \\
  $\mu_{ij}$ & $\Delta h_{i \rightarrow j}$ & mean message from $i$ to $j$ \\
  $\mu_{ii}$ & $h_i$ & prior mean of node $i$\\
  $P_{ii}$ & $J_{ii}$ & prior precision of node $i$\\
  $P_{i}$ & $(P_{ii})^{-1}$ & posterior precision of node $i$\\
  $\mu_i$ & $\mu_i$ & posterior mean of node $i$ \\
  $b$ & $J_{ji}$ & covariance of nodes $i$ and $j$\\
  $b'$ & $J_{ij} $& covariance of nodes $i$ and $j$ \\
  $a$ & $0$ & variance of node $i$ in the pairwise covariance matrix $V_{ij}$ \\
  $c$ & $0$ & variance of node $j$ in the pairwise covariance matrix $V_{ij}$ \\
  \hline
\end{tabular}
\end{center}
Using this fact we can derive again the BP equations for the scalar case (above are
the same equation using Weiss' notation).

\ignore{ In the Gaussian graphical models we send messages
$m_{ij}(x_i)$ of two fields signifying the mean and variance of
the computed results. The notation $m_{ij}(x_i)$ means that node
$i$ is sending a message to node $j$. We will further note $\Delta
J$ and $\Delta \mu$ as the message two scalar values. Note that
where $J_{ij} = 0$ no message is sent between node $i$ and $j$,
assuming the matrix $J$ is sparse the algorithm sends only
messages between neighbors where their covariance is not zero.

First we consider a simpler version of the algorithm in case all
the random variables are independent:
\begin{equation}
\label{BP_rule1}
 \Delta J_{i \rightarrow j} = J_{ii} + \Sigma_{k \in N(i) \backslash j} \Delta J_{k \rightarrow i}
\end{equation}
\begin{equation}
\label{BP_rule2}
 \Delta \mu_{i \rightarrow j} = \frac{J_{ii}\mu_i + \Sigma_{k \in N(i) \backslash j} \Delta
 J_{k \rightarrow i}
\Delta \mu_{k \rightarrow i}}{\Delta J_{i \rightarrow j}}
\end{equation}
Note that~\ref{BP_rule1} can be interpreted as the size of the
set, while~\ref{BP_rule2} can be interpreted as the weighted
average of the set (weighted sum of all subsets divided by the
total weight).

 Finally, the marginal probabilities are calculated:
\[ P_{ii}^{-1} = J_{ii} + \Sigma_{k \in N(i)} \Delta J_{k \rightarrow i} \]
\[ \mu_{i} = \mu_i + \Sigma_{k \in N(i)} \Delta \mu_{k \rightarrow i} \]

As shown in ~\ref{calc} this conforms exactly to multiplication of
several Gaussians, which makes sense since the random variables
are independent then $p(x_1, ..., x_k) = \Pi p(x_i)$. }

\begin{equation}
\label{GMRF_BP1} \overbrace{h_{i \backslash j}}^{\mu_0} =
\overbrace{h_i}^{\mu_{ii}} + \sum_{k \in N(i) \backslash j}
\overbrace{\Delta h_{k \rightarrow i}}^{\mu_{ki}} \mbox{ , }
\overbrace{J_{i \backslash j}}^{P_0} = \overbrace{J_{ii}}^{P_{ii}}
+ \sum_{k \in N(i) \backslash j} \overbrace{\Delta J_{k
\rightarrow i}}^{P_{ki}},
\end{equation}
\begin{equation}
\label{GMRF_BP2} \overbrace{\Delta h_{i \rightarrow j}}^{\mu_{ij}}
= -\overbrace{J_{ji}}^{b}\overbrace{(J_{i \backslash j})^{-1}}^{(a
+ P_{0})^{-1}}\overbrace{h_{i \backslash j}}^{\mu_{0}} \mbox{ , }
\overbrace{\Delta J_{i \rightarrow j}}^{P_{ij}} =
\overbrace{0}^{c} - \overbrace{J_{ji}}^{b}\overbrace{( 0 + J_{i
\backslash j})^{-1}}^{(a + P_0)^{-1}}\overbrace{J_{ij}}^{b'}.
\end{equation}

Finally:
\[ \hat{h_i} = h_{ii} + \sum_{k \in N(i)}
\Delta h_{k \rightarrow i} \mbox{   ,   } \hat{J}_{i} = J_{ii} +
\sum_{k \in N(i)} \Delta J_{k \rightarrow i},
\]
\[ \mu_i = \hat{J}_i^{-1}h_i \mbox{   ,   } P_{ii} = \hat{J}_i^{-1}. \]
\\

\section{GaBP code in Matlab}
Latest code appears on the web on: \cite{MatlabGABP}.
\subsection{The file gabp.m}
\small
\begin{verbatim}
% Implementation of the Gaussian BP algorithm, as given in:
% Linear Detection via Belief Propagation
% By Danny Bickson, Danny Dolev, Ori Shental, Paul H. Siegel and Jack K. Wolf.
% In the 45th Annual Allerton Conference on Communication, Control and Computing,
% Allerton House, Illinois, Sept. 07'
%
%
% Written by Danny Bickson.
% updated: 24-Aug-2008
%
% input: A - square matrix nxn
% b - vector nx1
% max_iter - maximal number of iterations
% epsilon - convergence threshold
% output: x - vector of size nx1, which is the solution to linear systems
              of equations A x = b
%         Pf - vector of size nx1, which is an approximation to the main
%         diagonal of inv(A)
function [x,Pf] = gabp(A, b, max_iter, epsilon)

% Stage 1 - initialize
P = diag(diag(A));
U = diag(b./diag(A));
n = length(A);

% Stage 2 - iterate
for l=1:max_iter
  % record last round messages for convergence detection
     old_U = U;

     for i=1:n
       for j=1:n
         % Compute P i\j - line 2
         if (i~=j && A(i,j) ~= 0)
             p_i_minus_j = sum(P(:,i)) - P(j,i); %
             assert(p_i_minus_j ~= 0);
             %iterate - line 3
             P(i,j) = -A(i,j) * A(j,i) / p_i_minus_j;
             % Compute U i\j - line 2
             h_i_minus_j = (sum(P(:,i).*U(:,i)) - P(j,i)*U(j,i)) / p_i_minus_j;
             %iterate - line 3
             U(i,j) = - A(i,j) * h_i_minus_j / P(i,j);
         end
       end
 end



 % Stage 3 - convergence detection
 if (sum(sum((U - old_U).^2)) < epsilon)
     disp(['GABP converged in round ', num2str(l)]);
     break;
 end

end % iterate

% Stage 4 - infer
Pf = zeros(1,n);
x = zeros(1,n);
for i = 1:n
   Pf(i) = sum(P(:,i));
   x(i) = sum(U(:,i).*P(:,i))./Pf(i);
end


end
\end{verbatim}
\subsection{The file run\_gabp.m}
\begin{verbatim}
% example for running the gabp algorithm, for computing the solution to Ax = b
% Written by Danny Bickson

% Initialize

%format long;
n = 3; A = [1 0.3 0.1;0.3 1 0.1;0.1 0.1 1]; 
b = [1 1 1]'; 
x = inv(A)*b; 
max_iter = 20; 
epsilon = 0.000001;

[x1, p] = gabp(A, b, max_iter, epsilon);

disp('x computed by gabp is: ');

x1

disp('x computed by matrix inversion is : ');

x'

disp('diag(inv(A)) computed by gabp is: (this is an
approximation!) ');

p

disp('diag(inv(A)) computed by matrix inverse is: ');

diag(inv(A))'
\end{verbatim}


\bibliographystyle{IEEEtran}   
\footnotesize
\bibliography{IEEEabrv,GaBP_J_v6,phd-thesis,CDMA,LDviaBP_Allerton07,kalman,lp,ISIT09-1,ISIT09-3,ls_num}       



\end{document}